\documentclass[a4paper,11pt]{article}
\usepackage{jheppub}
\usepackage[utf8x]{inputenc}
\usepackage[T1]{fontenc}
\usepackage{braket}
\usepackage{color}
\usepackage{amsfonts}
\usepackage{amssymb}
\usepackage{amsmath}
\usepackage{graphicx}
\usepackage{epstopdf}
\usepackage{enumitem}
\usepackage{comment}
\usepackage{mathtools}
\usepackage{ragged2e}
\hypersetup{
	colorlinks=true,
	linkcolor=blue,
	filecolor=cyan, 
	urlcolor=blue,
	citecolor=red,
		pdftitle={PEE threads and bit threads in gravitational subregions of AdS},
	pdfauthor={Debarshi Basu and Qiang Wen},
} 
\usepackage[titletoc,toc,title]{appendix}
\numberwithin{equation}{section}
\usepackage[nameinlink]{cleveref}
\usepackage{float}

\makeatletter
\newcommand{\vast}{\bBigg@{3}}
\newcommand{\Vast}{\bBigg@{5}}
\makeatother

\newcommand{\zbar}{\raisebox{0.2ex}{--}\kern-0.6em Z}
\newcommand{\dd}{\mathop{}\!\mathrm{d}}
\DeclareMathOperator{\sech}{sech}
\DeclareMathOperator{\csch}{csch}

\def\CA{{\cal A}}

\def\CE{{\cal E}}

\def\CI{{\cal I}}

\def\CL{{\cal L}}

\def\CM{{\cal M}}
\def\CN{{\cal N}}
\def\CO{{\cal O}}

\def\CQ{{\cal Q}}

\def\CW{{\cal W}}

\def\BR{\mathbb{R}}

\def\d{\textrm{d}}
\def\del{\partial}

\title{PEE threads and bit threads in gravitational subregions of AdS}

\author[a]{Debarshi Basu}
\author[a]{ Qiang Wen}

\affiliation[a]{Shing-Tung Yau Center and School of Physics, Southeast University, Nanjing 210096, China}

\emailAdd{debarshi.128@gmail.com, wenqiang@seu.edu.cn}

\abstract{
	Motivated by the kinematic-space description of gravitational subregions \cite{Basu:2026hbg}, we allow partial entanglement entropy (PEE) threads to be sourced not only from the asymptotic boundary, but also from boundaries of subregions. Based on this setup, we develop a framework to construct configurations for PEE threads and bit threads sourced from surfaces in the AdS bulk. The situations we have analyzed include the Poincar\'e AdS$_3$ and  the planar BTZ black brane with an end-of-the-world (EOW) brane, and the entanglement wedges of boundary multi-intervals. We construct novel configurations of PEE threads and bit threads in these situations, which shed new light in our understanding of the AdS$_3$/BCFT$_2$ correspondence and the holography defined in the entanglement wedge motivated by surface/state correspondence. A notable feature we found is that, although the elementary PEE threads are geodesics, the integral curves of the coarse-grained bit-thread current are generically not geodesics. Across the explicit constructions, the integrated source currents reduce to endpoint distance-difference potentials, which make divergencelessness, the norm bound and bottleneck saturation geometrically transparent. Our results clarify how boundary, brane, horizon and RT-surface degrees of freedom participate in holographic entanglement and provide a concrete link between PEE, bit threads, subregion kinematic space and the surface/state correspondence.
}

\begin{document} 
	\maketitle
	\flushbottom

	\section{Introduction}
	\label{sec:introduction}
	
	Quantum entanglement provides one of the sharpest available probes of the
	emergence of spacetime in holographic quantum gravity.  The AdS/CFT
	correspondence identifies a gravitational theory in an asymptotically AdS
	spacetime with a non-gravitational quantum field theory on its conformal
	boundary \cite{Maldacena:1997re,Gubser:1998bc,Witten:1998qj}.  The
	Ryu--Takayanagi prescription and its covariant, replica and quantum
	generalizations then associate the entropy of a boundary subregion with a
	distinguished extremal surface in the bulk
	\cite{Ryu:2006bv,Ryu:2006ef,Hubeny:2007xt,Lewkowycz:2013nqa,
		Dong:2016hjy,Faulkner:2013ana,Engelhardt:2014gca}.  Beyond furnishing an
	efficient method for computing entropies, this relation suggests that bulk
	geometry is organized by the entanglement structure of the dual state
	\cite{VanRaamsdonk:2009ar,VanRaamsdonk:2010pw,Rangamani:2016dms,
		Chen:2021lnq}.  This viewpoint is reinforced by differential entropy,
	entanglement-based reconstruction, the derivation of gravitational dynamics
	from entanglement, and tensor-network models of holography
	\cite{Balasubramanian:2013lsa,Headrick:2014eia,Czech:2014wka,
		Czech:2014ppa,Czech:2015qta,Czech:2015kbp,Lashkari:2013koa,
		Faulkner:2013ica,Swingle:2009bg,Haegeman:2011uy,Swingle:2012wq,
		Qi:2013caa,Pastawski:2015qua,Hayden:2016cfa}.
	
	A complementary and particularly geometric reformulation of holographic
	entropy is provided by bit threads \cite{Freedman:2016zud,
		Headrick:2020gyq,Headrick:2022nbe}.  Instead of regarding the RT surface only
	as a minimal cut, one describes the entropy as the maximal flux of a
	divergenceless vector field obeying a local norm bound.  The RT surface is then the bottleneck of a maximal flow. This formulation shifts the emphasis
	from the value of an entropy to the manner in which the corresponding
	entanglement flux is distributed through the bulk.  It makes nesting and
	simultaneous optimization geometrically transparent and gives a continuum
	analogue of cut-counting in tensor networks.
	
	The additional information contained in a bit-thread configuration is,
	however, accompanied by a substantial ambiguity.  For a fixed entangling
	region there are generally infinitely many maximal flows with the same flux.
	The minimal surface determines the bottleneck but does not select a unique
	thread configuration away from it.  This raises a basic physical question:
	is there an intrinsic fine-grained entanglement structure from which a
	preferred, or at least distinguished, class of bit-thread flows can be
	constructed?  Addressing this question requires resolving the entanglement more
	finely than through the single number $S_A$.
	
	The partial entanglement entropy (PEE) and the entanglement contour provide
	such a refinement.  They decompose the entropy of a region into additive
	contributions associated with its constituent degrees of freedom
	\cite{Wen:2018whg,Wen:2019iyq,Wen:2020ech,Han:2019scu,
		Kudler-Flam:2019oru,Wen:2018mev,Abt:2018ywl,Rolph:2021nan,
		Han:2021ycp,Lin:2021hqs,Ageev:2021ipd,Lin:2023orb}. Unlike the total entropy, a two-point PEE keeps
	track of which degrees of freedom are correlated with one another.  It
	therefore contains information that is naturally adapted to a thread
	description.
	
		The geometrization of two-point PEE makes this relation precise.  In a
	holographic CFT, a two-point PEE is represented by the bulk geodesic
	joining the corresponding endpoints, whose density is determined by the
	value of the two-point PEE; these geodesics are called PEE threads
	\cite{Lin:2023rxc}.  A source-labelled vector field describes the density
	and direction of the geodesic PEE-thread bundle emitted from a given site.
	For intervals in AdS$_3$, and for spherical regions in higher dimensions,
	superposing these elementary flows produces a distinguished bit-thread
	configuration.  More generally, the full set of PEE threads forms a
	continuous network that tessellates vacuum AdS and retains the geometric
	information needed to reconstruct bulk areas
	\cite{Lin:2024fze,Wen:2025gui}.  PEE threads therefore connect four
	apparently different descriptions: the additive fine structure of boundary
	entanglement, integral geometry in the bulk, tensor network models, and macroscopic max-flow
	configurations.
	
	This relation is closely tied to kinematic space, which is the
	space of bulk geodesics equipped with its natural invariant measure.  In
	vacuum AdS, the Crofton formula reconstructs the area of a bulk surface by
	counting its intersections with these geodesics
	\cite{Czech:2015qta,Czech:2016xec}. In this case, the PEE threads are just the geodesics in kinematic space, and the PEE structure supplies a physical
	interpretation of the kinematic measure \cite{Lin:2024fze}: geodesics are not merely elements of an abstract integral-geometric ensemble, but carriers of two-point
	entanglement.  The network of PEE threads may thus be viewed as the physical
	realization of the perfect covering of AdS space \cite{Wen:2025gui}.
	
	A recent development extends this connection from the complete vacuum AdS
	slice to its gravitational subregions
	\cite{Basu:2026hbg}.  Given a subregion bounded by a codimension-one
	surface, one may restrict the ambient geodesic network to the geodesic
	segments that cover that subregion.  The resulting subregion kinematic space
	continues to reconstruct internal surfaces through intersection counting.
	Moreover, PEE threads emitted from the bounding surface furnish a continuous
	network within the subregion. By defining tensor network models on this network \cite{Wen:2025gui}, one obtains a concrete realization of
	surface/state ideas \cite{Miyaji:2015yva,Miyaji:2015fia} and of generalized entanglement wedges for gravitational
	regions \cite{Bousso:2022hlz}.  This development supplies an important motivation for the present work.  It suggests that the PEE-thread construction should not be regarded
	as intrinsically tied to the asymptotic AdS boundary.  Rather, any surface
	that supports an effective gravitational state should be capable of acting
	as a source or endpoint of PEE data. The results of \cite{Basu:2026hbg} also serve as the workhorse for constructing the PEE-thread configurations in this paper.
	
    According to the surface/\allowbreak state correspondence
	\cite{Miyaji:2015yva,Miyaji:2015fia}, a convex codimension-two bulk surface is associated
	with a quantum state, while suitable deformations of the surface correspond
	to unitary transformations of that state.  An open or topologically non-trivial
	surface naturally supports a mixed state, and an RT-like prescription
	computes entropies of its subregions.  The proposal is conceptually
	appealing, but by itself it does not specify the detailed entanglement
	structure carried by the surface.  PEE threads provide precisely such
	additional data: they identify which infinitesimal parts of an extended
	surface are correlated and represent those correlations geometrically.
	Bit thread configurations can then be constructed from the coarse-grained flow generated by this microscopic PEE structure.
	
	The combined viewpoint leads to a sharper formulation of subregion
	holography.  Kinematic space specifies which geodesic families cover a
	gravitational region; PEE assigns quantum-information-theoretic weights to
	those geodesics; the surface/state correspondence interprets their endpoints
	as degrees of freedom of the state associated with the bounding surface; and
	bit threads organize the resulting data into norm-bounded maximal flows.
	These structures are complementary rather than redundant.  Kinematic-space
	intersection counting is unoriented and reconstructs areas, whereas
	bit-thread flux is oriented and solves a max-flow problem.  PEE threads
	provide the common microscopic input from which both descriptions can be
	understood.
	
	The goal of this paper is to develop this combined framework in situations
	where the relevant surface has several geometrically and physically distinct
	components.  Such configurations test the idea much more stringently than
	the vacuum AdS boundary, because the sources of PEE are no longer all
	equivalent under a global isometry.  The extended surface may include an
	asymptotic boundary, an end-of-the-world brane, a black-hole horizon or an RT
	surface used as a purifier. Using the scheme of \cite{Basu:2026hbg}, we will construct explicit PEE-thread configurations, and from these we will build explicit and novel bit-thread configurations via the superposition method of \cite{Lin:2023rxc}. The central question is then not simply whether the RT entropy can be reproduced, but what the corresponding PEE-thread network teaches us about the division of entanglement among these components and about the physical meaning of the resulting bit-thread flow.
	
	A natural laboratory to test this hypothesis is a holographic boundary conformal field theory (BCFT).
	BCFTs possess boundary degrees of freedom and a universal boundary entropy
	\cite{Cardy:1984bb,Cardy:1989ir,Affleck:1991tk,
		Callan:1994ub,McAvity:1995zd}.  In the AdS/BCFT framework, the bulk spacetime is truncated
	by an end-of-the-world brane $\CQ$
	\cite{Takayanagi:2011zk,Fujita:2011fp,Karch:2000gx}.  From the
	surface/state perspective, the EOW brane is not merely a wall that imposes a
	boundary condition.  It is part of the extended support of the state,
	\[
	\Sigma_{\rm ext}=\partial \CM\cup \CQ .
	\]
	It should therefore participate in the PEE decomposition alongside the
	asymptotic boundary.  This requires boundary--brane and brane--brane PEE
	sectors in addition to the familiar boundary--boundary sector. The physical usefulness of this extension is phase dependent.  When an RT
	surface ends on the EOW brane, brane-supported PEE channels account for the
	part of the entropy assigned to the brane component.  In a doubly
	holographic completion this component may be interpreted as an island.  When
	the dominant RT surface remains entirely boundary anchored, those same
	channels reorganize so that the physical answer reduces to the ordinary CFT
	result.  The extended PEE structure thus does not simply add a universal
	brane correction.  It resolves how different geometric components
	participate in purification in different homology chambers. Finite temperature introduces another qualitatively new component.  On a
	BTZ exterior slice the natural extended surface contains the horizon in
	addition to the asymptotic boundary and the EOW brane.  The horizon is not a
	second physical BCFT boundary; it is the geometric carrier of the thermal
	purification, in accordance with the thermofield-double interpretation of
	eternal black holes \cite{Maldacena:2001kr}.  The corresponding PEE network
	therefore contains boundary-, brane- and horizon-sourced sectors.  This
	provides a local geometric decomposition of thermal entanglement and
	clarifies how horizon degrees of freedom enter the coarse-grained
	bit-thread flow.

	AdS/BCFT also serves as a natural framework for studying entanglement islands \cite{Penington:2019npb,Almheiri:2019psf,Almheiri:2020cfm}. In this context, the PEE structure of the island phase was analyzed in \cite{Basu:2023wmv}, where it was shown that the entanglement entropy of a boundary region receives contributions from the island degrees of freedom through boundary-brane and brane-brane PEE channels, and that this decomposition captures the transition between the no-island and island saddles. This analysis will guide us in the PEE-thread picture: it tells us which subsets of PEE threads, among those sourced from the boundary, the brane, and the horizon, actually contribute to the entanglement entropy and to the bit-thread current in each homology phase.

	A second major application concerns mixed-state correlations.  The
	entanglement wedge cross section (EWCS) has been proposed as the holographic dual of a plethora of mixed state entanglement and correlation measures including
	the entanglement of purification, reflected entropy, odd
	entropy, entanglement negativity 	\cite{Takayanagi:2017knl,Nguyen:2017yqw,Bao:2017nhh, 	Umemoto:2018jpc,Tamaoka:2018ned,Dutta:2019gen,Kudler-Flam:2018qjo,Kusuki:2019zsp,Du:2019emy}.  Based on the PEE construction, the so-called balanced PEE was also proposed as a holographic dual of the EWCS, and this correspondence has been precisely verified in a variety of configurations
	\cite{Wen:2021qgx,Camargo:2022mme,Wen:2022jxr,Basu:2022nyl, Basu:2023wmv,Lin:2023ajt}. We formulate this problem by promoting the RT surface to an auxiliary state-supporting surface. Splitting it between the two subsystems defines a geometric purification;
	the EWCS then becomes the RT surface of the corresponding purified subsystem.
	This is a canonical implementation of the surface/state correspondence: an
	RT surface changes from a passive homology boundary into an active purifying
	component.
	
	The PEE-thread description gives this construction a microscopic content.
	The relevant flow is sourced not only from the original boundary subsystem
	\(A\), but from the extended purified subsystem \(A\cup \CE_A\), where
	\(\CE_A\) is the portion of the RT surface assigned to \(A\).  The PEE flux is
	therefore the combined flux through \(A\) and its geometric purifier.  For
	disjoint intervals, different splits of the RT surfaces describe different
	purifications, and minimizing the bottleneck reproduces the optimization
	intrinsic to the entanglement of purification.
	
	One of the most interesting phenomena uncovered by this framework is a
	separation between the geometry of the elementary PEE threads and that of the
	macroscopic bit-thread flow.  Every elementary PEE thread is a bulk
	geodesic.  Nevertheless, unlike the bit-thread configurations constructed in Poincar\'e AdS \cite{Lin:2023rxc}, once the PEE threads sourced from different
	components of an extended surface are superposed, the integral curves of the resulting current, i.e.\ the bit threads, are generically non-geodesic.
	They remain smooth, divergenceless and norm bounded, and their streamlines
	form a regular non-crossing foliation.  They
	saturate the norm bound on the appropriate RT surface or EWCS.  Thus
	geodesicity is a property of the microscopic carriers of two-point PEE, not
	a universal requirement on the coarse-grained information current.

	The planar-BTZ ingredients overlap with the endpoint-resolved analysis of \cite{BasuWenChandra}, where the emphasis is on intrinsic planar-BTZ sectors and their finite-cutoff extension; here those ingredients are incorporated into the broader extended-surface framework involving an EOW brane, a horizon and RT-surface purifiers.  Our explicit max-flow constructions cover the adjacent brane-ending and connected phases analyzed below.  In disconnected BCFT phases we establish the intersection-weighted PEE normalization and formulate the corresponding flow problem, but we do not claim a closed-form PEE-generated max-flow representative.

	The remainder of the paper is organized as follows.  In
	section~\ref{sec:review} we review the required ingredients of PEE threads,
	bit threads, subregion kinematic space, the surface/state correspondence, AdS/BCFT and
	minimal purification.  In section~\ref{sec:two-point-pee} we formulate the
	two-point PEE on the extended surfaces relevant to Poincar\'e AdS/BCFT and BTZ
	geometries.  Section~\ref{sec:pee-threads} constructs the corresponding
	PEE-thread flows and analyzes their bit-thread superpositions.  In
	section~\ref{sec:eop} we apply the minimal-purification framework to
	the entanglement wedge cross-section for adjacent and disjoint
	subsystems in Poincar\'e AdS and BTZ.  The BTZ boundary-, brane-, and
	horizon-sourced sectors are treated intrinsically in BTZ coordinates in
	sections~\ref{sec:PEE-BTZ} and \ref{sec:pee-threads}.  We conclude in
	section~\ref{sec:summary} with a critical discussion of our results and
	comments on possible future directions.

	\section{Review and framework}
	\label{sec:review}
	\subsection{PEE threads and bit threads}
	\label{subsec:pee-bit-review}
	The partial entanglement entropy (PEE), the PEE-thread network and a
	bit-thread flow describe three related but logically distinct levels of
	information.  The first is an additive decomposition of boundary
	entanglement, the second is a geodesic representation of its two-point
	structure, and the third is an oriented, norm-bounded bulk current.  Since
	the present work extends the sources of this current from the asymptotic
	boundary to EOW branes, horizons and RT surfaces, it is useful to state
	carefully the input and the limitations of the construction.

	\paragraph{PEE and the entanglement contour.}
	For a spatial region $A$, an entanglement contour assigns a density
	$s_A(\mathbf{x})$ to its constituent degrees of freedom such that
	\cite{Vidal:2014aal,Wen:2018whg,Wen:2019iyq,Wen:2020ech}
	\begin{align}
		S_A=\int_A\dd\sigma_{\mathbf{x}}\,s_A(\mathbf{x})\,.
		\label{eq:contour-normalization-review}
	\end{align}
	A two-point PEE $\CI(X,Y)$ refines this decomposition by resolving the
	correlation assigned to two non-overlapping sets of degrees of freedom.
	The minimal properties relevant here are symmetry, additivity and
	normalization,
	\begin{align}
		\CI(X,Y)&=\CI(Y,X),
		&
		\CI(X,Y\cup Z)&=\CI(X,Y)+\CI(X,Z),
		&
		\CI(A,A^c)&=S_A,
		\label{eq:pee-properties-review}
	\end{align}
	where $Y\cap Z=\varnothing$.  Positivity, an upper bound by the entropies of
	the participating regions, invariance under local unitaries within $X$ and
	$Y$, and covariance under symmetries of the state provide further physical
	constraints \cite{Wen:2018whg,Wen:2019iyq,Han:2019scu}.  When additivity
	may be implemented locally, one writes
	\begin{align}
		\CI(X,Y)
		=\int_X\dd\sigma_{\mathbf{x}}
		 \int_Y\dd\sigma_{\mathbf{y}}\,
		 \CI(\mathbf{x},\mathbf{y}),
		\qquad
		s_A(\mathbf{x})
		=\int_{A^c}\dd\sigma_{\mathbf{y}}\,
			 \CI(\mathbf{x},\mathbf{y}).
		\label{eq:two-point-pee-review}
	\end{align}
	Thus the contour is the one-endpoint marginal of the two-point PEE, and its
	normalization follows immediately from \eqref{eq:pee-properties-review}.

	For an ordered one-dimensional system, a particularly useful realization is
	the additive linear combination (ALC) proposal \cite{Wen:2018whg,Wen:2019iyq}.  If
	$A=\alpha_L\cup\alpha\cup\alpha_R$, with $\alpha_L$ and $\alpha_R$ lying to
	the left and right of $\alpha$, respectively, the contribution of $\alpha$
	to $S_A$ is
	\begin{align}
		s_A(\alpha)=\frac{1}{2}\left(
		S_{\alpha_L\cup\alpha}+S_{\alpha\cup\alpha_R}
		-S_{\alpha_L}-S_{\alpha_R}\right).
		\label{eq:alc-review}
	\end{align}
	This formula makes the additive structure manifest and can be applied to generic theories with one spatial dimension. 

	For the vacuum of a CFT$_2$ on the line, the physical requirements of additivity and normalization, together with other properties shared by the mutual information, determine the two-point PEE as \cite{Wen:2019iyq,Casini:2008wt},
	\begin{align}
		\CI(x,y)=\frac{c}{6}\frac{1}{(x-y)^2}
		=\frac{1}{4G}\frac{1}{(x-y)^2},
		\label{eq:vacuum-pee-kernel-review}
	\end{align}
	where the second equality uses the Brown--Henneaux relation for unit AdS
	radius.  In entropy decompositions, $\CI(x,y)|\dd x\,\dd y|$ is a positive
	density. 

	\paragraph{Bit threads and the max-flow problem.}
	Let $\CM$ be a static bulk time slice and let $\gamma_A$ be the RT surface
	homologous to $A$.  The Riemannian max-flow--min-cut theorem rewrites the RT
	prescription as \cite{Freedman:2016zud,Headrick:2017ucz}
	\begin{align}
		S_A
		&=\frac{\operatorname{Area}(\gamma_A)}{4G}
		=\max_{v_A}\Phi_A(v_A),
		\\
		\Phi_A(v_A)&=\int_A\dd\Sigma\,n_\mu v_A^\mu,
		\qquad
		\nabla_\mu v_A^\mu=0,
		\qquad
		|v_A|\leq\frac{1}{4G}.
		\label{eq:bit-thread-mfmc-review}
	\end{align}
	A vector field obeying the last
	line is an admissible flow.  A maximizing flow is normal to $\gamma_A$ and
	saturates the norm bound there,
	\begin{align}
		v_A^\mu\big|_{\gamma_A}
		=\frac{1}{4G}n_{\gamma_A}^\mu,
		\label{eq:bit-thread-saturation-review}
	\end{align}
	with the sign fixed by the chosen flux orientation. See \cite{Freedman:2016zud,Headrick:2017ucz,Agon:2018lwq,Headrick:2020gyq,Headrick:2022nbe,Du:2019emy,Lin:2023rxc,Caggioli:2024uza,Agon:2021tia,Lin:2022flo,Lin:2022aqf} for the construction of bit-thread configurations in various situations.  The RT surface is thus
	the bottleneck of the flow.  Away from the bottleneck, a maximal flow is
	highly non-unique.  Moreover, its integral curves need not be geodesics: the
	defining conditions are divergencelessness, the local norm bound, the
	appropriate boundary conditions and maximal flux.  In geometries with an
	EOW brane or another internal boundary, relative homology also requires the
	corresponding no-flux or explicitly prescribed source condition.  We will
	state those conditions separately in each application.

	\paragraph{From two-point PEE to PEE threads.}
	The construction of \cite{Lin:2023rxc} geometrizes the two-point PEE by the
	bulk geodesic $\Gamma(\mathbf{x},\mathbf{y})$ joining its two endpoint
	degrees of freedom.  These weighted geodesics are called \emph{PEE
	threads}.  For a fixed source $\mathbf{x}$, the geodesic through
	$\mathbf{x}$ and a generic bulk point fixes a unit tangent
	$\tau_{\mathbf{x}}^\mu$.  The corresponding source-labelled PEE-thread
	flow is
	\begin{align}
		V_{\mathbf{x}}^\mu
		=|V_{\mathbf{x}}|\,\tau_{\mathbf{x}}^\mu,
		\qquad
		\nabla_\mu V_{\mathbf{x}}^\mu=0
		\qquad\text{in the bulk interior away from the source.}
		\label{eq:source-pee-flow-review}
	\end{align}
	The direction is fixed by the geodesic family, while its norm is determined
	by matching bulk flux to the endpoint PEE density.  Consider the bundle of
	geodesics connecting the source point $\mathbf{x}$ to all points $\mathbf{y}$
	in an infinitesimal element $\dd\sigma_{\mathbf{y}}$.  We require that the
	flux of the PEE-thread flow through any cross-section of this bundle equals
	the PEE between $\mathbf{x}$ and $\dd\sigma_{\mathbf{y}}$:
	\begin{align}
		\left(V_{\mathbf{x}}\cdot n\right)_P
		\dd\Sigma_P
		={\CI}(\mathbf{x},\mathbf{y})
		\dd\sigma_{\mathbf{y}},
		\label{eq:pee-flux-matching-review}
	\end{align}
	where $P$ is the location of the cross-section $\dd \Sigma_P$ with normal $n$ and area element $\dd\Sigma_P$. This local condition, together
	with the geodesic direction $\tau_{\mathbf{x}}^\mu$, fully determines the
	source-labelled PEE-thread vector field $V_{\mathbf{x}}^\mu$ (see \cite{Lin:2023rxc} for more details).

	PEE threads emitted from different sources generally intersect.  They should
	not be confused with the streamlines of a single smooth vector field.  Each
	$V_{\mathbf{x}}$ describes one source-labelled geodesic bundle; the
	macroscopic current associated with a region $A$ is their oriented
	superposition,
	\begin{align}
		v_A^\mu
		=\int_A\dd\sigma_{\mathbf{x}}\,
		V_{\mathbf{x}}^\mu.
		\label{eq:pee-superposition-review}
	\end{align}
	Linearity immediately implies $\nabla_\mu v_A^\mu=0$.  A thread with both
	endpoints in $A$ is counted once from each endpoint with opposite
	orientations, so such ``inner'' contributions cancel in the current.  The
	remaining threads connect $A$ to $A^c$, and
	\begin{align}
		\int_{\gamma_A}\dd\Sigma\,n_\mu v_A^\mu
		=\int_A\dd\sigma_{\mathbf{x}}
		 \int_{A^c}\dd\sigma_{\mathbf{y}}\,
		 \CI(\mathbf{x},\mathbf{y})
		=S_A.
		\label{eq:pee-superposition-flux-review}
	\end{align}
	This establishes the correct flux, but it does \emph{not} by itself prove
	that \eqref{eq:pee-superposition-review} is a bit-thread flow.  The local norm
	bound in \eqref{eq:bit-thread-mfmc-review} must still be checked.  For a
	static interval in a CFT$_2$, and for a spherical region in higher
	dimensions, the symmetry of the configuration makes the superposition an
	admissible maximal flow \cite{Lin:2023rxc}.  For disconnected regions or
	multi-component state-supporting surfaces, a naive endpoint sum can instead
	overcount geodesics belonging to the wrong homology chamber.  In those cases
	the physically relevant object is an intersection-weighted, or
	chamber-projected, superposition.  This qualification will be important for
	the disconnected phases considered below.

	For static spherical regions and intervals, the integral curves of $v_A^\mu$ are geodesics, but there is no general reason for them to be so.  Later, we will find that the AdS/BCFT and minimal-purification examples provide explicit cases in which the streamlines of the maximal current $v_A^\mu$ are non-geodesic.

		A useful geometric reorganization of these examples is developed in appendices~\ref{app:distance-difference-potential} and~\ref{app:fermi-extended-flows}.  In two bulk dimensions, the source integrals over an oriented extended subsystem telescope: endpoints internal to the oriented chain cancel, while the final stream function is determined only by the endpoints of the RT surface or candidate EWCS that is being calibrated.  A finite endpoint contributes an ordinary geodesic-distance function and an asymptotic endpoint contributes a Busemann function.  Appendix~\ref{app:distance-difference-potential} derives this endpoint reduction directly from the PEE source current, while appendix~\ref{app:fermi-extended-flows} rewrites the same flows in Fermi normal coordinates and compares the PEE-selected current with the independent normal-geodesic comparator.

	\paragraph{From PEE threads to bit threads.}
	On a constant-time slice of Poincar\'e AdS$_3$,
	$\dd s^2=(\dd x^2+\dd z^2)/z^2$, the source-labelled field emitted from the
	boundary point $x=x_0$ is \cite{Lin:2023rxc}
	\begin{align}
		V_{x_0}^\mu(\bar x,\bar z)=\frac{1}{4G}
		\frac{2\bar{z}^2(\bar{x}-x_0)}
		{\left[(\bar{x}-x_0)^2+\bar{z}^2\right]^2}
		\left(\bar{z},
		\frac{\bar{z}^2-\left(\bar{x}-x_0\right)^2}
		{2\left(\bar{x}-x_0\right)}\right).
		\label{PEE-thread-flow}
	\end{align}
	The apparent pole in the second entry at $\bar x=x_0$ cancels against the
	prefactor.  Directly using the hyperbolic metric, one finds
	$\nabla_\mu V_{x_0}^\mu=0$, and its flux through a reference geodesic agrees
	with the kernel \eqref{eq:vacuum-pee-kernel-review}.  
	
	Based on the PEE-thread configuration one can construct the bit-thread configuration using the superposing scheme.
	\begin{itemize}
		\item \textbf{The superposing scheme:} \textit{we first orient the PEE threads to point from $A$ to its complement $A^c$, then superpose the vector fields representing all such oriented threads, and eventually get the vector field describing the bit threads.}
	\end{itemize}
	For the interval $A=[-b,b]$ on the boundary of Poincar\'e AdS$_3$, the resulting bit-thread current is
	\begin{align}
		v_A^{\mu}(\bar x,\bar z)
		&=\int_{-b}^{b}\dd x_0\,V^\mu_{x_0}(\bar x,\bar z)
		\notag\\
		&=\frac{\bar{z}^2}{4G}
		\frac{2b}
		{\left((b-\bar{x})^2+\bar{z}^2\right)
		 \left((b+\bar{x})^2+\bar{z}^2\right)}
		\left(2\bar{z}\bar{x},b^2-\bar{x}^2+\bar{z}^2\right),
		\label{bit-threads-CFT}
	\end{align}
	whose integral curves are the bit threads. Its norm is
	\begin{align}
		|v_A|
		=\frac{1}{4G}
		\frac{2b\bar z}
		{\sqrt{\left((b-\bar x)^2+\bar z^2\right)
		\left((b+\bar x)^2+\bar z^2\right)}}
		\leq\frac{1}{4G}\,,
		\label{eq:vacuum-bit-thread-norm-review}
	\end{align}
	which is saturated precisely on the RT semicircle
	$\gamma_A:\bar x^2+\bar z^2=b^2$, where the field is normal to
	$\gamma_A$.  Thus \eqref{bit-threads-CFT} is an admissible maximal flow and
	coincides with the standard symmetric bit-thread configuration in $d=2$
	\cite{Agon:2018lwq,Caggioli:2024uza}.  In the Fermi chart of appendix~\ref{app:fermi-extended-flows}, this benchmark has two ideal foci: the endpoint-difference potential reduces to $\Psi=-\eta$, and \eqref{bit-threads-CFT} becomes the universal normal-geodesic flow $v_{\rm geo}=(4G)^{-1}\operatorname{sech}\lambda\,\partial_\lambda$.  Thus the coincidence between the PEE-selected and geodesic bit-thread representatives in the vacuum interval is the ideal--ideal member of the more general construction used below.  The case of higher-dimensional static spherical regions was also discussed in \cite{Lin:2023rxc}.

		   \subsection{Subregion kinematic space and subregion reconstruction}
		   \label{subsec:subregion-kinematic-space}
		   
		   Kinematic space provides an integral-geometric description of a static bulk
		   slice.  For a maximally symmetric Riemannian manifold \(\mathcal M\), its
		   kinematic space \(\mathbb K_{\mathcal M}\) is the space of oriented
		   geodesics equipped with the invariant measure \(d\Gamma\).  The Crofton
		   formula reconstructs the area of a codimension-one surface
		   \(\Sigma\subset\mathcal M\) by counting its intersections with the geodesics,
		   \begin{equation}
		   	\operatorname{Area}(\Sigma)
		   	=
		   	\frac{1}{2}\frac{d-1}{\Omega_{d-2}}
		   	\int_{\mathbb K_{\mathcal M}}
		   	\#(\Gamma\cap\Sigma)\,d\Gamma ,
		   	\label{eq:crofton-review}
		   \end{equation}
		   where \(\#(\Gamma\cap\Sigma)\) denotes the unsigned intersection number
		   \cite{Czech:2015qta,Czech:2016xec}.  On a static slice of vacuum AdS, all the
		   geodesics are anchored on the asymptotic boundary and can be parametrized by
		   their endpoints.  In AdS$_3$, for example,
		   \begin{equation}
		   	\dd\Gamma=\frac{1}{2}\left|\frac{\partial^2\CL(x_1,x_2)}{\partial x_1\,\partial x_2}\right|\,\dd x_1\wedge \dd x_2 \,,\label{eq:ads3-kinematic-measure}
		   \end{equation}
		   with \(\CL(x_1,x_2)\) the regulated geodesic length connecting the two boundary points $x_1$ and $x_2$. It was shown in \cite{Lin:2024fze} that, the integrand of \eqref{eq:ads3-kinematic-measure} coincide with the two-point PEE $\CI(x_1,x_2)$ \eqref{eq:vacuum-pee-kernel-review}.  Geometrizing the two-point PEE between \(x_1\) and \(x_2\) as the geodesic joining them identifies the geodesics of kinematic space with PEE threads. Their continuous superposition forms a network that uniformly covers vacuum AdS, while the Crofton intersection count becomes an entanglement-based reconstruction of bulk areas \cite{Wen:2025gui}.
		   
		   The construction of \cite{Basu:2026hbg} extends this
		   picture from the full vacuum AdS slice to a connected gravitational subregion
		   \(\mathbf a\) bounded by a codimension-one surface
		   \(\partial\mathbf a\) and possibly a totally geodesic submanifold.  The subregion kinematic space
		   \(\mathbb K_{\mathbf a}\) consists of the geodesic chords contained in
		   \(\mathbf a\).  A chord is the portion of a complete ambient geodesic
		   between successive intersections with \(\partial\mathbf a\); its endpoints
		   therefore lie on the subregion boundary rather than necessarily on the
		   asymptotic AdS boundary.  For a non-convex subregion, a single ambient
		   geodesic may contribute several distinct chords, which must be treated as
		   separate elements of \(\mathbb K_{\mathbf a}\). 
		   
		   If the geodesic connecting any pair of boundary points is unique and \(\lambda_1\) and \(\lambda_2\) parametrize the endpoints of a chord \(\partial\mathbf a\) , the intrinsic kinematic measure for the subregion is \cite{Basu:2026hbg}
		   \begin{equation}
		   	\dd\Gamma_{\partial\mathbf a}
		   	=
		   	\frac{1}{2}
		   	\left|\det\!\left(
		   	\frac{\partial^2\CL(\lambda_1,\lambda_2)}
		   	{\partial \lambda_1\,\partial \lambda_2}
		   	\right)\right|
		   	\dd\lambda_1\wedge \dd\lambda_2 ,
		   	\label{eq:subregion-kinematic-measure}
		   \end{equation}
		   with the determinant understood over the transverse endpoint coordinates in
		   higher dimensions and the displayed endpoint differentials replaced there
		   by the corresponding endpoint volume forms.  This measure is equivalently characterized by requiring
		   the intersections of the chord family to be uniformly distributed over
		   \(\partial\mathbf a\) and over the allowed directions.  The area of any
		   codimension-one surface \(\Sigma\subset\mathbf a\) can then be reconstructed
		   intrinsically through
		   \begin{equation}
		   	\operatorname{Area}(\Sigma)
		   	=
		   	\frac{1}{2}\frac{d-1}{\Omega_{d-2}}
		   	\int_{\mathbb K_{\mathbf a}}
		   	\#(\Gamma_{\mathbf a}\cap\Sigma)\,
		   	\dd\Gamma_{\partial\mathbf a}.
		   	\label{eq:subregion-crofton}
		   \end{equation}
		   Thus the geometry inside \(\mathbf a\) may be reconstructed using only the
		   chords emitted from its bounding surface.  
		   
		   In vacuum AdS with $\partial \mathbf a$ being the AdS boundary, the kinematic measure \eqref{eq:subregion-kinematic-measure} can be identified with the PEE structure \cite{Lin:2023rxc,Lin:2024fze}, 
		   \begin{align}
		   	\dd\Gamma_{\partial\mathbf a}=\frac{6}{c}\CI(\lambda_1,\lambda_2)\,\dd\lambda_1\wedge\dd\lambda_2\label{PEE}
		   \end{align}
		   equivalently
		   \begin{align}
		   	\CI(\lambda_1,\lambda_2)=\frac{c}{12}\left|\det\!\left(
		   	\frac{\partial^2\CL(\lambda_1,\lambda_2)}
		   	{\partial \lambda_1\,\partial \lambda_2}
		   	\right)\right|.
		   	\label{eq:PEEPa}
		   \end{align}
		   In this paper we extend this relation to a generic subregion in AdS, thereby defining the PEE structure on its boundary $\partial\mathbf a$. This identification furnishes the basic dictionary between integral geometry and quantum
		   information: the endpoints \(\lambda_1,\lambda_2\) specify the elementary
		   degrees of freedom on the state-supporting surface, the associated chord is
		   the geometric representative of their two-point PEE, and the measure
		   \(\dd\Gamma_{\partial\mathbf a}\) determines the density with which such
		   elementary threads occur.  
		   
		  Although the examples of subregion reconstruction discussed in \cite{Basu:2026hbg} are confined to subregions of vacuum AdS, the same ideas apply to more general settings. In this paper we consider situations that can be viewed as subregions of vacuum AdS with different choices of \(\partial\mathbf a\), so that all the results of \cite{Basu:2026hbg} apply directly. In AdS/BCFT the supporting surface comprises the asymptotic boundary and the
		   EOW brane; in the planar BTZ black brane it additionally contains the
		   horizon; and in the minimal purification it contains the RT surface that
		   supplies the auxiliary purifying degrees of freedom.  Integrating the density in
		   \eqref{PEE} over one endpoint produces the source-labelled PEE-thread flow,
		   superposing these source flows produces the macroscopic bit-thread current,
		   and intersection-weighting the same elementary chords reconstructs areas in
		   non-trivial homology chambers.  Thus \eqref{PEE} is not merely a convenient
		   definition: it is the fundamental local input from which the two-point PEE
		   kernels, PEE-thread networks, bit-thread flows, entropy normalization
		   conditions and minimal-purification constructions developed throughout this
		   paper all follow.


\subsection{Surface/state correspondence}
\begin{figure}[ht]
	\centering
	\includegraphics[width=0.65\linewidth]{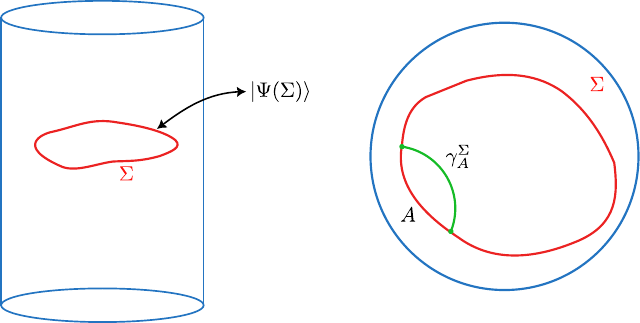}
	\caption{Surface-state correspondence}
	\label{fig:rt-ss}
\end{figure}

The surface/state correspondence extends the AdS/CFT dictionary by associating quantum states not only with the asymptotic boundary, but with arbitrary convex codimension-two surfaces $\Sigma$ in the bulk \cite{Miyaji:2015yva}; see figure \ref{fig:rt-ss} for an illustration. A closed surface homologous to a point is assigned a pure state, whereas an open surface or a surface wrapping a nontrivial cycle is assigned a mixed state.
The entanglement structure of the state on $\Sigma$ is determined geometrically: for a subregion $A\subset\Sigma$, the leading semiclassical entropy is
\begin{align}
	S_A^\Sigma
	=
	\frac{\operatorname{Area}(\gamma_A^\Sigma)}{4G},
	\qquad
	\del\gamma_A^\Sigma=\del A,
	\label{eq:ss-entropy}
\end{align}
where $\gamma_A^\Sigma$ is the minimal surface contained in the region enclosed by $\Sigma$ and homologous to $A$; when $\Sigma$ is the asymptotic boundary this reduces to the usual Ryu--Takayanagi prescription. See \cite{Miyaji:2015yva,Miyaji:2015fia} for more details of the correspondence.

For the purposes of the present work, the surface/state correspondence provides the conceptual basis for treating the extended boundary $\partial\mathbf a$ of a gravitational subregion as the support of an effective quantum state. Its connected components may lie on the asymptotic boundary, an EOW brane, a horizon section, or an RT surface. The corresponding components should not necessarily be interpreted as independent microscopic boundaries: the horizon sector, for example, represents the thermal purifying degrees of freedom inherited from the thermofield-double description, while an RT-surface component serves as an auxiliary purifier of a boundary mixed state \cite{Bao:2023til}. Once this effective state-supporting surface is specified, its elementary degrees of freedom are labelled by points $\lambda\in\partial\mathbf a$, and the two-point PEE density in \eqref{PEE} assigns a geodesic chord to each pair $(\lambda_1,\lambda_2)$. The resulting restricted PEE network simultaneously represents the correlation structure of the surface state, reconstructs the geometry inside $\mathbf a$, and supplies the elementary threads whose oriented superposition produces the bit-thread flows studied below.
		   \subsection{AdS/BCFT and islands}
		   \label{subsec:review-adsbcft-islands}
		   
		   A boundary conformal field theory (BCFT) is a conformal field theory defined on a
		   manifold $\CM$ with a codimension-one boundary
		   $\partial \CM$, endowed with a conformal boundary condition. In two dimensions the
		   boundary condition contributes a universal constant to the entanglement entropy of an
		   interval adjacent to the boundary, the Affleck--Ludwig boundary entropy
		   $\log g$ \cite{Cardy:1984bb,Cardy:1989ir,Cardy:2004hm,Calabrese:2004eu}.
		   
		   The AdS/BCFT construction provides a holographic description of such
		   theories \cite{Takayanagi:2011zk,Fujita:2011fp,Nozaki:2012qd}: the dual is a portion
		   $\CN$ of an asymptotically AdS spacetime whose boundary contains both the usual
		   asymptotic component $\CM$ and an end-of-the-world (EOW) brane $\CQ$,
		   \begin{align}
		   	\partial \CN=\CM\cup \CQ.
		   \end{align}
		   For Einstein gravity and a constant-tension brane, the brane embedding is fixed by the
		   Neumann condition
		   \begin{align}
		   	K_{\mu\nu}=(K-T)h_{\mu\nu}\implies K=\frac{d}{d-1}T\label{NBC}
		   \end{align}
		   where $h_{ab}$, $K_{ab}$ and $T$ are the induced metric, extrinsic curvature and tension
		   of $\CQ$. In the AdS$_2$ foliation of Poincar\'e AdS$_3$,
		   \begin{align}
		   	\dd s^2=\dd\rho^2+\cosh^2\rho\,\frac{-\dd t^2+\dd y^2}{y^2}\,,
		   \end{align}
		   the brane sits on a constant
		   $\rho$-slice, $\rho=\rho_0$, so that $T=\tanh\rho_0$, and the holographic boundary
		   entropy is
		   \begin{align}
		   	\log g_b=\frac{\rho_0}{4G}=\frac{c}{6}\rho_0,
		   \end{align}
		   up to the sign convention used for the retained side of the brane
		   \cite{Azeyanagi:2007qj,Takayanagi:2011zk,Fujita:2011fp}.
		   
		   The Ryu--Takayanagi prescription is modified only through the allowed
		   homology condition: an extremal surface anchored at the entangling surface
		   on $\CM$ may terminate on $\CQ$,
		   \begin{align}
		   	S_A=\frac{\operatorname{Area}(\gamma_A)}{4G}~~,~~
		   	\partial\gamma_A=\partial A\cup(\gamma_A\cap \CQ),
		   \end{align}
		   and $\gamma_A$ meets $\CQ$ orthogonally. For an interval away from the BCFT boundary one
		   generally finds competing saddles -- a connected surface joining the two boundary
		   endpoints and a disconnected pair of surfaces ending on $\CQ$ -- whose exchange of
		   dominance is the bulk counterpart of a transition between bulk and boundary OPE channels
		   \cite{Fujita:2011fp,Sully:2020pza}.
		   
		   In a braneworld (or doubly holographic) regime \cite{Randall:1999vf,Karch:2000ct,Chen:2020uac,Chen:2020hmv}, the brane supports gravitational modes
		   and a region of $\CQ$ may be interpreted as an entanglement island \cite{Penington:2019npb,
		   		Almheiri:2019psf,Almheiri:2019hni,Penington:2019kki,
		   		Almheiri:2019qdq,Geng:2020qvw,Geng:2021iyq}, whose endpoint is
		   fixed by the quantum extremal-surface condition
		   \cite{Faulkner:2013ana,Engelhardt:2014gca}. From the
		   surface/state viewpoint, the EOW brane is an additional state-supporting component of the
		   extended surface: PEE threads may connect the asymptotic boundary to $\CQ$, or connect
		   different portions of $\CQ$, and in a brane-ending (island) phase these sectors resolve
		   the entanglement carried by the degrees of freedom assigned to the brane. The PEE-thread
		   framework therefore does not alter the island prescription; rather, it refines it by
		   resolving how the entanglement flux is redistributed between bath, brane and purifying
		   degrees of freedom across the island transition.
		   
		   \subsection{The general scheme}
		   Before moving on, it is important to separate a geometric statement from the physical interpretation that we attach to it.  On the asymptotic CFT boundary, the mixed endpoint derivative has the usual two-point-PEE interpretation.  On an EOW brane, a horizon section, or an RT surface promoted to a purifier, the quantity obtained from the subregion Crofton measure is first of all a classical endpoint density on the extended state-supporting surface.  Calling this density a two-point PEE on a non-asymptotic component is the additional surface/state assumption of our framework.  Accordingly, the Crofton identities, flux matching, endpoint reduction and max-flow checks below are geometric statements at leading semiclassical order; their interpretation in terms of microscopic brane-, horizon-, or RT-surface degrees of freedom is conditional on that extended surface/state dictionary.

		   With this qualification understood, let us summarize the general procedure, which will be applied to the situations considered in the rest of the paper:
		   \begin{enumerate}
		   	\item \textit{construct the PEE structure on the boundary $\partial \mathbf{a}$ of a subregion via \eqref{eq:PEEPa};}
		   	\item \textit{construct the PEE-thread configuration from the two-point PEE by extending the flux-matching condition \eqref{eq:pee-flux-matching-review} to gravitational subregions;}
		   	\item \textit{superpose the oriented PEE threads following the superposing scheme to obtain the macroscopic current;}
		   	\item \textit{perform the divergencelessness, norm-bound and bottleneck checks to verify that the macroscopic current is a genuine bit-thread flow.}
		   \end{enumerate}

		   \section{Two-point PEE in AdS/BCFT}
		   \label{sec:two-point-pee}
		   The aim of this section is to formulate the two-point partial entanglement entropy in holographic BCFT \cite{Takayanagi:2011zk}. In an ordinary holographic CFT, the two-point PEE between infinitesimal boundary elements may be regarded as a Crofton density on the space of boundary anchored geodesics. In AdS/BCFT, however, the asymptotic boundary is not the only geometric component of the holographic state. The bulk is truncated by an end-of-the-world brane, and the surface/state point of view naturally suggests that the relevant ``kinematic support'' of the state is the extended convex surface
		   \begin{align}
		   	\Sigma_e=\del \CM\cup\CQ\label{extended-surface}
		   \end{align}
		   where $\del\CM$ is the asymptotic boundary and $\CQ$ is the EOW brane. Consequently, one should allow PEE densities not only between two asymptotic boundary points, but also between an asymptotic boundary point and a brane point, and between two brane points.
		   
		   This extension is not merely a formal device. It has two related physical interpretations. First, in an \emph{island phase} \cite{Sully:2020pza,Rozali:2019day,Suzuki:2022xwv}, a boundary region is purified not only by its asymptotic complement but also by appropriate degrees of freedom represented geometrically on the EOW brane. Secondly, the mixed and brane--brane sectors participate in the normalization that reproduces the constant brane-endpoint contribution to the RT entropy. Mixed endpoint derivatives alone cannot determine an endpoint-independent constant such as the boundary entropy. The extended kernels reproduce that contribution only after the global normalization, the oriented endpoint ranges, the regulator/Weyl frame and the brane branch have been fixed.
		   
		   Given a regulated geodesic length $\CL(p,q)$ between two points
		   $p,q\in\Sigma_e$, the relation between the two-point PEE and the geodesic
		   length is given by \eqref{eq:PEEPa} (together with $c=3/2G$).  We distinguish the positive PEE density from its oriented Crofton representative:
		   \begin{align}
		   	\widehat{\CI}(p,q)&=\frac{1}{8G}
		   	\frac{\del^2\CL(p,q)}{\del p\,\del q}\,,
		   	&
		   	\CI(p,q)&=\left|\widehat{\CI}(p,q)\right|\,.
		   	\label{PEE-oriented-convention}
		   \end{align}
		   This relation bridges geodesic geometry and PEE data on the extended
		   surface, and supplies the fundamental local input from which all
		   PEE-thread and bit-thread constructions in this paper follow. Strictly speaking, the two-point PEE $\mathcal I(p,q)$ and its
		   geometric representative geodesic $\Gamma_{p\to q}$ are unoriented:
		   $\mathcal I(p,q)=\mathcal I(q,p)$. An orientation is introduced only
		   when one endpoint is marked as the source in order to describe the
		   threads by a vector field. We orient $\Gamma_{p\to q}$ from the
		   source $p$ toward its partner $q$, so that
		   $\tau^\mu_{q\to p}=-\tau^\mu_{p\to q}$ while the PEE weight remains
		   unchanged. Accordingly, the flow associated with a region $A$ is
		   oriented from $A$ toward its complement in the extended purified
		   boundary; oppositely oriented contributions from pairs whose two
		   endpoints both lie in $A$ cancel.
		   
		   The positive density $\CI(p,q)\,|\dd p\,\dd q|$ is used in entropy
		   decompositions, whereas the signed density
		   $\widehat{\CI}(p,q)\,\dd p\wedge\dd q$ is used in oriented source-current
		   and flux integrals.  This distinction is essential whenever an extended
		   boundary component is traversed with reversed limits.  Equivalently, one
		   may keep a positive kernel and place the sign entirely in the oriented
		   endpoint measure, but the two conventions must not be mixed.
		   Once orientations are fixed, we also use
		   \begin{align}
		   	\widehat{\CI}(X,Y)
		   	=\int_X\dd p\int_Y\dd q\,\widehat{\CI}(p,q)
		   \end{align}
		   for an integrated oriented sector. Such a sector contribution may be
		   negative even though the underlying physical PEE density $\CI(p,q)$ is
		   non-negative. The entropy is the positive total outward flux; individual
		   terms in its oriented extended-boundary decomposition need not be positive.
		   
		   In particular, the corresponding positive two-point PEE is
		   \begin{align}
		   	\CI(p,q)=\frac{1}{8G}\left|\frac{\del^2\CL(p,q)}{\del p\,\del q}\right|\label{PEE-BCFT-general}
		   \end{align}
		   where the derivatives are taken with respect to the coordinates used on the two components of $\Sigma_e$. This expression should be interpreted as a density. Under a coordinate transformation on the brane, the kernel transforms with the appropriate Jacobian. Therefore, any apparent non-symmetry of a mixed kernel written in non-canonical variables is not by itself a physical problem. The invariant statements are that $\CI(p,q)\,|\dd p\,\dd q|$ and $\widehat{\CI}(p,q)\,\dd p\wedge\dd q$ are independent of the coordinate used to parametrize the EOW brane.
\subsection{Poincar\'e AdS$_3$}
		   We begin with a BCFT ground state on the half-line with a conformal boundary at  $x=0$. The bulk dual geometry is given by a portion of Poincar\'e AdS$_3$ truncated by an EOW brane with profile\footnote{The angle $\theta$ is related to the brane tension as 
		   \begin{align}
		   	\cos\theta=\sech\rho_0=\sqrt{1-T^2}\,.
		   	\end{align}}
		   \begin{align}
		   	\CQ: x=-z\tan\theta\,,\label{brane}
		   \end{align}
		   where $\theta$ is the angle made by the EOW brane with the holographic direction, as depicted in figure~\ref{fig:pee-1}. The explicit Poincar\'e calculations below use the positive-tension branch $0\leq\theta<\pi/2$; on the negative-tension branch the same formulae are read with the oriented kernel $\widehat{\CI}$ and the positive density $\CI=|\widehat{\CI}|$. It is useful to parametrize points on the brane by their distance from the conformal boundary $\hat y(>0)$,
		   \begin{align}
		   	(x_\CQ,z_\CQ)=(-\hat y\sin\theta,\hat y\cos\theta)
		   \end{align}
		   For concreteness, we consider the subsystem $A=[0,b]$ adjacent to the boundary. There are three types of two-point PEE relevant to our study, corresponding to the three elementary classes of geodesics anchored on the extended surface $\Sigma_e$:
		   	\begin{figure}[ht]
		   	\centering
		   	\includegraphics[width=0.55\linewidth]{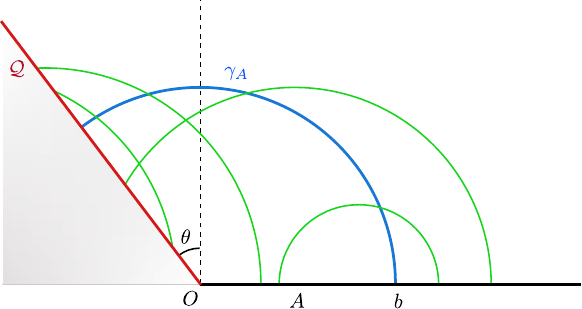}
		   	\caption{Different classes of geodesics concerning two-point PEEs relevant to a subsystem $A=[0,b]$ in the ground state of a BCFT$_2$.}
		   	\label{fig:pee-1}
		   \end{figure}
		   \begin{itemize}
		   	\item {\bf Boundary-boundary:} For two points $x,y$ on the asymptotic boundary, the standard vacuum kernel is
		   	\begin{align}
		   		\CI_{\del\del}(x,y)=\frac{c}{12}\frac{2}{(x-y)^2}=\frac{1}{4G(x-y)^2}\,.\label{PEE-I}
		   	\end{align}
		   	\item {\bf Boundary-brane:} Consider a geodesic connecting a point $(x,0)$ on the asymptotic boundary and another point on the EOW brane at a distance $\hat y$ from the boundary. The length of this geodesic may be easily computed through the embedding formula \eqref{Poincare-length} as follows
			   	\begin{align}
			   		\cosh\CL(x,\hat y)
			   		&=\frac{(x+\hat y \sin\theta)^2+\hat y^2\cos^2\theta+\epsilon^2}{2\epsilon\,\hat y\cos\theta}\,,\notag\\
			   		\CL(x,\hat y)
			   		&=\log\left(\frac{x^2+\hat y^2+2x\hat y\sin\theta}{\epsilon\,\hat y\cos\theta}\right)+\CO(\epsilon^2)
			   	\end{align}
			   	Therefore, from \eqref{PEE-BCFT-general}, we may obtain the two-point PEE as follows
			   	\begin{align}
			   		\CI_{\del \CQ}(x,\hat y)
			   		&=\frac{1}{8G}\left|\frac{\del^2\CL(x,\hat y)}{\del x\,\del\hat y}\right|\notag\\
			   		&=\frac{1}{4G}\frac{2x\hat y+(x^2+\hat y^2)\sin\theta}{\left(x^2+\hat y^2+2x\hat y\sin\theta\right)^2}\,.
			   		\label{PEE-II}
			   	\end{align}
		   	This kernel is positive on the physical domain. Its role is to transfer part of the entanglement contour from the asymptotic boundary to the EOW brane.
		   	\item {\bf Brane-brane:} Consider a geodesic joining the points on the EOW brane at distances $\hat y_1$ and $\hat y_2$ from the boundary. The length of this geodesic may again be obtained from the embedding formula \eqref{Poincare-length} as follows
		   	\begin{align}
		   		\CL(\hat y_1,\hat y_2)&=\textrm{arccosh}\left[\frac{\left(\hat y_1-\hat y_2\right)^2\sin^2\theta+(\hat y_1^2+\hat y_2^2)\cos^2\theta}{2\hat y_1\hat y_2\cos^2\theta}\right]\notag\\
		   		&=\textrm{arccosh}\left(1+\frac{\left(\hat y_1-\hat y_2\right)^2}{2\hat y_1\hat y_2}\sec^2\theta\right)
		   	\end{align}
		   	Therefore, the two-point PEE between two brane points is obtained as
		   	\begin{align}
		   		\CI_{\CQ\CQ}(\hat y_1,\hat y_2)=\frac{1}{8G}\left|\frac{\del^2\CL(\hat y_1,\hat y_2)}{\del \hat y_1\,\del\hat y_2}\right|=\frac{1}{4G}\frac{\left|\hat y_1-\hat y_2\right|\sin^2\theta}{\left[\left(\hat y_1-\hat y_2\right)^2+4\hat y_1\hat y_2\cos^2\theta\right]^{3/2}}\,.\label{PEE-III}
		   	\end{align}
		   \end{itemize}
		   See figure \ref{fig:pee-1} for an illustration. Note that all these two-point PEE kernels are symmetric in their arguments, which reflects on the natural choice of the coordinate along the brane. Interestingly, \cref{PEE-II} reduces to
		   \begin{align}
		   	\CI_{\del \CQ}(x,\hat y)\to \frac{1}{4G(x+\hat y)^2}
		   \end{align}
		   in the limit $\theta\to \frac{\pi}{2}$, whence the EOW brane approaches the (reflected) asymptotic boundary. This is nothing but the ordinary boundary-boundary kernel between $x$ and the mirror point $-\hat y$. Similarly, $\CI_{\CQ\CQ}$ reduces to the mirror boundary kernel.
		   \subsubsection*{Entanglement entropy and normalization property}
		   We first consider the interval $A=[0,b]$ adjacent to the conformal boundary $x=0$. The RT surface is the geodesic ending on the EOW brane,
		   \begin{align}
		   	\gamma_A:~x^2+z^2=b^2\,,
		   \end{align}
		   whose endpoint on the EOW brane is $(x,z)=(-b\sin\theta,b\cos\theta)$. The brane component assigned to $A$ is ${\rm I}_A:0<\hat y<b$. The complement has as asymptotic component $B=[b,\infty)$ and a brane component assigned to $B$, denoted by $I_B:b<\hat y<\infty$. In a doubly holographic realization these brane components admit the usual island interpretation; in bottom-up AdS/BCFT they encode the boundary-condition degrees of freedom.
		   Assuming that the normalization property holds true for the extended boundary surface $\Sigma_e$, the entanglement entropy may be obtained from the various partial entanglement entropies as follows (see \cite{Basu:2023wmv} for more details in the normalization property in the island phase)
		   \begin{align}
		   	S_A=\CI(A\cup I_A,B\cup I_{B})=\CI(A, B)+\CI(A,I_{B})+\CI(I_A,B)+\CI(I_A, I_{B})\,.
		   \end{align}
		   Utilizing the forms of the two-point PEEs discussed earlier, we may now obtain
			   {\small
			   \begin{align}
			   	S_A&=\int_{0}^{b}\d x\int_{b}^{\infty}\d y\,\CI_{\del\del}(x,y)\notag\\
			   	&\quad+\left(\int_{b}^{\infty}\d x\int_{0}^{b}\d\hat y
			   	+\int_{0}^{b}\d x\int_{b}^{\infty}\d\hat y\right)
			   	\CI_{\del \CQ}(x,\hat y)\notag\\
			   	&\quad+\int_{0}^{b}\d \hat x\int_{b}^{\infty}\d\hat y\,
			   	\CI_{\CQ\CQ}(\hat x,\hat y)\notag\\
			   	&=\frac{c}{12}\left[2\log\left(\frac{b}{\epsilon}\right)+2\log\left[2\left(1+\sin\theta\right)\right]+2\log\sec\theta\right]\notag\\
			   	&=\frac{c}{6}\log\left(\frac{2b}{\epsilon}\right)+\frac{c}{6}\log\left(\sec\theta+\tan\theta\right)\label{EE}
			   \end{align}
			   }
		   We may identify the constant term in the above expression as the boundary entropy
		   \begin{align}
		   	S_\textrm{bdy}=\frac{c}{6}\textrm{arctanh}\left(\sin\theta\right)=\frac{\rho^{}_0}{4G}\,.
		   \end{align}
		   where we have used the relation $\sec\theta=\cosh\rho^{}_0$ and the Brown--Henneaux formula \cite{Brown:1986nw} in the last equality.
			   Appendix~\ref{app:fermi-extended-flows} gives a complementary geometric interpretation of this constant.  In Fermi coordinates adapted to the complete RT geodesic, the brane endpoint sits at proper-length coordinate $\eta_Q=-\rho_0$.  Relative to the zero-tension position $\eta_Q=0$, the EOW-brane endpoint is therefore shifted by the proper Fermi length $\rho_0$, whose contribution to the RT entropy is exactly $S_{\rm bdy}=\rho_0/(4G)$.
		    \begin{figure}[ht]
		   	\centering
		   	\includegraphics[width=0.55\linewidth]{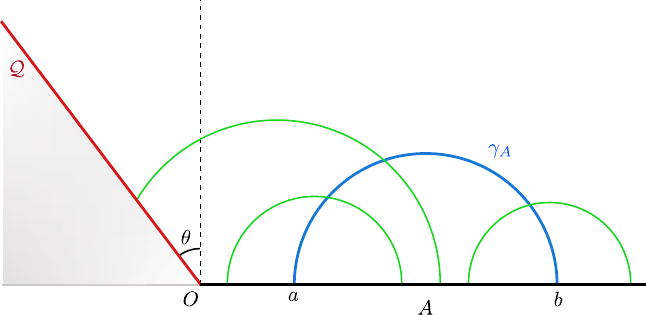}
		   	\caption{Entanglement entropy from PEE in BCFT$_2$ for the interval $A=[a,b]$ away from the boundary, in the connected phase.}
		   	\label{fig:pee-2}
		   \end{figure}
		   
		   Next, we consider the subsystem $A=[a,b]\,,\,0<a<b$ away from the boundary. In the connected phase, the RT surface is the usual boundary anchored geodesic connecting $a$ and $b$. The relevant decomposition is $S_A^{\rm con}=\CI(A,B)+\CI(A,{\rm I}_B)$, where $B=[0,a]\cup[b,\infty)$ is the asymptotic component and ${\rm I}_B$ corresponds to the entire EOW brane (see \cite{Basu:2023wmv}). The boundary-boundary term alone contains residual dependence on the endpoints $a,b$, while the
		   mixed boundary--brane term cancels precisely the image contribution induced by the BCFT boundary. The final result is (cf. \cref{fig:pee-2} for an illustration)
		   \begin{align}
		   	S^\textrm{con}_A&=\int_{a}^{b}\d x\left(\int_{0}^{a}+\int_{b}^{\infty}\right)\d y\,\CI_{\del\del}(x,y)+\int_{a}^{b}\d x\int_{0}^{\infty}\d \hat y\,\CI_{\del \CQ}(x,\hat{y})\notag\\
		   	&=\frac{c}{12}\left(2\log\left[\frac{a(b-a)^2}{\epsilon^2\,b}\right]+2\log\left(\frac{b}{a}\right)\right)\notag\\
		   	&=\frac{c}{3}\log\left(\frac{b-a}{\epsilon}\right)\,.
		   \end{align}
		   This is an important consistency check: in the connected phase the EOW brane is present in the PEE decomposition, but the physical answer is the ordinary CFT interval entropy.
		   
\begin{figure}[ht]
		   	\centering
		   	\includegraphics[width=0.65\textwidth]{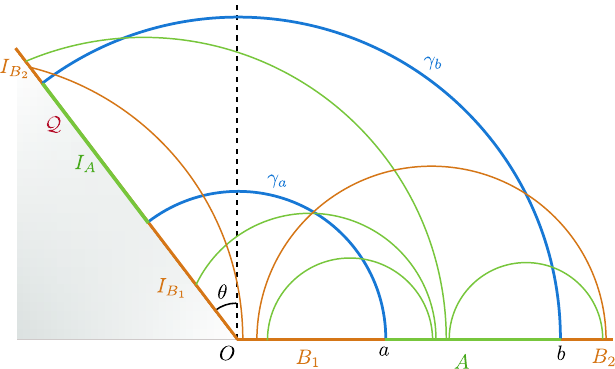}
		   	\caption{The disconnected phase of the entanglement entropy of $A=[a,b]$ away from the boundary. The blue circular arcs denote the RT segments $\gamma_{a}\,,\,\gamma_b$. The oriented extended boundary components $A_e$ and $B_{1e},B_{2e}$ are shown by green and orange segments respectively. Some illustrative PEE threads are sketched in green (with weight $\omega=1$) and orange (with weight $\omega=2)$.}
		   	\label{fig:PEE-disc-Poincare}
		   \end{figure}

		   The disconnected phase requires a more careful statement. The RT surface consists of two brane-ending geodesics, one anchored at $a$ and one at $b$, as depicted in figure \ref{fig:PEE-disc-Poincare},
		   \begin{align}
		   	\gamma_A^{\rm disc}=\gamma_a\cup\gamma_b~~,~~\gamma_\ell:~x^2+z^2=\ell^2~~(\ell=a,b)
		   \end{align}
		   In this case, a naive pairwise sum over all endpoint pairs fails to reproduce the correct entanglement entropy.
		   As described in \cite{Lin:2023rxc}, for disconnected sub-regions, naive application of the PEE threads flow proposal is unable to reproduce the entanglement entropy. Upon utilizing the doubling trick \cite{Takayanagi:2011zk}, the disconnected configuration corresponds to two disjoint intervals in some auxiliary chiral CFT$_2$. Hence, one must take into account the extended normalization property discussed in \cite{Lin:2023rxc}, namely for a union of disconnected subregions $A=\cup_i A_i$ and $A^c\equiv B=\cup_i B_i$, the entanglement entropy is obtained through minimization of the following functional
		   \begin{align}
		   	S_A=\min_{\Sigma_A}\sum_{i,j}\left[\omega_{A_iA_j}\CI(A_i,A_j)+\omega_{B_iB_j}\CI(B_i,B_j)+\omega_{A_iB_j}\CI(A_i,B_j)\right]
		   \end{align}
		   where $\Sigma_{A}$ is a codimension-two hypersurface homologous to $A$, and each weight is the intersection number of the corresponding thread class with $\Sigma_A$. To make this prescription explicit, the doubling trick maps the two brane-ending RT components to boundary-anchored geodesics in the doubled slice. Cutting the doubled supporting curve along them gives three oriented components,
		   \begin{align}
		   	\mathcal A_e=A\cup I_A\,,\qquad
		   	\mathcal B_{1e}=B_1\cup I_{B_1}\,,\qquad
		   	\mathcal B_{2e}=B_2\cup I_{B_2}\,.
		   \end{align}
		   The relevant thread classes, intersection numbers, and weights are
		   \begin{center}
		   	\begin{tabular}{c|c|c}
		   		endpoint class & $N_{\Sigma_A}$ & $\omega$\\ \hline
		   		$\mathcal A_e$--$\mathcal B_{1e}$ or
		   		$\mathcal A_e$--$\mathcal B_{2e}$ & $1$ & $1$\\
		   		$\mathcal B_{1e}$--$\mathcal B_{2e}$ & $2$ & $2$\\
		   		within any one component & $0$ & $0$
		   	\end{tabular}
		   \end{center}
		   Thus the second complement--complement class must be counted twice; it cannot be obtained from an unweighted superposition of ordinary vector flows. In the present configuration, the intersection-weighted proposal gives
		   \begin{align}
		   	S_A^\textrm{disc}&=\CI(A\cup I_A,B_1\cup B_2\cup I_{B_1}\cup I_{B_2})+2\CI(B_1\cup I_{B_1},B_2\cup I_{B_2})\notag\\
		   	&=\frac{1}{4G}\left(\log\left(\frac{2a}{\epsilon}\right)+\log\left(\frac{2b}{\epsilon}\right)+2\log\left(\sec\theta+\tan\theta\right)\right)
		   \end{align}
		   The contributions from different types of two-point PEEs to the entanglement entropy are listed in \cref{appB}.
		   Hence, with our extended definitions for the two-point PEEs, the normalization property of the two-point PEE is satisfied for different configurations of RT surfaces in AdS$_3$/BCFT$_2$. 
		   \subsection{BTZ black brane}\label{sec:PEE-BTZ}
\begin{figure}[ht]
		   	\centering
		   	\includegraphics[width=0.55\linewidth]{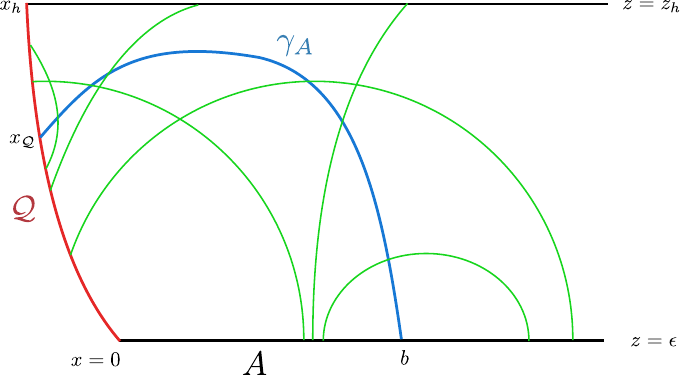}
		   	\caption{PEE threads in BTZ black brane truncated by an EOW brane. The reference RT surface is shown in blue, while some illustrative PEE threads are sketched in green.}
		   	\label{fig:bcft-btz}
		   \end{figure}

		   Next we consider a BCFT in the thermal state with a boundary at $x=x_B$, defined on the cylinder $S^1_\beta\times \BR$.
		   The bulk dual theory is given by the BTZ black brane
		   \begin{align}
		   	\d s^2=\frac{1}{z^2}\left(-h(z)\d t^2+\frac{\d z^2}{h(z)}+\d x^2\right)~~,~~h(z)=1-\frac{z^2}{z_h^2}\label{BTZ-metric}
		   \end{align}
		   where the horizon radius $z_h$ is related to the temperature of the dual field theory as $\beta=2\pi z_h$. In the high temperature phase, the spatial direction $x$ is uncompactified.
		   The Neumann boundary conditions \eqref{NBC} lead to the differential equation\footnote{The subcriticality condition is $|T|<1$.}
		   \begin{align}
		   	x'(z)=\pm\frac{T}{\sqrt{1-T^2h(z)}}
		   \end{align}
		   which may be solved to obtain the following profile for the EOW brane
		   \begin{align}
		   	z=\sigma\kappa z_h \sinh\left(\frac{x-x_B}{z_h}\right)\,,\qquad
		   	\kappa=\frac{\sqrt{1-T^2}}{|T|}>0\,,\qquad
		   	\sigma=\operatorname{sgn}(T)=\pm1\,.\label{brane-BTZ}
		   \end{align}
		   The sign $\sigma$ labels the branch bending toward larger or smaller $x$; the magnitude $\kappa$ is always positive. Equivalently,
		   \begin{align}
		   	T=\frac{\sigma}{\sqrt{1+\kappa^2}}\,.
		   \end{align}
		   A schematic of the brane configuration is shown in figure~\ref{fig:bcft-btz}.
		   Note that the EOW brane anchors on the horizon at the location
		   \begin{align}
		   	x_h=x_B+\sigma z_h\textrm{arcsinh}\left(\frac{1}{\kappa}\right)\,.
		   \end{align} 
		   In the following, we set $x_B=0$ for simplicity and display most intermediate formulae on the $\sigma=+1$ branch. The other branch follows by retaining the explicit $\sigma$ and reversing the brane orientation.
		   
		   A useful refinement of the surface/state viewpoint becomes necessary in the BTZ black-brane geometry.  In the vacuum AdS/BCFT construction, the natural extended surface supporting the PEE data is \eqref{extended-surface}. In the thermal BTZ case this is no longer the complete support.  A constant-time exterior slice terminates not only on the asymptotic boundary and on the EOW brane, but also on the black-brane horizon.  Since the dual BCFT state is thermal, and hence mixed, the horizon should be regarded as a geometric purifier of the exterior thermal density matrix, in close analogy with the thermofield-double interpretation of eternal black holes \cite{Maldacena:2001kr}.  Thus the appropriate extended surface is
		   \begin{align}
		   	\Sigma_e^{\rm BTZ}=\partial M\cup \CQ\cup h ,
		   \end{align}
		   where $h$ denotes the horizon segment on the spatial slice.  This statement should not be interpreted as saying that the horizon is an additional physical BCFT boundary.  Rather, it is an auxiliary purifying component of the surface/state data: PEE threads may end on $h$ because thermal correlations of the exterior region can be purified by degrees of freedom behind, or equivalently represented at, the horizon.  This one-sided description has a direct two-sided origin.  In the reflection-symmetric thermofield-double purification \cite{Maldacena:2001kr}, let $\widetilde x$ denote the endpoint of a cross-boundary geodesic on the second asymptotic boundary and let $y$ be its crossing point on the bifurcation line.  The complete geodesic is cut into two equal halves at
\begin{align}
 y=\frac{x+\widetilde x}{2},\qquad \widetilde x=2y-x,
\end{align}
and the endpoint measure is preserved under this midpoint map,
\begin{align}
 \CI_{\partial h}(x,y)\,\dd x\wedge\dd y
 =\CI_{\partial\widetilde\partial}(x,\widetilde x)\,\dd x\wedge\dd\widetilde x,
 \qquad
 \CI_{\partial h}(x,y)=2\CI_{\partial\widetilde\partial}(x,2y-x).
 \label{eq:BCFT-TFD-horizon-pushforward}
\end{align}
This is the origin of the relative factor of two in the boundary--horizon kernel below; a detailed derivation is given in \cite{BasuWenChandra}.  The map depends on the right endpoint $x$, so a point $y$ on the horizon is a relational label for a complete two-ended TFD chord, not a source-independent local factor of the left-CFT Hilbert space.  The localization is also purification dependent: it is natural for the reflection-symmetric TFD selected by the eternal geometry, and no literal pointwise factorization of the gravitational Hilbert space on the horizon is assumed.  Consequently, the two-point PEE decomposition in the BTZ geometry contains new sectors absent in the
		   Poincar\'e vacuum case, namely boundary--horizon and brane--horizon channels, in addition to the boundary--boundary, boundary--brane and brane--brane sectors.  These horizon-supported kernels are essential for the normalization property of the extended PEE.  In connected phases they combine with the boundary and brane channels so that the final entropy reduces to the standard thermal CFT result, while in adjacent or brane-ending phases they encode the thermal part of the purification together with the EOW-brane contribution. From the bit-thread perspective, this means that the maximal flow is allowed to have flux not only through $A\subset\partial M$ and the brane component
		   $I_A\subset \CQ$, but also through the appropriate portion of the horizon. The BTZ horizon therefore plays for thermal purification the same structural role that the EOW brane plays for island purification: it enlarges the kinematic support of the PEE-thread construction and makes the thermal entropy decomposition compatible with the surface/state correspondence.
		   
		   In the presence of the EOW brane there are six endpoint sectors, of which only five have non-trivial smooth kernels:
		   \begin{itemize}
		   	\item {\bf Boundary-boundary:} The length of the geodesic connecting two points $x$ and $y$ on the asymptotic boundary is given by
		   	\begin{align}
			   	\CL_{\del\del}(x,y)=2\log\left[\frac{2z_h}{\epsilon}\left|\sinh\left(\frac{x-y}{2z_h}\right)\right|\right]
		   	\end{align}
		   	Therefore, the two-point PEE is obtained as
		   	\begin{align}
		   		\CI_{\del\del}(x,y)=\frac{1}{16G z_h^2}\csch^2\left(\frac{x-y}{2z_h}\right)\label{bdy-bdy-PEE-BTZ}
		   	\end{align}
		   	\item {\bf Boundary-horizon}: Similarly consider a geodesic joining a point $x$ on the boundary and another point $y$ on the horizon. Its length is readily computed using the embedding formula \eqref{length-BTZ} as
		   	\begin{align}
		   		\CL_{\del h}(x,y)=\log\left[\frac{2z_h}{\epsilon}\cosh\left(\frac{x-y}{z_h}\right)\right]
		   	\end{align}
			   Therefore, the oriented kernel and its positive density are
			   \begin{align}
			   	\widehat{\CI}_{\del h}(x,y)
			   	&=-\frac{1}{8G z_h^2}\sech^2\left(\frac{x-y}{z_h}\right),&
			   	\CI_{\del h}(x,y)&=\left|\widehat{\CI}_{\del h}(x,y)\right|\,.
			   	\label{bdy-hor-PEE-BTZ}
			   \end{align}
			   \item {\bf Boundary-brane:} The length of a geodesic joining a point $(x,\epsilon)$ on the asymptotic boundary and another point $\left(y,\sigma\kappa z_h\sinh\left(\frac{y}{z_h}\right)\right)$ on the EOW brane may be obtained from \eqref{length-BTZ} as follows
		   	\begin{align}
			   	\CL_{\del \CQ}(x,y)=\log\Bigg[\frac{2z_h}{\epsilon\kappa}\frac{\cosh\left(\frac{x-y}{z_h}\right)-s_\kappa(y)}{\sigma\sinh\left(\frac{y}{z_h}\right)}\Bigg]
		   	\end{align}
		   	where we have defined 
		   	\begin{align}
		   		s_\kappa(y)=\sqrt{1-\kappa^2\sinh^2\left(\frac{y}{z_h}\right)}
		   	\end{align}
		   	Hence the mixed derivative between points on the asymptotic boundary and EOW brane gives the oriented kernel
		   	\begin{align}
		   	\widehat{\CI}_{\del \CQ}(x,y)
		   	&=\frac{1}{8Gz_h^2s_\kappa(y)}
		   	\frac{\cosh\left(\frac{x-y}{z_h}\right)-s_\kappa(y)
		   	-\kappa^2\sinh\left(\frac{x}{z_h}\right)\sinh\left(\frac{y}{z_h}\right)}
		   	{\left[\cosh\left(\frac{x-y}{z_h}\right)-s_\kappa(y)\right]^2}\,.
		   		\label{I-del-B}
		   	\end{align}
		   	Unlike the earlier cases, the coordinate expression of this density does not look symmetric when the ambient BTZ coordinate $y$ is used on the brane. This is a statement about endpoint densities in different coordinates, not a failure of exchange symmetry: the invariant exchange statement $\CI_{\del\CQ}(x,y)\,|\dd x\,\dd y|=\CI_{\CQ\del}(y,x)\,|\dd y\,\dd x|$ is restored once the brane-coordinate Jacobian is included. In the large-tension limit $\kappa\to 0$ the brane approaches the reflected asymptotic boundary, and \eqref{I-del-B} reduces to the boundary kernel \eqref{bdy-bdy-PEE-BTZ}.
		   	\item {\bf Brane-horizon:} Next we consider a geodesic connecting a point $\left(x,\sigma\kappa z_h\sinh\left(\frac{x}{z_h}\right)\right)$ on the EOW brane and another $(y,z_h)$ on the horizon. The length of this geodesic may be computed from \eqref{length-BTZ} as follows:
		   	\begin{align}
			   		\CL_{\CQ h}=\textrm{arccosh}\left[\frac{\cosh\left(\frac{x-y}{z_h}\right)}{\sigma\kappa\sinh\left(\frac{x}{z_h}\right)}\right]
		   	\end{align}
		   	Therefore, the oriented mixed kernel and its positive density are
		   	\begin{align}
			   		\widehat{\CI}_{\CQ h}(x,y)=-\frac{1}{8G}\frac{\cosh \left(\frac{x-y}{z_h}\right)-\kappa ^2 \sinh \left(\frac{x}{z_h}\right) \sinh \left(\frac{y}{z_h}\right)}{z_h^2 \left[\cosh ^2\left(\frac{x-y}{z_h}\right)-\kappa ^2 \sinh ^2\left(\frac{x}{z_h}\right)\right]^{3/2}}\,,\qquad
		   		\CI_{\CQ h}=\left|\widehat{\CI}_{\CQ h}\right|\,.
		   	\end{align}
			   	The apparent asymmetry of its coordinate expression is again removed only after the exchanged endpoint density and the relevant Jacobian are included. In the large-tension limit $\kappa\to 0$, the oriented kernel above reduces to $\widehat{\CI}_{\del h}$ in \eqref{bdy-hor-PEE-BTZ}.
		   	\item{\bf Brane-brane:} Finally, consider a geodesic joining two points on the EOW brane with length given by
		   	\begin{align}
		   		\CL_{\CQ\CQ}=\textrm{arccosh}\left[\frac{\cosh \left(\frac{x-y}{z_h}\right)-s_\kappa(x)s_\kappa(y)}{\kappa ^2 \sinh \left(\frac{x}{z_h}\right) \sinh \left(\frac{y}{z_h}\right)}\right]
		   	\end{align}
		   	from which the corresponding two-point PEE is readily determined. To simplify notations, we make changes to the following variables
		   	\begin{align}
		   		\xi=\sqrt{1+\kappa^2}\tanh\left(\frac{x}{z_h}\right)~~,~~\chi=\sqrt{1+\kappa^2}\tanh\left(\frac{y}{z_h}\right)\label{var}
		   	\end{align}
		   	in terms of which the two-point PEE has the following form 
		   	{\footnotesize
		   	\begin{align}
		   		\widehat{\CI}_{\CQ\CQ}(\xi,\chi)
		   		&=-\frac{(1+\kappa^2)^2}{8G}
		   		\left[2(1-\xi\chi)
		   		+\frac{(\xi+\chi)^2-2(\xi^2\chi^2+1)}
		   		{\sqrt{(1-\xi^2)(1-\chi^2)}}\right]\notag\\
		   		&\quad\times\left[\left\{\xi\chi+(1+\kappa^2)
		   		\left(\sqrt{(1-\xi^2)(1-\chi^2)}-1\right)\right\}^2
		   		-\kappa^4\xi^2\chi^2\right]^{-3/2},\notag\\
		   		\CI_{\CQ\CQ}(\xi,\chi)
		   		&=\left|\widehat{\CI}_{\CQ\CQ}(\xi,\chi)\right|\,.
		   	\end{align}
		   	}
		   	In the large-tension limit $\kappa\to 0$, the above two-point PEE agrees with \eqref{bdy-bdy-PEE-BTZ} once the Jacobians of the transformations \eqref{var} are taken into account.
		   	\item{\bf Horizon-horizon:} At equal time the horizon distance is
		   	\begin{align}
		   		\CL_{hh}(x,y)=\frac{|x-y|}{z_h}\,.
		   	\end{align}
		   	Its smooth mixed derivative vanishes for $x\neq y$, while distributionally
		   	\begin{align}
		   		\partial_x\partial_y\CL_{hh}(x,y)=-\frac{2}{z_h}\delta(x-y)\,,\qquad
		   		\widehat{\CI}_{hh}(x,y)=-\frac{1}{4Gz_h}\delta(x-y)\,.
		   	\end{align}
		   	Thus there is no independent smooth horizon--horizon source congruence. The contact term records coincident endpoint data and is not used as a separate bulk source field away from the source.
		   \end{itemize}
		   	\subsubsection*{Entanglement entropy and normalization property}
		   	Now we discuss the normalization property of the extended two-point PEE and determine the entanglement entropy for various subregions in AdS/BCFT. We begin with the subsystem $A=[0,b]$ adjacent to the boundary. The brane-ending RT surface is the appropriate segment of the complete BTZ geodesic
		   	\begin{align}
		    		\sqrt{1-\frac{z^2}{z_h^2}}=\cosh\left(\frac{x}{z_h}\right)\sech\left(\frac{b}{z_h}\right),
		   	\end{align}
		   	which is rederived intrinsically in appendix~\ref{app:geodesics}. Its intersection with \eqref{brane-BTZ} gives
		   	\begin{align}
		   		x_\CQ=\sigma z_h \,\textrm{arctanh}\left[\frac{1}{\sqrt{1+\kappa^2}}\tanh\left(\frac{b}{z_h}\right)\right]\label{x-brane}
		   	\end{align}
		   	Assuming that the normalization property holds true, one may find the entanglement entropy in terms of various PEEs as follows \cite{Basu:2023wmv}
		   	\begin{align}
		   		S_A
		   		&=\widehat{\CI}(A,B)+\widehat{\CI}(A,I_B)+\widehat{\CI}(A,h)\notag\\
		   		&\quad+\widehat{\CI}(I_A,B)+\widehat{\CI}(I_A,I_B)+\widehat{\CI}(I_A,h)
		   	\end{align}
		   	where $B$ denotes the complement of the subsystem $A$ on the asymptotic boundary, $h$ denotes the horizon, and $I_A$ and $I_B$ denote the brane components assigned to $A$ and $B$, respectively. They may be interpreted as island regions only in a doubly holographic completion.

		   	The contributions from different types of two-point PEEs to the entanglement entropy are listed in appendix \ref{appB}.
		   	Finally, including all these contributions, we find 
		   	\begin{align}
		   		S_A
		   		&=\frac{1}{4G}\log \left[\frac{2z_h}{\epsilon }\sinh \left(\frac{b}{z_h}\right)\right]+\frac{\sigma}{4G}\operatorname{arcsinh}\left(\frac{1}{\kappa}\right)
		   	\end{align}
		   	where we may identify the additive constant term as the boundary entropy
		   	\begin{align}
		   		S_\textrm{bdy}=\frac{\sigma}{4G}\operatorname{arcsinh}\left(\frac{1}{\kappa}\right)=\frac{\sigma\rho_0}{4G}\,,\qquad
		   		\rho_0=\operatorname{arctanh}|T|\,.\label{S-bdy-BTZ}
		   	\end{align}
		   	The sign is fixed by the branch/orientation label $\sigma$; it is not carried by $\kappa$.
		   	
		   	\begin{figure}[ht]
		   		\centering
		   		\includegraphics[width=0.55\linewidth]{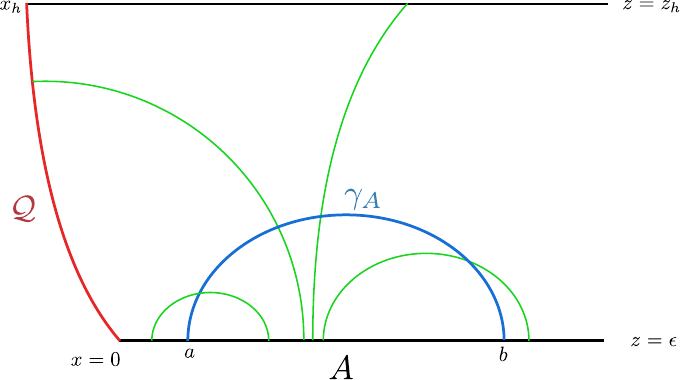}
		   		\caption{Entanglement entropy from two-point PEE in a thermal BCFT. The subsystem $A=[a,b]$ is away from the boundary, leading to the connected configuration.}
		   		\label{fig:bcft-pee2}
		   	\end{figure}
		   	Next we consider the subsystem $A=[a,b]$ away from the boundary; the connected thermal configuration is illustrated in figure~\ref{fig:bcft-pee2}. The disconnected thermal phase requires the same intersection-weighted construction as its Poincar\'e counterpart and is not evaluated explicitly here. Accordingly, the explicit BTZ normalization and flow analysis in this paper is restricted to the adjacent brane-ending phase and the connected phase. For the latter,
		   	\begin{align}
		   		S_A&=\int_{a}^{b}\d x\left(\int_{0}^{a}+\int_{b}^{\infty}\right)\d y\,\CI_{\del\del}(x,y)+\int_{a}^{b}\d x\int_{\infty}^{x_h}\d y\, \widehat{\CI}_{\del h}(x,y)+\int_{a}^{b}\d x\int_{x_h}^{0}\d y\, \widehat{\CI}_{\del \CQ}(x,y)\notag\\
		   		&=\frac{1}{8G}\Bigg(2 \log \left[\frac{4 z_h^2 \sinh \left(\frac{a}{2 z_h}\right) \sinh ^2\left(\frac{b-a}{2 z_h}\right)}{\epsilon ^2 \sinh \left(\frac{b}{2 z_h}\right)}\right]+\log \left[\frac{\sqrt{\kappa ^2+1} \cosh \left(\frac{b}{z_h}\right)-\sinh \left(\frac{b}{z_h}\right)}{\sqrt{\kappa ^2+1} \cosh \left(\frac{a}{z_h}\right)-\sinh \left(\frac{a}{z_h}\right)}\right]
		   		\notag\\
		   		&\qquad\qquad+\log \left[\frac{\left(\cosh \left(\frac{b}{z_h}\right)-1\right) \left(\sqrt{\kappa ^2+1} \cosh \left(\frac{a}{z_h}\right)-\sinh \left(\frac{a}{z_h}\right)\right)}{\left(\cosh \left(\frac{a}{z_h}\right)-1\right) \left(\sqrt{\kappa ^2+1} \cosh \left(\frac{b}{z_h}\right)-\sinh \left(\frac{b}{z_h}\right)\right)}\right]\Bigg)\notag\\
		   		&=\frac{1}{2G}\log\left[\frac{2z_h}{\epsilon}\sinh\left(\frac{b-a}{2z_h}\right)\right]
		   	\end{align}
		   	This cancellation is a useful check of the extended PEE kernels. It also clarifies their physical role: the extended kernels do not modify the connected RT answer; they provide a decomposition of it into boundary, brane, and horizon channels. Hence, with our extended definitions for the two-point PEEs, the normalization property of the two-point PEE is satisfied for different configurations of RT surfaces in AdS$_3$/BCFT$_2$.
		   	
		   	The computations above show that the two-point PEE construction extends consistently to AdS/BCFT once the EOW brane is treated as part of the extended surface supporting the surface/state data. In adjacent brane-ending phases, the mixed and brane-brane PEE kernels reproduce the Affleck--Ludwig boundary entropy, thereby giving a microscopic PEE interpretation of the brane contribution to the RT formula. In
		   	connected phases for intervals away from the boundary, the same kernels reorganize themselves so that all brane-dependent terms cancel, leaving the ordinary CFT result. For disconnected phases, however, the normalization property must be understood in its intersection-weighted form: the entropy is not obtained by summing all endpoint pairings, but by counting PEE threads according to their intersections with the relevant homologous RT surface. This distinction will be important in the construction of PEE-thread and bit-thread configurations in the following sections.

		   \section{PEE threads and bit threads in AdS/BCFT}
		   \label{sec:pee-threads}
		   \subsection{Poincar\'e AdS$_3$}
		   In this section, we will derive expressions for various PEE-thread flows in the Poincar\'e AdS$_3$ truncated by the EOW brane \eqref{brane}
		   dual to the vacuum state of the BCFT$_2$, following the procedure outlined in \cite{Lin:2023rxc}. 
		   
		   We consider the simple choice of the reference subsystem $A=[0,b]$ adjacent to the boundary $x=0$. The reference RT surface has the following semicircular profile
		   \begin{align}
		   	\Sigma~:~x^2+z^2=b^2\,,\label{RT-profile}
		   \end{align}
		   which joins $(x,z):(b,0)$ on the boundary to $(-b\sin\theta,b\cos\theta)$ on the brane. In the following analysis, we shall require the unit normal to this RT surface:
		   \begin{align}
		   	n^{}_{\Sigma,\mu}=\frac{1}{z\sqrt{x^2+z^2}}(x,z)\,.\label{normal-RT}
		   \end{align}
		   
		   Along with the PEE threads connecting points on the asymptotic boundary \eqref{PEE-thread-flow}, we have the following two classes of threads specific to the BCFT$_2$:
		   \begin{itemize}
		   	\item \textbf{Class-I:} Threads joining the EOW brane to the asymptotic boundary. These have the following semicircular profile
		   	\begin{align}
		   		(x-x_c)^2+z^2=r^2\label{circle-eqn}
		   	\end{align}
		   	with the center and radius given by
		   	\begin{align}
		   		x_c=\frac{x_0^2-\hat y^2}{2 (x_0+\hat y \sin\theta)}~~,~~r=x_0-x_c\,.\label{center-radius-class-I}
		   	\end{align}

		   	\item \textbf{Class-II:} Threads joining two points on the EOW brane. These have the same profile \eqref{circle-eqn}, with the center and radius given by
		   	\begin{align}
		   		x_c=-\frac{\hat y+\hat y_0}{2\sin\theta}~~,~~r^2=\left(\frac{\hat y+\hat y_0}{2\sin\theta}\right)^2-\hat y\,\hat y_0\,.\label{center-radius-class-II}
		   	\end{align}
		   \end{itemize}
		   \subsection{PEE threads in Class-I}
		   We may sub-classify the PEE threads in class-I as follows
		   \begin{enumerate}
		   	\item Threads emanating from the EOW brane, crossing the reference RT surface and ending on the asymptotic boundary.
		   	\item Threads emanating from the asymptotic boundary and ending on the EOW brane.
		   \end{enumerate}
		   Although the threads belonging to these two sub-classes have the same topology, the crucial difference lies in the expression for the PEE-thread flow vector at arbitrary bulk points. We clarify this statement towards the end of this subsection.
		   
		   We begin with the computation of the norm of the PEE-thread flow for the threads emanating from the EOW brane and crossing the reference RT surface at the location $P_m:(x_m,z_m)$.
		   Solving \eqref{circle-eqn} and \eqref{RT-profile}, we may obtain the location $\hat y$ where the PEE thread anchors on the EOW brane as follows
		   \begin{align}
		   	x_m=\frac{x_0 \left(b^2-\hat y^2\right)+\hat y \left(b^2-x_0^2\right) \sin \theta}{x_0^2-\hat y^2}\label{crossing-point}
		   \end{align}
		   The hyperbolically unit-normalized tangent vector to this PEE thread is
		   \begin{align}
		   	\tau^\mu
		   	=\frac{z^2}{\sqrt{(x-x_c)^2+z^2}}\left(1,\frac{x_c-x}{z}\right)
		   	=\frac{z}{r}(z,x_c-x)\,,\qquad
		   	g_{\mu\nu}\tau^\mu\tau^\nu=1\,.
		   	\label{tangent-circle}
		   \end{align}
		   Here $x_c$ is given in \eqref{center-radius-class-I}.
		   Now, from \cref{normal-RT} we have the inner product
		   \begin{align}
		   	n^{}_\Sigma\cdot\tau\Big|_{(x_m,z_m)}&=\frac{x_c z_m}{\sqrt{(x_m^2+z_m^2)\left((x_m-x_c)^2+z_m^2\right)}}\notag\\&=\frac{z_m\left(x_m^2+z_m^2-\hat y^2\right)}{\sqrt{x_m^2+z_m^2} \sqrt{\left(x_m^2+z_m^2+\hat y^2+2 x_m \hat y \sin \theta\right)^2-4 \hat y^2 z_m^2 \cos ^2\theta}}
		   \end{align}
		   where, in the second equality, we have used \eqref{crossing-point}.
		   Furthermore, the induced volume element on the RT surface is obtained as
		   \begin{align}
		   	\d\Sigma=\sqrt{\frac{\d x_m^2+\d z_m^2}{z_m^2}}=\frac{\sqrt{x_m^2+z_m^2}}{z_m^2}\d x_m
		   \end{align}
		   Finally, we utilize the PEE-thread flow proposal described in \cite{Lin:2023rxc}
		   \begin{align}
		   	\int_\Sigma\widehat f_{\hat y}(x_m,z_m)\left(n^{}_\Sigma\cdot\tau\right)\d \Sigma=\int_{A^c}\widehat{\CI}_{\del\CQ}(x_0,\hat{y})\,\d x_0
		   \end{align}
		   to obtain the signed scalar amplitude
		   \begin{align}
		   	\widehat f_{\hat y}(x_m,z_m)=\frac{1}{4G}\frac{2 z_m \left(2 x_m \hat y+\left(x_m^2+z_m^2+\hat y^2\right)\sin \theta\right)}{\left(x_m^2+z_m^2+\hat y^2+2 x_m\hat y \sin \theta\right)^2-4 \hat y^2 z_m^2 \cos ^2\theta}\,,\qquad
		   	\left|V_{\hat y}\right|=\left|\widehat f_{\hat y}\right|\,.
		   \end{align}
		   Here we have used \eqref{crossing-point} to replace the area element on $A^c$ in terms of $\d x_m$ along the RT surface, and have written
		   \begin{align}
		   	\widehat{\CI}_{\del\CQ}(x_0,\hat{y})\,\d x_0&=\frac{1}{4G}\frac{2x_0\hat y+(x_0^2+\hat y^2)\sin\theta}{\left(x_0^2+\hat y^2+2x_0\hat y\sin\theta\right)^2}\d x_0\notag\\&=\frac{1}{4G}\frac{2 \left(x_m^2+z_m^2-\hat y^2\right) \left(2 x_m\hat y+\left(x_m^2+z_m^2+\hat y^2\right)\sin\theta\right)}{\left[\left(x_m^2+z_m^2+\hat y^2+2 x_m \hat y \sin \theta\right)^2-4 \hat y^2 z_m^2 \cos ^2\theta \right]^{3/2}}\d x_m
		   \end{align}
		   Finally, noting that on the reference RT surface, we have
			   {\small
			   \begin{align}
			   	\tau^\mu\big|_\Sigma
			   	&=\left(\frac{2 z_m^2 (x_m-x_0)}{(x_m-x_0)^2+z_m^2},
			   	\frac{z_m\left[z_m^2-(x_m-x_0)^2\right]}{(x_m-x_0)^2+z_m^2}\right)\notag\\
			   	&=\frac{2z_m^2}{\sqrt{\left(x_m^2+z_m^2+\hat y^2+2 x_m \hat y \sin \theta \right)^2-4 \hat y^2 z_m^2 \cos ^2\theta}}\notag\\
			   	&\qquad\times\left(x_m+\hat y \sin\theta,
			   	-\frac{x_m^2-z_m^2+\hat y^2+2 x_m \hat y \sin \theta }{2z_m}\right)
			   \end{align}
			   }
			   we may write down the expression for the PEE-thread flow emanating from a point $\hat y$ on the EOW brane and passing through a bulk point $(x,z)$ as follows
			    {\small
			    \begin{align}
			   	V^\mu_{\hat y}(x,z)
			   	&=\frac{1}{2G}\frac{z^3 \left(2 x \hat y+\left(x^2+\hat y^2+z^2\right)\sin\theta \right)}
			   	{\left[\left(x^2+\hat y^2+z^2+2 x \hat y \sin\theta\right)^2-4 \hat y^2 z^2 \cos ^2\theta\right]^{3/2}}\notag\\
			   	&\qquad\times\left(x+\hat y \sin\theta,
			   	-\frac{x^2+\hat y^2-z^2+2 x \hat y \sin\theta}{2 z}\right)
			   	\label{PEE-thread-flow-class-I}
			   \end{align}
			   }
		   In the limit $\theta\to\frac{\pi}{2}$, the above expression reduces to
		   \begin{align}
		    \lim_{\theta\to\frac{\pi}{2}}V^\mu_{\hat y}(x,z)=\frac{1}{4G}\frac{2z^2(x+\hat y)}{\left[(x+\hat y)^2+z^2\right]^2}\left(z,\frac{z^2-(x+\hat y)^2}{2(x+\hat y)}\right)
		   \end{align}
		   Incidentally, this is identical to \cref{PEE-thread-flow} with the replacement $x_0\to -\hat y$. Once again, this may be attributed to the fact that as $\theta\to\frac{\pi}{2}$, the EOW brane approaches the asymptotic boundary.
		   
		   Next, we consider the PEE threads emanating from the asymptotic boundary. It is natural to expect the flow vector to have the form \eqref{PEE-thread-flow} irrespective of its endpoint on the EOW brane. We utilize a natural extension of the PEE-thread flow proposal described in \cite{Lin:2023rxc}
		   \begin{align}
		    \int_\Sigma\widehat f_{x_0}\left(n^{}_\Sigma\cdot\tau\right)\d \Sigma=\int_{\textrm{I}_{A^c}}\widehat{\CI}_{\del\CQ}(x_0,\hat{y})\,\d \hat y\,,\qquad |V_{x_0}|=|\widehat f_{x_0}|
		   \end{align}
		   where $\textrm{I}_{A^c}$ is the brane component assigned to the complement of $A$; it is an island region only in a doubly holographic interpretation.
		    \begin{figure}[ht]
		   	\centering
		   	\includegraphics[width=0.9\linewidth]{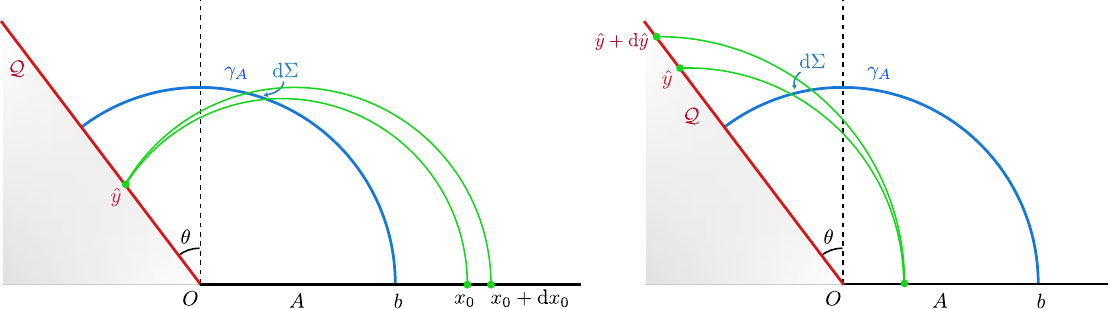}
		   	\caption{Mapping the area elements on the EOW brane onto that on the asymptotic boundary.}
		   	\label{fig:thread-i-computation}
		   \end{figure}
		   As described earlier, the PEE threads for the two sub-classes have the same topology and we expect the corresponding thread flow vectors to be related to each other. It is important to note that a naive substitution of \eqref{crossing-point} to the expression \eqref{PEE-thread-flow-class-I} does not lead to \eqref{PEE-thread-flow}. In order to remedy this, we note that the extended PEE threads proposal may be written in the following two forms (cf. \cref{fig:thread-i-computation})
		   \begin{align}
		   	\left(V_{\hat y}\cdot n^{}_\Sigma\right)\,\d\Sigma=\widehat{\CI}_{\del\CQ}(x_0,\hat y)\d x_0~~,~~\left(V_{x_0}\cdot n^{}_\Sigma\right)\,\d\Sigma=\widehat{\CI}_{\del\CQ}(x_0,\hat y)\d \hat y
		   \end{align}
		   These relations imply
		   \begin{align}
		   	\frac{V_{\hat y}\cdot n^{}_\Sigma}{V_{x_0}\cdot n^{}_\Sigma}=\del_{\hat y}x_0
		   \end{align}
		   It is straightforward to verify that the two expressions of the PEE threads flow are related through this Jacobian factor. This explains why the brane-sourced and boundary-sourced vector fields have the same geodesic support but different local densities. The difference is not physical; it reflects the choice of which endpoint is used as the source coordinate
		   for the PEE-thread bundle.
		   \subsection{PEE threads in Class-II}
		   In this subsection, we consider the PEE threads belonging to class-II. To be precise, we consider a PEE thread joining a source point $\hat y_0\in [0,b]$ to another point $\hat y$ on the EOW brane. The center and radius of this semicircular PEE thread are given in \eqref{center-radius-class-II}. Utilizing \eqref{tangent-circle}, we may obtain the unit tangent vector to this thread and subsequently compute the inner product
		   \begin{align}
		    n^{}_\Sigma\cdot \tau\big|_{(x_m,z_m)}=\frac{x_c z_m}{\sqrt{x_m^2+z_m^2} \sqrt{(x_m-x_c)^2+z_m^2}}
		   \end{align}
		   where $P_m:(x_m,z_m)$ denotes the point where the PEE thread crosses the reference RT surface \eqref{RT-profile}. We may obtain the target location $\hat y$ after the thread crosses the reference surface by solving \eqref{RT-profile} and \eqref{circle-eqn}:
		   \begin{align}
		   	\hat y=-\frac{(x_m^2+z_m^2)\sin\theta+x_m\hat y_0}{x_m+\hat y_0\sin\theta}
		   \end{align}
		   Finally, utilizing the flux matching condition, we may write down the expression for the PEE-thread flow vector in class-II as follows
			   {\small
			   \begin{align}
			   	V^\mu_{\hat y_0}(x,z)
			   	&=\frac{1}{2G}\frac{z^3 \left(2 x \hat y_0+\left(x^2+\hat y_0^2+z^2\right)\sin\theta \right)}
			   	{\left[\left(x^2+\hat y_0^2+z^2+2 x \hat y_0 \sin\theta\right)^2-4 \hat y_0^2 z^2 \cos ^2\theta\right]^{3/2}}\notag\\
			   	&\qquad\times\left(x+\hat y_0 \sin\theta,
			   	-\frac{x^2+\hat y_0^2-z^2+2 x \hat y_0 \sin\theta}{2 z}\right)
			   \end{align}
			   }
		   Remarkably, the above expression is identical to the PEE-thread flow in class-I. Therefore, the elementary flow sourced from a brane point has the same functional form as the Class-I brane-sourced flow. In other words, the PEE-thread vector field depends only on the source point on the EOW brane, not on whether the other endpoint of the corresponding geodesic lies on the asymptotic boundary or on the EOW brane.

		   \subsection{Bit threads}
		   The bit-thread vector field for the subsystem $A=[0,b]$ is obtained by integrating the respective PEE-thread flows over the asymptotic region $A$ and over the brane component ${\rm I}_A$, parametrized by $0\leq\hat y\leq b$. The brane is oppositely oriented in the closed extended boundary, so its source-current integral runs from $\hat y=b$ to $\hat y=0$:
		   \begin{align}
		   	v^\mu_A=\int_0^b \d x_0\,V^\mu_{x_0}+\int_b^0\d \hat y\,V^\mu_{\hat y}\,.
		   \end{align}
		   Upon performing the integrals explicitly, we find
		   \begin{align}
		   	v^\mu_A=\frac{1}{8 G}\Bigg(-z+\frac{2z^3}{D(x,z)}&+\frac{z (\CA_\theta(x,z)-2z^2)}{\sqrt{\CA_\theta(x,z)^2-4b^2z^2\cos^2\theta}},\notag\\&2z^2\left(\frac{b-x}{D(x,z)}+\frac{x+b\sin\theta}{\sqrt{\CA_\theta(x,z)^2-4b^2z^2\cos^2\theta}}\right)\Bigg)\label{bit-Poincare-BCFT}
		   \end{align}
		   where we have defined
		   \begin{align}
		   	D(x,z)=(b-x)^2+z^2~~,~~\CA_\theta(x,z)=b^2+x^2+z^2+2bx\sin\theta
		   \end{align}
		   This is one of the main results in this manuscript. On the RT surface \eqref{RT-profile}, we have
		   \begin{align}
		   	v^\mu_A=\frac{1}{4G\,b }\left(x\sqrt{b^2-x^2},b^2-x^2\right)\,,\label{BCFT-bit threads-on-RT}
		   \end{align}
		   from which it is easy to verify that $v^\mu_A$ is normal to the RT surface. Furthermore, we may find that the norm of $v^\mu_A$ satisfies the following inequality
		   \begin{align}
		   	\left(4G\left|v^{}_A\right|\right)^2=\frac{1}{2} \left(1+\sqrt{1-\frac{4 b^2 z^2 (1+\sin \theta)^2 \left(-b^2+x^2+z^2\right)^2}{D(x,z)^2 \left(\CA_\theta(x,z)^2-4 b^2 z^2 \cos ^2\theta \right)}}\right)\leq 1\label{norm-bound}
		   \end{align}
		   which follows readily from the fact that for points away from the RT surface, we have $\CA_\theta(x,z)>2 b z \cos\theta$.
		   For points on the RT surface, the inequality is saturated: $\left|v_A\right|_\Sigma=\frac{1}{4G}$, which is also seen readily from \eqref{BCFT-bit threads-on-RT}. 
		   Interestingly, in the limit $\theta\to\frac{\pi}{2}$, we recover the geodesic bit threads for the subsystem $A=[-b,b]$ in the CFT$_2$, reported in \eqref{bit-threads-CFT}. This may be attributed to the fact that in the limit $\theta\to\frac{\pi}{2}$, the EOW brane approaches the asymptotic boundary and the system mimics the CFT$_2$ on the whole complex plane. 
		   
		   \paragraph{Integral curves of the flow:} For a legitimate bit-thread candidate, the flow \eqref{bit-Poincare-BCFT} must be divergenceless and obey the norm bound and boundary conditions. We additionally investigate whether its integral curves furnish a regular non-crossing foliation. This is distinct from the simultaneous-locking notion of bit-thread nesting for a family of nested boundary regions. The integral curves satisfy the first-order differential equations
		   \begin{align}
		   	\frac{\dd x(s)}{\dd s}=v^x(x(s),z(s))~~,~~\frac{\dd z(s)}{\dd s}=v^z(x(s),z(s))
		   \end{align}
		   Solving this set of coupled differential equations is cumbersome. Instead, we introduce a stream function $\Psi(x,z)$ through the one-form relation $v_A=(8G)^{-1}\star\dd \Psi$, where the $\star$ denotes the Hodge dual operation (see appendix \ref{app:distance-difference-potential}). Such a stream function is readily obtained by looking for a first integral of motion. Integrating the relation $\frac{\dd z}{\dd x}=\frac{v^z_A}{v^x_A}$, we obtain the first integral as
		   \begin{align}\label{Psipp}
		   	\Psi(x,z)=\log\frac{D(x,z)}{z\ell_\star}-\operatorname{arccosh}\left(\frac{\CA_\theta(x,z)}{2bz \cos\theta}\right)\,,
		   \end{align}
		   where $\ell_\star$ is an arbitrary reference scale. Changing $\ell_\star$ only shifts $\Psi$ by a constant and therefore leaves the vector field and its level-set foliation unchanged. One may check directly that
		   \begin{align}
		   	v^x_A=\frac{z^2}{8G}\del_z \Psi~~,~~v^z_A=-\frac{z^2}{8G}\del_x \Psi\label{v-del-Psi}
		   \end{align}
		   from which the divergenceless condition follows readily:
		   \begin{align}
		   	\nabla\cdot v_A=\frac{1}{\sqrt{g}}\del_\mu(\sqrt{g}v_A^\mu)=0\,.
		   \end{align}
		   The integral curves of the bit-thread flow are the level sets of $\Psi$. Using hyperbolic identities, we may write the level sets as
		   \begin{align}
		   	\chi_0^2D(x,z)^2-2\chi_0D(x,z)\CA_\theta(x,z)+4b^2z^2\cos^2\theta=0
		   \end{align}
		   with the branch condition
		   \begin{align}
		   	\chi_0D(x,z)-\CA_\theta(x,z)\geq 0
		   \end{align}
		   where
		   \begin{align}
		   	\chi_0=\frac{2b\cos\theta}{\ell_\star e^{\Psi_0}}>0
		   \end{align}
		   labels the level set $\Psi=\Psi_0$; the displayed branch condition removes the extraneous branch introduced by squaring the inverse-hyperbolic function. This is the most compact implicit equation for the integral curves.
		   
\begin{figure}[ht]
		   	\centering
		   	\includegraphics[width=0.49\linewidth]{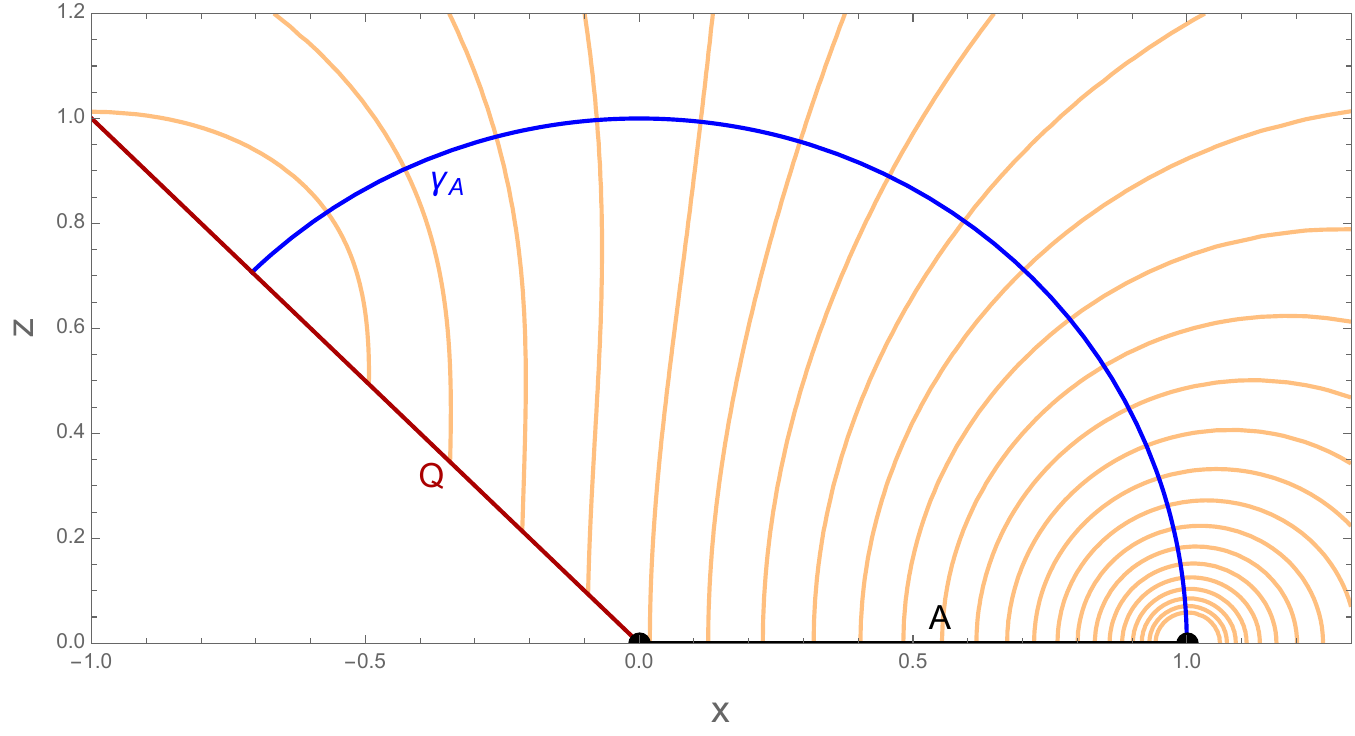}
		   	\includegraphics[width=0.49\linewidth]{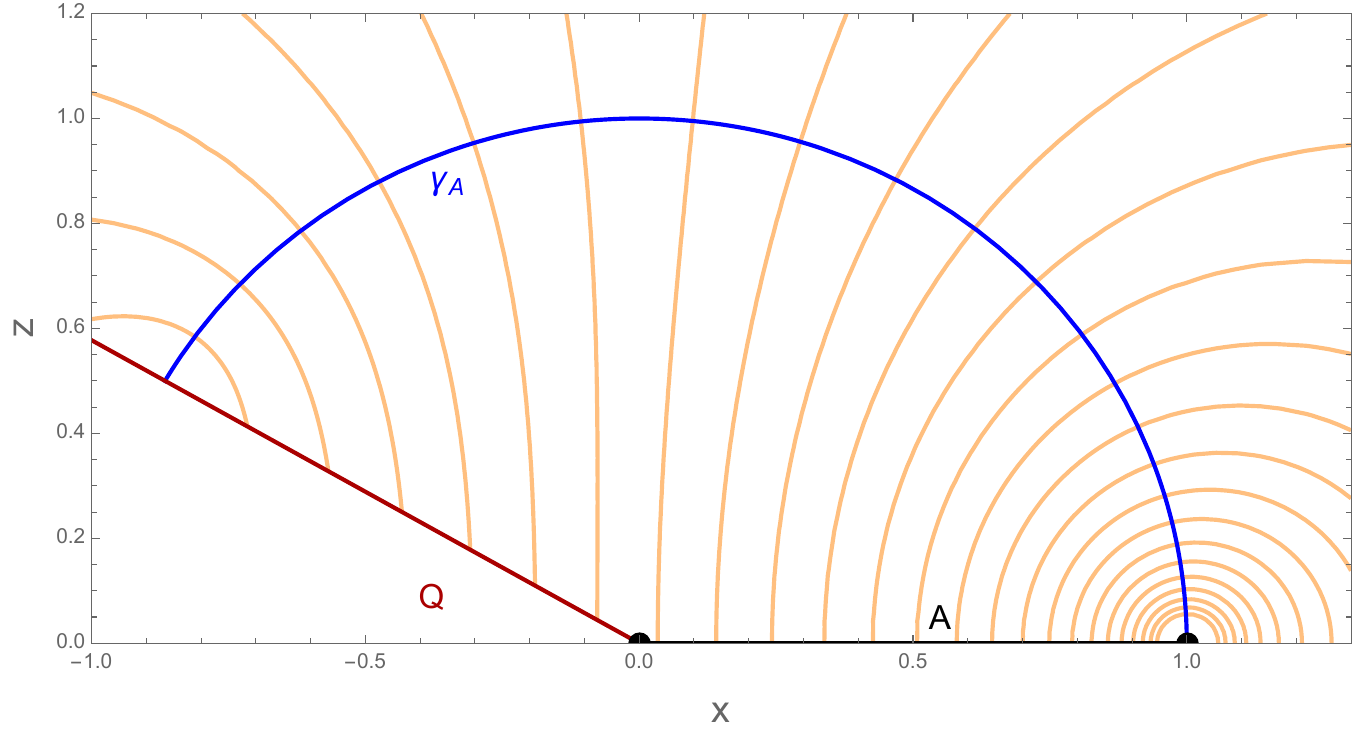}
		   	\includegraphics[width=0.49\linewidth]{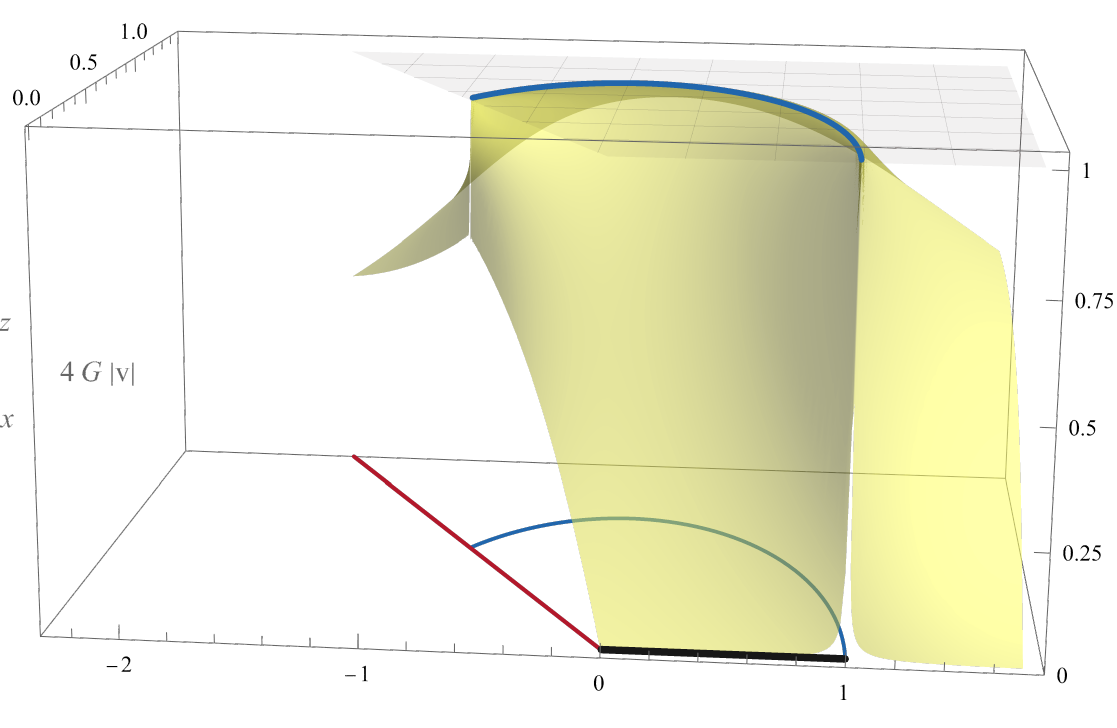}
		   	\includegraphics[width=0.49\linewidth]{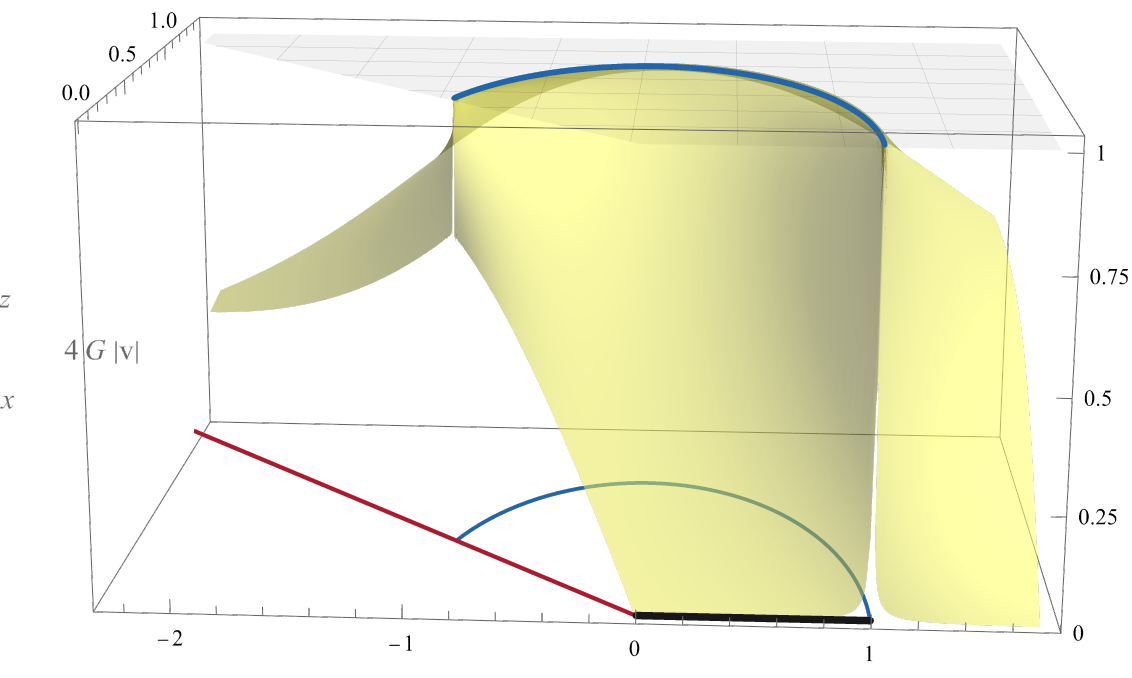}
		   	\caption{Bit-thread configuration in Poincar\'e AdS$_3$ truncated by an EOW brane, constructed from the PEE threads in AdS/BCFT. These are the integral lines of the bit-thread flow $v^\mu_A$. Left panel: $\theta=\frac{\pi}{4}$; right panel: $\theta=\frac{\pi}{3}$. We have set $b=1$. The lower panels depict the norm of the bit-thread flows, which saturate precisely on the RT surface.}
		   	\label{fig:bit-threads-poincare-bcft}
		   \end{figure}

		   Interestingly, for generic $\theta$, this is a quartic curve, not a semicircular geodesic. Thus the streamlines of this bit-thread flow are generically not geodesics, even though the elementary PEE threads being superposed are geodesics. This is in apparent tension with the construction in \cite{Lin:2023rxc}, where it was shown that for a static interval or spherical region
		   in holographic CFT, the superposition of geodesic PEE threads can produce a canonical geodesic bit-thread congruence. The crucial point, however, is that this geodesic closure is a special consequence of
		   the residual symmetry of an ordinary connected boundary region. In the present AdS/BCFT problem adapted to the surface/state interpretation, the relevant region is not merely the boundary interval $A=[0,b]$,
		   but the extended region $A\cup I_A$, with one component on the asymptotic boundary and one component on the EOW brane. Although the elementary PEE threads entering the construction are still geodesics, the final bit-thread current is a density-weighted sum of boundary-sourced and brane-sourced geodesic congruences. The integral curves of this summed current are therefore the level sets $\Psi(x,z)=\Psi_0$ which are generically quartic curves rather than semicircular geodesics. This shows that the integral curves of the macroscopic bit-thread flow need not coincide with the individual geodesic PEE threads from which the flow is assembled. The distinction is not caused simply by the presence of brane-anchored PEE threads. As we shall see later, for an interval $A=[a,b]$ away from the BCFT boundary in the connected RT phase, the same superposition procedure, even when rewritten using brane-anchored threads together with complementary boundary threads, collapses to the standard geodesic bit-thread congruence. Thus the correct distinction is between ordinary connected interval phases, where the PEE-thread superposition enjoys geodesic closure, and genuinely extended-boundary phases, where the brane component changes the chamber structure and the resulting max-flow streamlines are generically non-geodesic. This is also consistent with the broader lesson of \cite{Lin:2023rxc}: for general regions, such as disconnected intervals, the naive vector-field superposition of PEE threads is not the fundamental object, and one must instead count PEE threads with appropriate intersection weights with homologous surfaces. From this perspective, the AdS/BCFT result is not a contradiction to the
		   PEE-thread picture, but a refinement of it: geodesicity is a property of the elementary PEE threads, whereas the physical bit-thread flow is constrained only by divergencelessness, the norm bound, and saturation/normality on the RT surface. The non-geodesic AdS/BCFT flow \eqref{bit-Poincare-BCFT} satisfies these requirements and its streamlines form a regular non-crossing foliation of the physical bulk region. In figure \ref{fig:bit-threads-poincare-bcft}, we have shown plots of the integral lines for $\theta=\frac{\pi}{3}\,,\,\frac{\pi}{4}$.
		   
		   \begin{figure}[ht]
		   	\centering
		   	\includegraphics[width=0.95\textwidth]{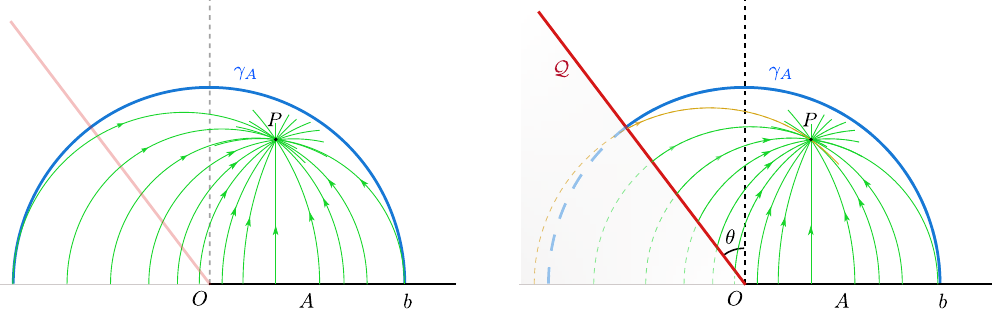}
		   	\caption{Superposition of PEE threads leading to bit threads. The left panel shows the usual AdS/CFT configuration, while the right panel shows the configuration in the presence of an EOW brane. In the right panel, the RT surface and the PEE threads inside the region bounded by the EOW brane are just portions of those in the left panel.}
		   	\label{fig:Non-geodesic-illustration}
		   \end{figure}
		   Figure~\ref{fig:Non-geodesic-illustration} illustrates the geometric origin of this distinction. In the ordinary AdS/CFT configuration shown in the left panel, we superpose the PEE flows emanating from all points of the boundary interval $[-b,b]$ to obtain the bit-thread flow at the point $P$. In the AdS/BCFT configuration shown in the right panel, the bit-thread flow at $P$ is the superposition of all the PEE flows emanating from $A\cup I_A$. Compared with the left panel, additional PEE flows, e.g., the brown PEE threads, contribute to the bit-thread flow, thereby producing a different bit-thread flow at $P$. Since the bit threads constructed in the left panel are geodesics \cite{Lin:2023rxc}, those in the right panel are generically non-geodesic. However, when $P$ lies on the RT surface $\gamma_A$, the PEE flows contributing to the bit-thread flow at $P$ are the same in the two configurations; consequently the bit-thread flow on $\gamma_A$ is the same in both cases, namely normal to $\gamma_A$ with norm $\frac{1}{4G}$.
		   
\begin{figure}[ht]
		   		\centering
		   		\includegraphics[width=0.54\linewidth]{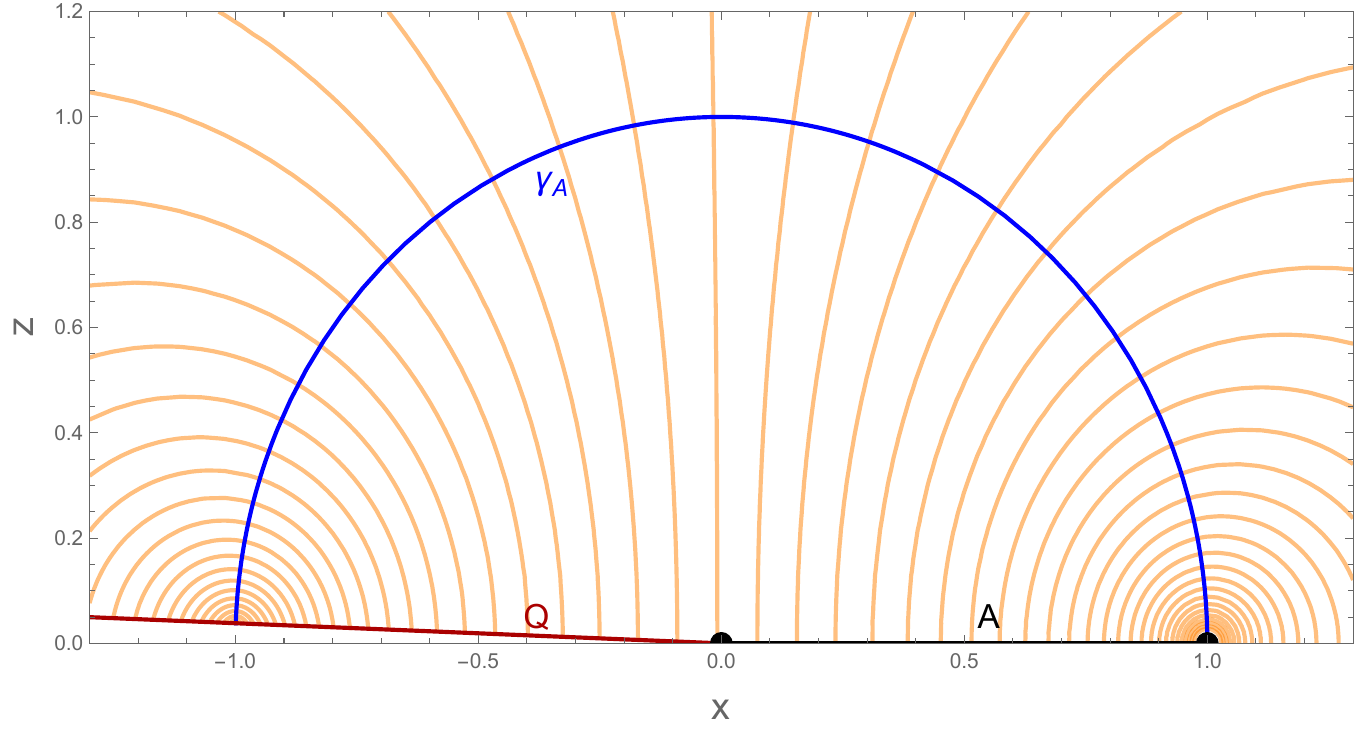}
		   		\includegraphics[width=0.45\linewidth]{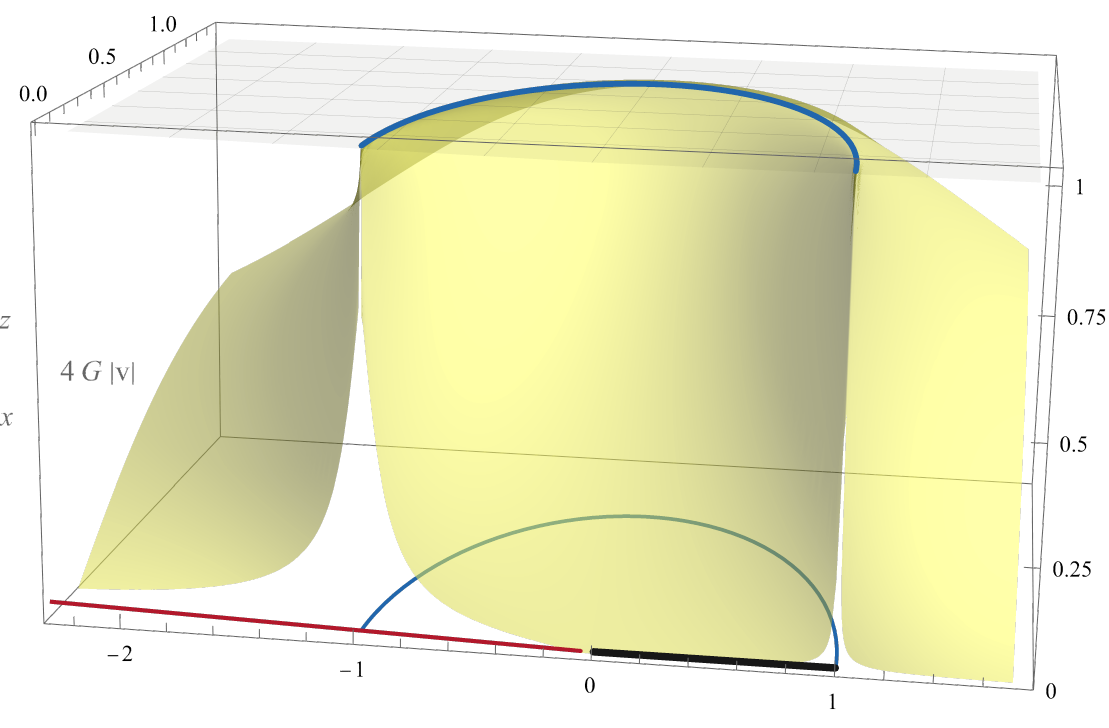}
		   		\caption{Bit-thread configuration for $\theta=\frac{\pi}{2.05}$, corresponding to $|T|=\sin\theta\simeq0.999$. As the EOW brane approaches the asymptotic boundary in this large-tension regime, the integral curves approach semicircular geodesics.}
		   		\label{fig:bit-bcft-pi2}
		   	\end{figure}

		   The special limit $\theta\to\frac{\pi}{2}$ is different. In this limit, the streamlines reduce to
		   \begin{align}
		   	x^2+z^2+2b\frac{2+\chi_0}{2-\chi_0}x+b^2=0
		   \end{align}
		   these are precisely semicircular geodesics in the Poincar\'e half-plane. Thus the standard geodesic bit-thread picture is recovered only in the $\theta\to\frac{\pi}{2}$ limit, where the EOW brane approaches the reflected asymptotic boundary and the system becomes equivalent to an ordinary CFT interval $[-b,b]$. See figure \ref{fig:bit-bcft-pi2} for an illustration.
		   	
		   	It is also instructive to note the geometric interpretation of the stream function $\Psi$. The first term in \eqref{Psipp} (we set $\ell_\star=1$ for simplicity)
		   	\begin{align}
		   		\Phi_b=\log\frac{(x-b)^2+z^2}{z}\label{Busemann}
		   	\end{align}
		   	is a Busemann function\footnote{For a unit-speed geodesic ray $\gamma_\xi(s)$ ending at an ideal-boundary point $\xi$, the Busemann function is defined as $\Phi_\xi(p)=\lim_{s\to\infty}\left[d(p,\gamma_\xi(s))-s\right]$, up to an additive constant; its level sets are horocycles centered at $\xi$.
		   	} \cite{Amini:2018jhd,beran2024bonnetmyersrigiditytheoremglobally} associated with the boundary point $(b,0)$, while the second term is the hyperbolic distance from $(x,z)$ to the brane endpoint $q=(-b\sin\theta,b\cos\theta)$:
		   	\begin{align}
		   		\cosh\CL_q(x,z)=1+\frac{(x+b\sin\theta)^2+(z-b\cos\theta)^2}{2b z\cos\theta}=\frac{\CA_\theta(x,z)}{2b z\cos\theta}
		   	\end{align}
		   	The norm bound \eqref{norm-bound} also follows directly from this interpretation. Both $\Phi_b$ and $\CL_q$ have unit hyperbolic gradient norm: $\left|\nabla \Phi_b\right|=1=\left|\nabla \CL_q\right|$. Then from \eqref{v-del-Psi}, we have 
		   	\begin{align}
		   		\left|v_A\right|=\frac{1}{8G}\left|\nabla\Psi\right|\leq\frac{1}{8G}\left(\left|\nabla\Phi_b\right|+\left|\nabla\CL_q\right|\right)=\frac{1}{4G}
		   	\end{align}
		   	The RT surface is precisely the geodesic connecting the boundary point $(b,0)$ to $q$. On this geodesic, the two unit gradients are antiparallel, and the inequality is saturated.

			   This is the finite--ideal specialization of the general endpoint-reduction identity in appendix~\ref{app:distance-difference-potential}.  In the direct source integral, the endpoint at the BCFT boundary separating the asymptotic and brane source pieces is internal to the oriented extended subsystem $A\cup I_A$ and cancels; the surviving data are exactly the ideal endpoint $(b,0)$ and the finite brane endpoint $q$.  Appendix~\ref{app:fermi-extended-flows} places the same RT geodesic at $\lambda=0$ and shows explicitly that the PEE-selected flow and the normal-geodesic flow agree on the bottleneck but generically differ away from it.  The limit $\theta\to\pi/2$ is then transparent: the finite focus is pushed to the second ideal endpoint, and the two representatives coincide with the standard CFT interval flow.
		   \paragraph{Regular non-crossing foliation:} To prove global regularity, we must show that the level sets of $\Psi$ are regular everywhere in the open physical manifold
		   \begin{align}
		   	\CM_Q^0=\{(x,z)|z>0,x+z\tan\theta>0\}
		   \end{align}
		   In other words, we need $\nabla \Psi\neq 0$ throughout $\CM_Q^0$. Since $v_A$ is the Hodge dual of $\dd\Psi$, this is equivalent to proving $v_A^\mu\neq 0$ everywhere in $\CM_Q^0$. Geometrically, a critical point would require $\nabla\Phi_b=\nabla\CL_q$. By uniqueness of hyperbolic geodesics, the two equal unit vectors would then be tangent to the same complete geodesic with the same orientation. The point must therefore lie on the continuation of the RT geodesic beyond the finite focus $q$. That continuation is on the excised side of the EOW brane: immediately beyond $q$ it obeys $x+z\tan\theta<0$, whereas the physical region is $\CM_Q^0=\{z>0,\,x+z\tan\theta>0\}$. The other continuation terminates at the ideal point $(b,0)$ and leaves the regulated bulk. Hence $\dd\Psi$ has no zero in $\CM_Q^0$.

		   The implicit-function theorem now implies that every level set $\Psi(x,z)=\Psi_0$ is a smooth curve throughout the open physical manifold. Distinct levels cannot meet because $\Psi$ is single-valued. Two apparent branches with the same label are locally the same regular level set and therefore cannot cross. Thus the integral curves form a regular ordered non-crossing foliation of $\CM_Q^0$. Together with divergencelessness, the norm bound, the boundary condition and RT saturation, this establishes that $v_A^\mu$ is a valid bit-thread max-flow representative. The ordered foliation is a property of this representative and should not be confused with the simultaneous-locking theorem for a nested family of boundary regions.

		   \paragraph{Connected phase:} We may also find the bit-thread configuration for the subsystem $A=[a,b]$ away from the boundary for the connected phase of the RT surface. For threads emanating from inside the interval, the PEE threads are just given by \eqref{PEE-thread-flow} and hence the bit-thread configuration is obtained as
		   \begin{align}
		   	v_A^\mu&=\int_{a}^{b}\d x_0\, V^\mu_{x_0}=\frac{1}{4G}\frac{z^3 (b-a)}{ \left((a-x)^2+z^2\right) \left((b-x)^2+z^2\right)}\left(2 x-a-b,z+\frac{(x-a) (b-x)}{z}\right)\label{bit-threads-connected-phase}
		   \end{align}
		   It is straightforward to check that the integral curves of this bit-thread field are semicircular geodesics orthogonal to the asymptotic boundary and the RT surface. Therefore, we may conclude that, for an interval away from the boundary in the connected phase, geodesic PEE threads superpose to a geodesic bit-thread congruence.
		   
		   It is instructive to test the consistency of our PEE-thread configuration by computing the bit-thread flow corresponding to the connected phase from the PEE threads emanating outside the interval $A$. These sources comprise the complement of the subsystem $A$ together with the entire EOW brane. We may then find the collection of all these PEE threads as follows
		   \begin{align}
		   	v^\mu_A&=\int^{0}_{\infty}\d\hat y\,V^\mu_{\hat y}(x,z)+\left(\int_{\infty}^{b}+\int_{a}^{0}\right)\d x_0\,V^\mu_{x_0}(x,z)\notag\\
		   	&=\frac{z^2}{(4 G) \left(x^2+z^2\right)}(-z,x)+\frac{1}{4 G}\Bigg(\frac{z^3 (a-b) (a+b-2 x)}{\left((a-x)^2+z^2\right) \left((b-x)^2+z^2\right)}+\frac{z^3}{x^2+z^2},\notag\\
		   	&\qquad\qquad\qquad\qquad\qquad\qquad\qquad\quad\frac{z^2 (a-b) \left((a-x) (b-x)-z^2\right)}{\left((a-x)^2+z^2\right) \left((b-x)^2+z^2\right)}-\frac{x z^2}{x^2+z^2}\Bigg)\notag\\
		   	&=\frac{1}{4G}\frac{z^3 (b-a)}{ \left((a-x)^2+z^2\right) \left((b-x)^2+z^2\right)}\left(2 x-a-b,z+\frac{(x-a) (b-x)}{z}\right)\,,
		   	\end{align}
		   	Although this representation contains brane-anchored PEE threads, the total current is again exactly \eqref{bit-threads-connected-phase}. The brane contribution cancels into the same ordinary geodesic max-flow representative. Thus the distinction is not simply ``boundary-anchored threads give geodesic flows, brane-anchored threads give non-geodesic flows.'' The real distinction is between an ordinary connected interval phase and a genuinely extended-boundary phase. This provides another consistency check for our bit-thread configuration.
		   	
		   	\paragraph{Disconnected phase:} It is natural to ask whether one can find explicit bit threads for the disconnected phase of an interval
		   	away from the brane, or for two disjoint intervals in ordinary AdS$_3$/CFT$_2$. The short answer is: yes in principle, but the result is much less canonical, and closed-form expressions are harder.
		   	
		   	For $A=[a,b]$, the disconnected BCFT phase has the RT surface $\gamma_A^{\rm disc}=\gamma_a\cup\gamma_b$, where $\gamma_a$ and $\gamma_b$ are two brane-anchored geodesics from the boundary endpoints $a,b$ to the EOW brane. The entanglement entropy is given by
		   	\begin{align}
		   		S_A^{\rm disc}=\frac{1}{4G}\left(\log\frac{2a}{\epsilon}+\log\frac{2b}{\epsilon}+2\rho_0\right)
		   	\end{align}
		   	A max flow must saturate both components $\gamma_{a,b}$. Such a  flow exists by the max-flow/min-cut theorem, but it is not unique, because the minimal surface has two disconnected components. In two dimensions, a general construction can be formulated in terms of a stream function $\psi$:
		   	\begin{align}
		   		v^x=z^2\del_z\psi~,~v^z=-z^2\del_x\psi~,~\left|\nabla\psi\right|\leq\frac{1}{4G}\,.
		   	\end{align}
		   	The flux through any curve is the difference of $\psi$ between its endpoints. Therefore, to build a max flow for a disconnected cut, one prescribes boundary values of $\psi$ so that the jump across $\gamma_a$ equals ${\rm Area}(\gamma_a)/(4G)$ and the jump across $\gamma_b$ equals ${\rm Area}(\gamma_b)/(4G)$, and then searches for a Lipschitz extension satisfying the norm bound. Geometrically, this is a calibration problem. This gives a clean existence statement, but it does not by itself give a simple analytic formula.
		   \subsubsection*{Entanglement contour and flux through the boundary and brane}
		   
		   We now consider the flux of the bit-thread flow through the asymptotic boundary and the EOW brane. The flux through the RT surface has already been established by the norm saturation condition \eqref{BCFT-bit threads-on-RT}, which together with divergencelessness guarantees that it reproduces the entanglement entropy.

		   The normal vector to the boundary at $z=\epsilon$ is given by $n_\mu=\frac{1}{\epsilon}(0,1)$. We have
		   \begin{align}
		   	\lim_{z\to 0}\left(v^{}_A\cdot n\right)\d\Sigma=\frac{1}{4G}\left(\frac{1}{b-x}+\frac{x+b \sin \theta}{b^2+x^2+2 b x \sin\theta}\right)\d x
		   \end{align}
		   Following \cite{Caggioli:2024uza}, we may define the entanglement contour associated to the asymptotic boundary as 
		   \begin{align}
		   	s_A^{(\del)}(x)=\frac{1}{4G}\left(\frac{1}{b-x}+\frac{x+b \sin \theta}{b^2+x^2+2 b x \sin\theta}\right)~~,~~0<x<b\,.
		   \end{align}
		   Therefore, the flux of the bit-thread vector through the boundary interval is
		   \begin{align}
		   	\Phi_\del(A)&=\int_{0}^{b-\epsilon}\dd x\,s_A^{(\del)}(x)=\frac{1}{8G}\log\left[\frac{2b^2(1+\sin\theta)}{\epsilon^2}\right]
		   \end{align}
		   It is also instructive to consider the flux through the EOW brane \eqref{brane}. The unit normal vector to the EOW brane is given by
		   \begin{align}
		   	n^{}_{\mu,\CQ}=\frac{1}{z}\left(\cos\theta,\sin\theta\right)
		   \end{align}
		   The brane entanglement contour is the flux density through the brane component. With the parametrization used throughout, $(x,z)=(-\hat y\sin\theta,\hat y\cos\theta)$, one obtains
		   \begin{align}
		   	s_A^{(\CQ)}(\hat y)=\frac{1}{8G}\left(1+\frac{b}{\hat y}\right)\left(\frac{1}{\sqrt{b^2+\hat y^2+2b \hat y\cos2\theta}}-\frac{b-\hat y}{b^2+\hat y^2+2b \hat y\sin\theta}\right)
		   \end{align}
		   Finally, the flux through the brane component $I_A$, whose coordinate range is $0\leq\hat y\leq b$, is given by
		   \begin{align}
		   	\Phi_\CQ({\rm I}_A)&=\int_{0}^{\hat y=b}\dd \hat y\,s^{(\CQ)}_A(\hat y)=\frac{1}{8G}\log\left[2\sec\theta(\sec\theta+\tan\theta)\right]
		   \end{align}
		   It is easy to verify the contour normalization property
		   \begin{align}
		   	\Phi_\CQ({\rm I}_A)+\Phi_\del(A)=\frac{1}{4G}\left[\log\left(\frac{2b}{\epsilon}\right)+\log\left(\sec\theta+\tan\theta\right)\right]=S_A
		   \end{align}

		   		   \subsection{BTZ black brane}\label{app:pee-threads-BTZ}
		   Next, we consider the PEE threads in the BTZ black brane \eqref{BTZ-metric}.
		   As in the Poincar\'e case, once the source point is fixed, the local PEE-thread
		   vector field is determined by the unique bulk geodesic through that source and a
		   generic bulk point; the other endpoint only fixes the density of the thread bundle.
		   The distinction between the different endpoint sectors is thus encoded in the
		   integration range and in the two-point PEE density, not in a new tangent vector field.
		   It therefore suffices to construct the independent source-type flows --- sourced from
		   the asymptotic boundary $\partial M$, the EOW brane $\CQ$, and the horizon $h$, the
		   latter two being the new components of the extended surface
		   $\Sigma_e^{\rm BTZ}=\partial M\cup \CQ\cup h$. The full bit-thread configuration is
		   then obtained by integrating these elementary source flows over the portions of the
		   extended surface assigned to the subsystem.
		   \subsubsection{Boundary sourced PEE-thread flow}
		   The boundary-sourced PEE-thread flow admits a direct intrinsic derivation in planar BTZ, given in \cite{BasuWenChandra}.  For completeness, we derive the same field here independently by using the AdS$_3$ isometry that relates the BTZ spatial slice to the Poincar\'e half-plane.  We denote the Poincar\'e coordinates by $(\bar x,\bar z)$ throughout this derivation,
		   \begin{align}
		   	\bar{x}=z_h e^{\frac{x}{z_h}}\sqrt{1-\frac{z^2}{z_h^2}}~~,~~\bar{z}=z e^{\frac{x}{z_h}}
		   \end{align}
		   It is readily verified that
		   \begin{align}
		   	\frac{\dd \bar x^2+\dd \bar z^2}{\bar z^2}=\frac{1}{z^2}\left(\dd x^2+\frac{\dd z^2}{1-\frac{z^2}{z_h^2}}\right)
		   \end{align}
		   Let $x_0$ be a  source point on the asymptotic boundary. Under the above isometry, this boundary point maps to the Poincar\'e boundary point $\bar{x}_0=e^{x_0/z_h}$. The BTZ boundary-sourced PEE-thread flow is obtained by pushing forward the Poincar\'e PEE-thread flow \eqref{PEE-thread-flow} and including the source-coordinate Jacobian $\dd \bar{x}_0=\frac{1}{z_h}e^{x_0/z_h}\dd x_0$. In other words,
		   \begin{align}
		   	V^\mu_{x_0}\,\dd x_0=\chi_\star\left(V^{\bar\mu}_{\bar x_0}\,\dd\bar x_0\right)
		   \end{align}
		   where $\chi$ denotes the inverse map from Poincar\'e coordinates to BTZ coordinates:
		   \begin{align}
		   	x=z_h\log\frac{\sqrt{\bar x^2+\bar z^2}}{z_h}~~,~~z=z_h\frac{\bar z}{\bar x^2+\bar z^2}
		   \end{align}
		   Performing this push forward, we obtain the result
		   \begin{align}
		   	V^\mu_{x_0}(x,z)=&\frac{z^2}{8Gz_h^2\left[\cosh\left(\frac{x-x_0}{z_h}\right)-\sqrt{1-\frac{z^2}{z_h^2}}\right]^2}\notag\\
		   	&\times\left(\frac{z}{z_h}\sinh\left(\frac{x-x_0}{z_h}\right),\sqrt{1-\frac{z^2}{z_h^2}}\left(1-\sqrt{1-\frac{z^2}{z_h^2}}\cosh\left(\frac{x-x_0}{z_h}\right)\right)\right)\label{eq:BTZ-boundary-source-flow}
		   \end{align}
		   This is exactly the PEE-thread flow obtained in \cite{BasuWenChandra}. Several comments are in order. First, the direction of this vector field is the push-forward of a Poincar\'e geodesic tangent flow. Hence its integral curves are precisely the BTZ geodesics emanating from the boundary point $x_0$. The source-coordinate Jacobian changes only the density of the thread bundle, not the geodesic support. Second, the flow is divergenceless. This follows immediately because the map from the BTZ slice to the Poincar\'e half-plane is an isometry and the Poincar\'e PEE-thread flow is divergenceless. Third, the near-boundary limit reproduces the thermal two-point PEE kernel. As $z\to0$,
		   \begin{align}
		   	V_{x_0}^{z}(x,z)=-\frac{1}{16G}\left(\frac{z}{z_h}\right)^2\operatorname{csch}^2\left(\frac{x-x_0}{2z_h}
		   	\right)+O(z^4).
		   \end{align}
		   The outward normal to the cutoff boundary $z=\epsilon$ gives the flux
		   density
		   \begin{align}
		   	-\lim_{z\to 0}\frac{V_{x_0}^{z}}{z^2}=\frac{1}{16Gz_h^2}
		   	\operatorname{csch}^2\left(\frac{x-x_0}{2z_h}\right)=\CI_{\del\del}(x,x_0)\,,
		   \end{align}
		   which is precisely the boundary--boundary two-point PEE density in the BTZ
		   black-brane geometry. Thus the pushed-forward PEE-thread flow is normalized correctly.
		   \subsubsection{Brane sourced PEE-thread flow}
		   Next, we consider the PEE thread bundle emanating from a point $x=y$ on the EOW brane. The profile of such a PEE thread may be obtained from \eqref{general-solution-geodesics}, with the constants $x_c$ and $\hat c$ given by
		   \begin{align}
		   	x_c&=z_h\log\left[\frac{e^{x_0/z_h}+e^{y/z_h} s_\kappa(y)}{z_h \left(2+\kappa ^2 e^{-\frac{x_0-y}{z_h}} \sinh ^2\left(\frac{y}{z_h}\right) \csch\left(\frac{x_0-y}{z_h}\right)\right)}\right]\,,\notag\\
		   	\frac{\hat c}{z_h}&=\coth \left(\frac{x_0-y}{z_h}\right)-\csch\left(\frac{x_0-y}{z_h}\right) s_\kappa(y)
		   \end{align}
		   which leads to
		   \begin{align}
		   	\sqrt{h(z)}=\frac{\sinh \left(\frac{x-y}{z_h}\right)-\sinh \left(\frac{x-x_0}{z_h}\right) s_\kappa(y)}{\sinh \left(\frac{x_0-y}{z_h}\right)}
		   \end{align}
		   \begin{align}
		   	x_0=x_m+2z_h\,\textrm{arctanh}\left[\frac{\Xi+\sqrt{\Xi^2+\frac{z_m^2}{z_h^2}\sinh^2\left(\frac{x_m-y}{z_h}\right)}}{\left(1+\sqrt{h(z_m)}\right)\sinh\left(\frac{x_m-y}{z_h}\right)}\right]\label{x0-BTZ-bcft}
		   \end{align}
		   with
		   \begin{align}
		   	\Xi=s_\kappa(y)-\sqrt{h(z_m)} \cosh \left(\frac{x_m-y}{z_h}\right)
		   \end{align}
		   Define the positive normalization factor
		   \begin{align}
		   	\CN_y^2(x_m,z_m)
		   	&=\left[\sqrt{h(z_m)}\cosh\left(\frac{x_m-y}{z_h}\right)-s_\kappa(y)\right]^2
		   	+\frac{z_m^2}{z_h^2}\sinh^2\left(\frac{x_m-y}{z_h}\right).
		   \end{align}
		   The unit tangent vector to the PEE thread at $P_m:(x_m,z_m)$ is then
		   \begin{align}
		   	\tau_y^\mu
		   	&=\frac{z_m}{\CN_y(x_m,z_m)}\left(
		   	\frac{z_m}{z_h}\sinh\left(\frac{x_m-y}{z_h}\right),
		   	\sqrt{h(z_m)}\left[\sqrt{h(z_m)}\cosh\left(\frac{x_m-y}{z_h}\right)-s_\kappa(y)\right]\right)\,.
		   \end{align}
		   Flux matching fixes the signed scalar amplitude of the brane-sourced current:
		   \begin{align}
		   	\widehat f_y(x_m,z_m)
		   	&=\frac{z_m}{8Gz_h^2\CN_y^2(x_m,z_m)}
		   	\left[\sqrt{h(z_m)}
		   	+\frac{\kappa^2\sinh\left(\frac{x_m}{z_h}\right)\sinh\left(\frac{y}{z_h}\right)
		   		-\cosh\left(\frac{x_m-y}{z_h}\right)}{s_\kappa(y)}\right],\notag\\
		   	V_y^\mu&=\widehat f_y\tau_y^\mu\,,\qquad |V_y|=|\widehat f_y|\,.
		   \end{align}
		   This makes the orientation separate from the invariant norm. Direct substitution, with $\sqrt g=1/(z^2\sqrt h)$, gives the independent local check
		   \begin{align}
		   	\nabla_\mu V_y^\mu
		   	=z^2\sqrt h\left[\partial_x\left(\frac{V_y^x}{z^2\sqrt h}\right)+\partial_z\left(\frac{V_y^z}{z^2\sqrt h}\right)\right]=0
		   \end{align}
		   away from the brane source.
		   In the limit $\kappa\to 0^+$ the brane approaches the reflected asymptotic boundary and $V_y$ reduces to the boundary-sourced flow, while in the zero-temperature limit $z_h\to\infty$ it reproduces the Poincar\'e AdS$_3$ result \eqref{PEE-thread-flow-class-I}. These limits provide strong consistency checks of the construction.
		   \subsubsection{Auxiliary PEE threads: horizon sourced flow}
		   Let $y_h$ label a source point on the horizon. A geodesic joining $y_h$ to a boundary endpoint $x_0$ has the intrinsic profile \eqref{geod-class-II},
		   \begin{align}
		   	\sqrt{h(z)}=\frac{\sinh\left(\frac{x-y_h}{z_h}\right)}
		   	{\sinh\left(\frac{x_0-y_h}{z_h}\right)}\,.
		   	\label{eq:horizon-source-geodesic}
		   \end{align}
		   Eliminating $x_0$ in favor of $(x,z)$ gives
		   \begin{align}
		   	x_0=y_h+z_h\operatorname{arcsinh}\left[
		   	\frac{\sinh\left(\frac{x-y_h}{z_h}\right)}{s(z)}
		   	\right]\,,
		   \end{align}
		   with the opposite sign selecting the oppositely oriented complete geodesic. Defining
		   \begin{align}
		   	\CN_h^2=\cosh^2\left(\frac{x-y_h}{z_h}\right)-\frac{z^2}{z_h^2}\,,
		   \end{align}
		   the resulting horizon-sourced PEE-thread current may be obtained from the flux-matching condition as follows
		   \begin{align}
		   	V_{y_h}^{\mu}(x,z)
		   	=&\frac{z^2\sqrt{h(z)}}{8Gz_h^2\,\CN_h^3}
		   	\left(
		   	\frac{z}{z_h}\sinh \left(\frac{x-y_h}{z_h}\right),\,
		   	-h(z)\cosh \left(\frac{x-y_h}{z_h}\right)
		   	\right)\,.
		   	\label{eq:horizon-source-flow}
		   \end{align}
		   It passes two direct local checks:
		   \begin{align}
		   	\nabla_\mu V_{y_h}^{\mu}&=0\,,
		   	&
		   	-\lim_{z\to0}\frac{V_{y_h}^{z}}{z^2}
		   	&=\frac{1}{8Gz_h^2}\sech^2\left(\frac{x-y_h}{z_h}\right)
		   	=\CI_{\del h}(x,y_h)=-\widehat{\CI}_{\del h}(x,y_h)\,.
		   \end{align}
		   Its flux density through the $\sigma=+1$ brane,
		   $z_\CQ(q)=\kappa z_h\sinh(q/z_h)$, is likewise
		   \begin{align}
		   	\left.
		   	\sqrt g\left(V_{y_h}^{x}\frac{\dd z_\CQ}{\dd q}
		   	-V_{y_h}^{z}\right)\right|_{\CQ}
		   	=\CI_{\CQ h}(q,y_h)=-\widehat{\CI}_{\CQ h}(q,y_h)\,,
		   \end{align}
		   for the displayed endpoint ordering, including precisely the brane-coordinate Jacobian. Thus \eqref{eq:horizon-source-flow} has both the boundary--horizon and brane--horizon positive flux normalizations; the oriented kernels carry the opposite sign because the horizon component is oppositely oriented. The horizon--horizon channel does not produce an additional smooth field: as explained in section~\ref{sec:PEE-BTZ}, its mixed kernel vanishes away from coincidence.
		   \subsubsection{Bit threads}
		   Similar to the Poincar\'e AdS$_3$ case, we obtain the bit-thread flow from the collection of PEE threads emitted by the boundary subsystem $A=[0,b]$ and by the brane component $I_A$, whose coordinate range is $0\leq y\leq x_\CQ$ on the $\sigma=+1$ branch. With the orientation of the closed extended boundary, the brane integral runs from $x_\CQ$ to $0$:
		   \begin{align}
		   	v^\mu_A=\int_0^b \d x_0\,V^\mu_{x_0}+\int_{x_\CQ}^{0}\d y\,V^\mu_y\,.
		   	\label{BTZ-bit-threads}
		   \end{align}
		   If the chosen extended subsystem contains a horizon segment $h_A$, the additional term is $\int_{h_A}\dd y_h\,V^\mu_{y_h}$ with $V^\mu_{y_h}$ given in \eqref{eq:horizon-source-flow}. For the adjacent brane-ending RT phase considered here, $h_A=\varnothing$; the horizon is a sink component of the complementary extended region.
		   The two source integrals can be evaluated before introducing the final stream function. Let $\Phi_0$ and $\Phi_b$ denote the BTZ Busemann functions of the boundary points $0$ and $b$, respectively, and let $\CL_\CQ$ be the distance to the brane endpoint of the RT surface. Direct differentiation of the displayed endpoint functions reproduces the elementary fields above and gives the primitives
		   \begin{align}
		   	\int_0^b\dd x_0\,V_{x_0}
		   	&=\frac{1}{8G}\star\dd\left(\Phi_b-\Phi_0\right),\notag\\
		   	\int_{x_\CQ}^{0}\dd y\,V_y
		   	&=\frac{1}{8G}\star\dd\left(\Phi_0-\CL_\CQ\right).
		   \end{align}
		   The internal endpoint at the BCFT boundary therefore cancels explicitly. As in the Poincar\'e cases discussed earlier, the surviving boundary Busemann function is
		   \begin{align}
		   	\Phi_b=\log\left[\frac{z_h}{z}\left(\cosh\left(\frac{x-b}{z_h}\right)-\sqrt{1-\frac{z^2}{z_h^2}}\right)\right]
		   \end{align}
		   and the distance function to the brane endpoint $x_\CQ$ in \eqref{x-brane},
		   \begin{align}
		   	\CL_\CQ=\operatorname{arccosh}U_\CQ\,,\qquad
		   	U_\CQ=\frac{z_h^2}{z z_\CQ}\left(\cosh\left(\frac{x-x_\CQ}{z_h}\right)-\sqrt{\left(1-\frac{z^2}{z_h^2}\right)\left(1-\frac{z_\CQ^2}{z_h^2}\right)}\right)
		   \end{align}
		   Then the stream function is given by
		   \begin{align}
		   	\Psi_{\rm BTZ}&=\frac12\left(\Phi_b-\CL_\CQ\right)\,.\label{Psi-BTZ}
		   \end{align}
		   In the above expressions, we have set $z_\CQ=\sigma\kappa\,z_h\sinh\left(\frac{x_\CQ}{z_h}\right)$. The first term in \eqref{Psi-BTZ}
		   remembers the asymptotic endpoint of the RT surface, while the second term remembers the brane endpoint. The streamlines are equipotential curves between the two ends of the RT geodesic.
		   
		   The one-form and coordinate components of the flow are
		   \begin{align}
		   	v_A&=\frac{1}{4G}\star\dd\Psi_{\rm BTZ}\,,
		   	\label{eq:BTZ-distance-difference-flow}\\
		   	v_A^x&=\frac{z^2\sqrt{h(z)}}{4G}\,\partial_z\Psi_{\rm BTZ}\,,
		   	&
		   	v_A^z&=-\frac{z^2\sqrt{h(z)}}{4G}\,\partial_x\Psi_{\rm BTZ}\,.
		   	\label{eq:BTZ-distance-difference-components}
		   \end{align}
		   Since \eqref{eq:BTZ-distance-difference-flow} is a Hodge dual of an exact one-form, $\nabla_\mu v_A^\mu=0$ identically. Moreover, $\Phi_b$ and $\CL_\CQ$ obey the BTZ eikonal equations $|\nabla\Phi_b|=|\nabla\CL_\CQ|=1$, so
		   \begin{align}
		   	|v_A|=\frac{1}{4G}|\nabla\Psi_{\rm BTZ}|
		   	\leq\frac{1}{4G}\,.
		   \end{align}
		   On the RT geodesic joining $(b,0)$ to $(x_\CQ,z_\CQ)$ the two unit gradients are antiparallel and the bound is saturated.
		   
		   Equation~\eqref{eq:BTZ-distance-difference-flow} is therefore exactly the explicitly integrated source current \eqref{BTZ-bit-threads}. The general endpoint-reduction identity derived in appendix~\ref{app:distance-difference-potential} explains the cancellation structurally. When a horizon segment belongs to the extended subsystem, the same reduction uses \eqref{eq:horizon-source-flow}; its internal horizon endpoints cancel in the same way.
			   Since a constant-time planar-BTZ slice is locally $\mathbb H^2$, the same finite--ideal flow also admits the Fermi description of appendix~\ref{app:fermi-extended-flows}.  The BTZ Busemann function $\Phi_b$ and the distance $\CL_\CQ$ are simply the intrinsic BTZ representatives of the two endpoint functions entering the universal Fermi formula.  The local norm and saturation statements are therefore the same hyperbolic facts as in Poincar\'e AdS; the additional BTZ information resides in the choice of exterior component and geodesic chamber.  The same endpoint-resolved planar-BTZ construction is developed from a complementary viewpoint in \cite{BasuWenChandra}, where we also extend the analysis to finite-cutoff AdS$_3$ and compare the PEE-selected flow with independent normal-geodesic calibrations.
		   
\begin{figure}[ht]
		   	\centering
		   	\includegraphics[width=0.49\linewidth]{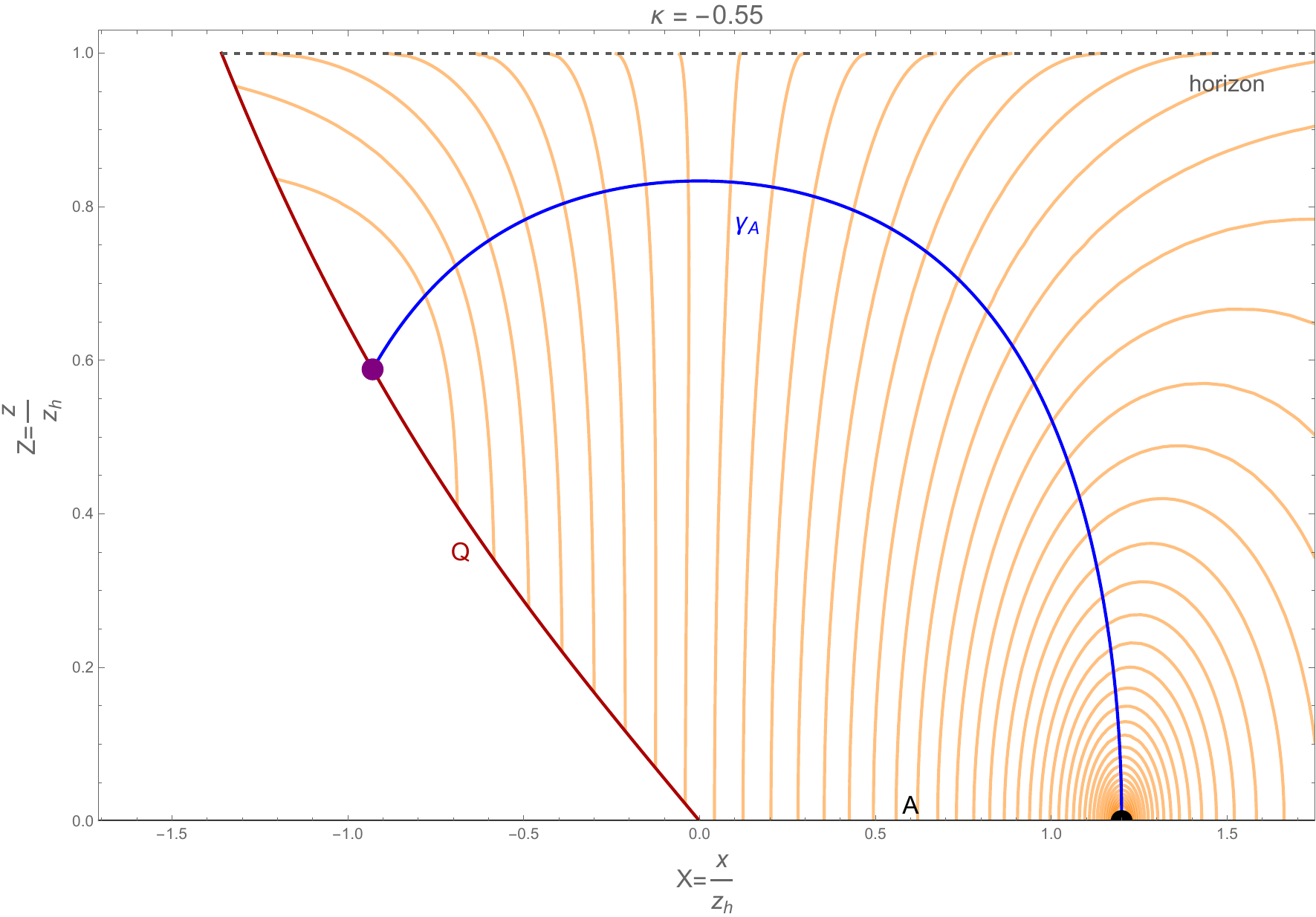}
		   	\includegraphics[width=0.49\linewidth]{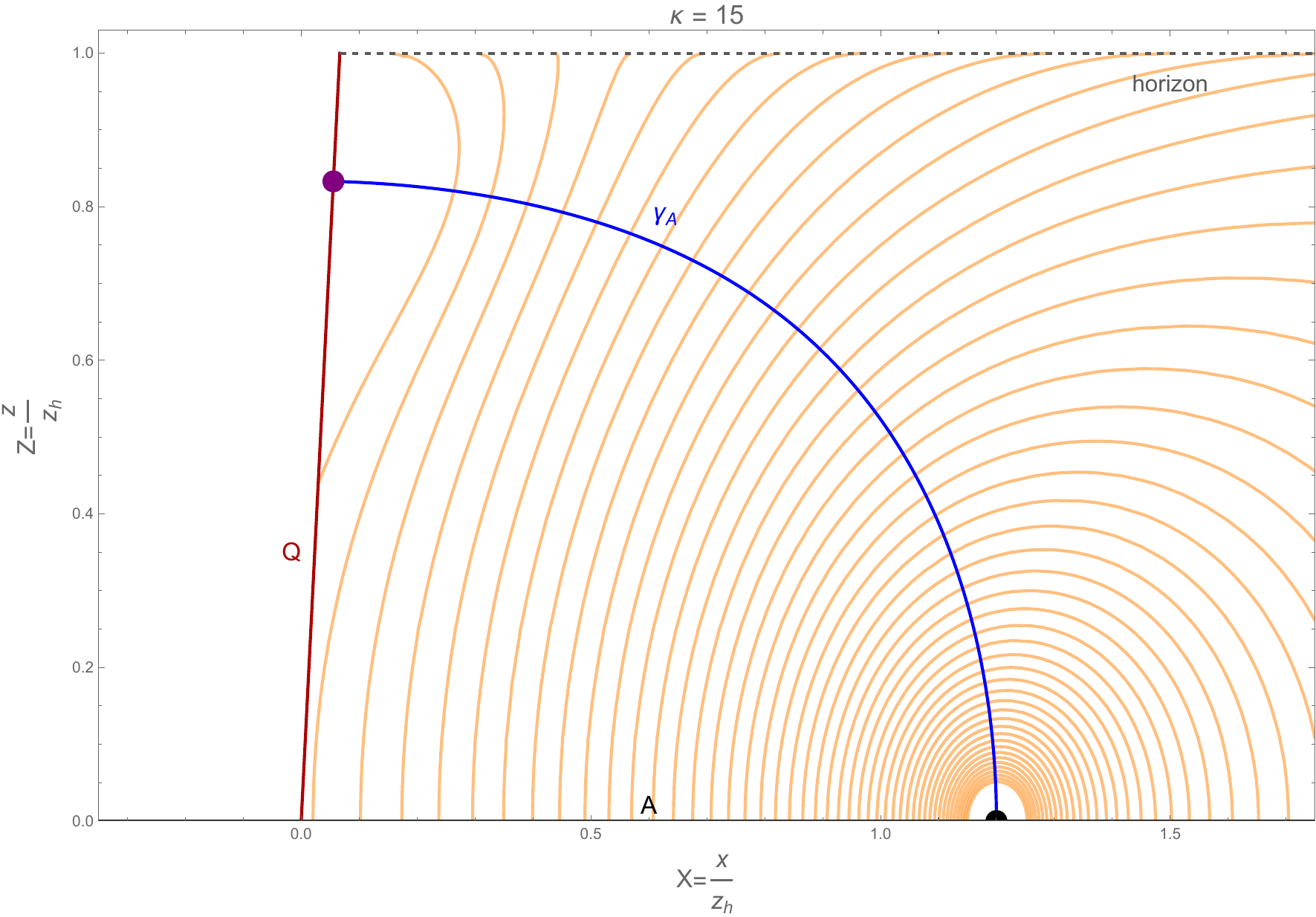}
		   	\caption{Integral curves of the bit-thread vector field \eqref{BTZ-bit-threads} for $b=1.2z_h$. Left: $\kappa=0.55$, $\sigma=-1$. Right: $\kappa=15$, $\sigma=+1$. The magnitude $\kappa$ is positive in both panels; $\sigma$ specifies the brane branch.}
		   	\label{fig:btz-bit-threads}
		   \end{figure}

\begin{figure}[ht]
		   	\centering
		   	\includegraphics[width=0.5\textwidth]{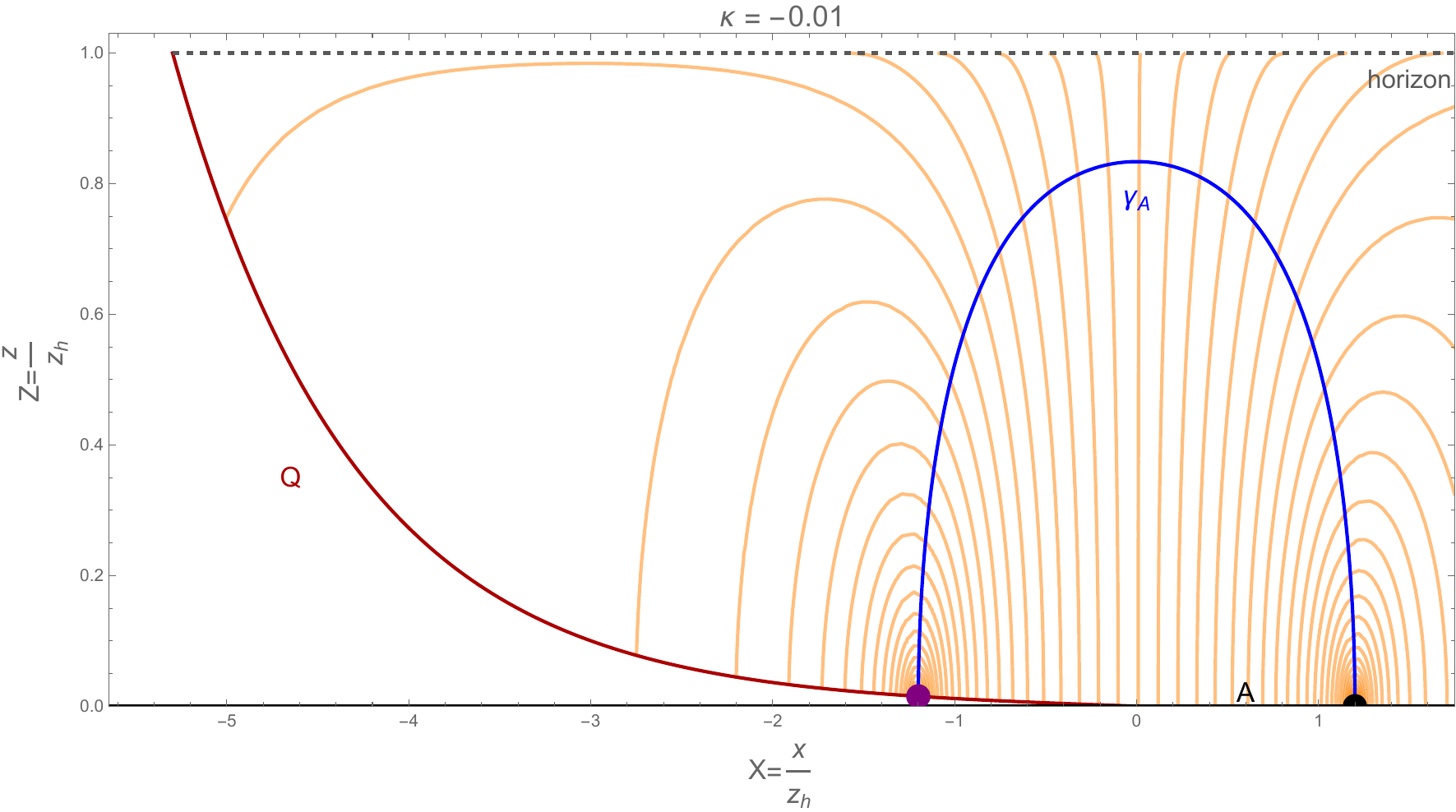}
		   	\caption{Integral curves for $b=1.2z_h$, $\kappa=0.01$ and $\sigma=-1$. As $\kappa\to0^+$ on this branch, the EOW brane approaches the reflected asymptotic boundary and the non-geodesic integral curves approach the geodesic bit threads.}
		   \label{fig:btz-bit-geodesic-limit}
		   \end{figure}

		   \paragraph{Integral curves and regularity.} The integral curves are the level sets $\Psi_{\rm BTZ}(x,z)=\Psi_0$. They need not be BTZ geodesics because the current is a density-weighted sum of different geodesic congruences. To establish that these level sets form a regular ordered non-crossing foliation, it remains to exclude critical points. A zero of $\dd\Psi_{\rm BTZ}$ would require $\nabla\Phi_b=\nabla\CL_\CQ$. By uniqueness of geodesics in the selected BTZ chamber, such a point must lie on the continuation of the complete RT geodesic beyond one of its two foci. It cannot lie on the RT segment itself, where the two unit gradients are antiparallel. The continuation through the ideal endpoint $(b,0)$ leaves the regulated exterior, while the continuation beyond $(x_\CQ,z_\CQ)$ crosses the EOW brane into the excised region. Explicitly, on the $\sigma=+1$ branch define
		   \begin{align}
		   	F_\CQ(x,z)=x-z_h\operatorname{arcsinh}\left(\frac{z}{\kappa z_h}\right).
		   \end{align}
		   The retained side has $F_\CQ\geq0$. The RT segment approaches the brane with $\dd F_\CQ/\dd\lambda>0$ when its unit tangent is oriented from the brane toward the boundary; its continuation beyond the brane has $F_\CQ<0$. The $\sigma=-1$ statement follows by reflection. Hence $\dd\Psi_{\rm BTZ}\neq0$ throughout the open physical region. The implicit-function theorem gives smooth level sets, and distinct levels cannot intersect because $\Psi_{\rm BTZ}$ is single-valued. This is an ordered non-crossing streamline foliation, not the stronger simultaneous-locking notion of bit-thread nesting. Representative integral curves for the two EOW-brane branches are shown in figure~\ref{fig:btz-bit-threads}. The approach to the geodesic bit-thread configuration in the small-$\kappa$ limit is displayed separately in figure~\ref{fig:btz-bit-geodesic-limit}.
		   
		   
		   \subsection*{Flux through the boundary, horizon and the EOW brane}
		   Recall that the anchor point of this RT surface on the EOW brane is given by \eqref{x-brane}. Therefore, the flux through the RT surface is computed as follows
		   \begin{align}
		   	\Phi_\textrm{RT}
		   	&=\frac{1}{4G}\CL_{\rm BTZ}\big((b,\epsilon),(x_\CQ,z_\CQ)\big)\notag\\
		   	&=\frac{1}{4G}\left(\log \left[\frac{2 z_h}{\epsilon }\sinh \left(\frac{b}{z_h}\right)\right]+\sigma\operatorname{arcsinh}\left(\frac{1}{\kappa}\right)\right)
		   \end{align}
		   where the second term is the boundary entropy \eqref{S-bdy-BTZ}, using
		   \begin{align}
		   	T=\frac{\sigma}{\sqrt{1+\kappa^2}}\,,
		   	\qquad
		   	\operatorname{arcsinh}\left(\frac1\kappa\right)
		   	=\operatorname{arctanh}|T|=\rho_0\,.
		   \end{align}
		   For evaluating the integral, we have utilized the entanglement wedge regularization to set
		   \begin{align}
		   	b_\epsilon=z_h \textrm{arccosh}\left[\frac{\sqrt{z_h^2-\epsilon ^2} }{z_h}\cosh \left(\frac{b}{z_h}\right)\right]=b-\frac{\epsilon ^2 }{2 z_h}\coth \left(\frac{b}{z_h}\right)+\CO(\epsilon^4)
		   \end{align}
		   Therefore, the flux through the RT surface correctly reproduces the entanglement entropy for the subsystem adjacent to the boundary of the thermal BCFT.
		   
		   The external fluxes follow either by direct pullback of \eqref{eq:BTZ-distance-difference-components} or, more economically, by endpoint differences of the stream function:
		   \begin{align}
		   	\Phi_{\mathcal C}
		   	=\frac{1}{4G}\left[\Psi_{\rm BTZ}\right]_{\partial\mathcal C}
		   \end{align}
		   for every oriented component $\mathcal C$ of the extended boundary. In terms of the integrated oriented sectors listed explicitly in appendix~\ref{appB}, the fluxes through the asymptotic interval and the brane component are
		   \begin{align}
		   	\Phi_{\partial M}(A)
		   	&=\widehat{\CI}(A,B)+\widehat{\CI}(A,I_B)+\widehat{\CI}(A,h)\,,
		   	\notag\\
		   	\Phi_{\CQ}(I_A)
		   	&=\widehat{\CI}(I_A,B)+\widehat{\CI}(I_A,I_B)+\widehat{\CI}(I_A,h)\,.
		   \end{align}
		   For the complementary components, we denote by $\Phi^{\rm sink}$ the positive inward sink-flux magnitude, namely $\Phi^{\rm sink}_{\mathcal C}=-\Phi^{\rm out}_{\mathcal C}$ with respect to the outward normal of the bulk region. These magnitudes are distributed among the complementary boundary, brane and horizon components as
		   \begin{align}
		   	\Phi^{\rm sink}_{\partial M}(B)
		   	&=\widehat{\CI}(A,B)+\widehat{\CI}(I_A,B)\,,
		   	\notag\\
		   	\Phi^{\rm sink}_{\CQ}(I_B)
		   	&=\widehat{\CI}(A,I_B)+\widehat{\CI}(I_A,I_B)\,,
		   	\notag\\
		   	\Phi^{\rm sink}_h
		   	&=\widehat{\CI}(A,h)+\widehat{\CI}(I_A,h)\,.
		   \end{align}
		   Consequently, the complete normalization and flux-conservation identities are
		   \begin{align}
		   	\Phi_{\partial M}(A)+\Phi_{\CQ}(I_A)
		   	&=\Phi_{\rm RT}=S_A\,,\notag\\
		   	\Phi^{\rm sink}_{\partial M}(B)+\Phi^{\rm sink}_{\CQ}(I_B)+\Phi^{\rm sink}_h
		   	&=\Phi_{\rm RT}=S_A\,,\notag\\
		   	\sum_{\mathcal C\subset\partial\mathcal M}\Phi^{\rm out}_{\mathcal C}&=0\,.
		   \end{align}
		   This completes the boundary, brane, horizon, and RT flux checks promised in the subsection heading.


\section{PEE threads and bit threads inside entanglement wedges: minimal purification}
		   \label{sec:eop}
\begin{figure}[ht]
		   	\centering
		   	\includegraphics[width=0.85\linewidth]{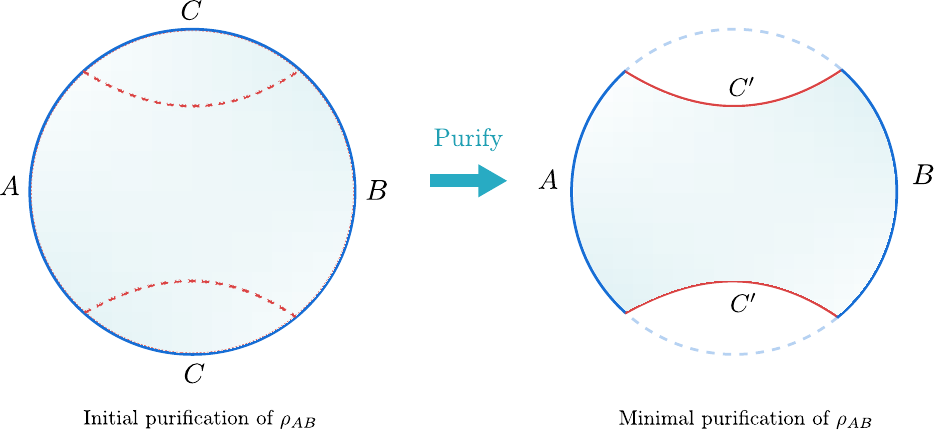}
		   	\caption{Minimal purification of the mixed state $\rho_{AB}$ obtained by tracing out the subsystem $C$ is furnished by introducing purifying degrees of freedom $C^\prime$ on the RT surface of $A\cup B$ \cite{Takayanagi:2017knl,Du:2019emy}.}
		   	\label{fig:rt-purification}
		   \end{figure}

		   In this section, we construct bit-thread configurations whose bottleneck is the entanglement wedge cross section (EWCS), using the PEE-thread network and the surface/state correspondence. The basic building block is a surface/state interpretation of the so-called minimal purification \cite{Takayanagi:2017knl,Dutta:2019gen,Du:2019emy}. See figure~\ref{fig:rt-purification} for an illustration.
		   
\begin{figure}[ht]
		   	\centering
		   	\includegraphics[width=0.45\textwidth]{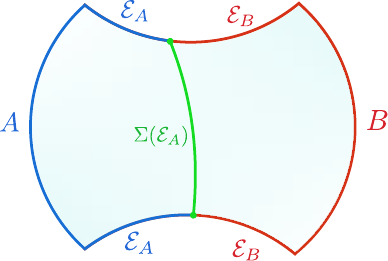}
		   	\caption{Geometric purification from the surface/state correspondence.
		   	The RT surface $\gamma_{AB}$ bounding the entanglement wedge of $AB$ is promoted to a purifying system and partitioned as
		   	in \eqref{minimal-purification}.
		   	For a fixed partition, $\Sigma(\mathcal{E}_{A})$ is the minimal surface homologous to $A\cup\mathcal{E}_{A}$, and therefore computes
		   	$S(A^\sharp)=S(B^\sharp)$.
		   	Minimizing its area over all partitions of $\gamma_{AB}$ reproduces the EWCS.}
		   	\label{fig:purification-partition}
		   \end{figure}

		   Applied to the entanglement wedge of a mixed boundary state $\rho_{AB}$, this principle gives a natural
		   interpretation of the RT surface $\gamma_{AB}$. The surface $\gamma_{AB}$ is not only the geometric boundary of the entanglement wedge; it also supplies the auxiliary Hilbert-space factor needed to purify the state on $A\cup B$. Within the surface/state proposal and for the connected contractible setting considered here, the closed extended convex surface
		   \begin{align}
		   	\Sigma_{\rm pur}=A\cup B\cup \gamma_{AB}
		   \end{align}
		   is associated with a pure state. A choice of oriented bipartition $\gamma_{AB}=\CE_A\cup \CE_B$, where $\CE_A$ is the segment beginning at the RT endpoint adjacent to $A$ and ending at the EWCS split point, then induces a purification
		   \begin{align}
		   	A^\sharp=A\cup \CE_A~~,~~ B^\sharp=B\cup \CE_B\,,\label{minimal-purification}
		   \end{align}
		   of the original mixed state $\rho_{AB}$; see figure \ref{fig:purification-partition} for an illustration. Before optimization, a choice of purifier segment $\CE_A$ determines a candidate cross section $\Sigma(\CE_A)$ and a candidate purified entropy
		   \begin{align}
		   	S(A^\sharp;\CE_A)=\frac{\mathrm{Area}\bigl(\Sigma(\CE_A)\bigr)}{4G}\,.
		   \end{align}
		   The canonical or minimal geometric purification is singled out only after minimizing over the split of $\gamma_{AB}$:
		   \begin{align}
		   	E_{\rm geom}(A:B)=\min_{\CE_A}S(A^\sharp;\CE_A)
		   	=\frac{\mathrm{Area}(\Sigma_{AB})}{4G}=E_W(A:B)\,.
		   \end{align}
		   Within the surface/state class of geometric purifications, this prescription promotes the RT surface from a passive homology boundary to an active purifying component. It reproduces the EWCS minimization, but it does not by itself prove minimization over the complete quantum-information-theoretic space of purifications.
		   
		   These considerations motivate a microscopic PEE-thread realization of the
		   holographic conjecture $E_P=E_W$ \cite{Takayanagi:2017knl} within the minimal purification. In the ordinary PEE-thread construction, the bit-thread flow for a subsystem is obtained by superposing the elementary PEE-thread flows sourced from all points in that subsystem \cite{Lin:2023rxc}. Applying the same rule to the purified subsystem $A^\sharp=A\cup\CE_A$ leads to the following proposal.
		   The physical bit-thread field whose bottleneck is the EWCS is obtained only after the flux/length optimization over the purifier split:
		   \begin{align}
		   	v_{AB}^\mu(x,z)=\int_{A^\sharp}\dd\lambda\, v_\lambda^\mu(x,z)=\int_A \d x_0 \,v^\mu_{x_0}(x,z)+\int_{\CE_A} \d y \,v^\mu_{y}(x,z)\label{EW-proposal}
		   \end{align}
		   where $\lambda$ denotes a collective coordinate on $A^\sharp$, and $v^\mu_\lambda$ denotes the PEE-thread flow emanating from $\lambda$. In the second equality, we choose $x_0$ to denote the points on the asymptotic boundary and $v^\mu_{x_0}$ describes the corresponding PEE-thread flow; similarly $y$ is a canonical choice of coordinate on the RT surface and $v^\mu_y$ the corresponding PEE-thread flow vector. On the other hand, if we regard the RT surface $\gamma_{AB}$ as the EOW brane, then $\mathcal{E}_A$ can be understood as the entanglement island $I_A$, and Eq.~\eqref{EW-proposal} can be understood in exactly the same way as in the AdS/BCFT configuration.
			   The endpoint-reduction formula of appendix~\ref{app:distance-difference-potential} is especially useful in this setting.  For a fixed purifier split, $A^\sharp=A\cup\CE_A$ is an oriented relative chain on the extended surface $A\cup B\cup\gamma_{AB}$.  The endpoints at which its boundary and RT-surface pieces are glued are internal and cancel in the integrated PEE current.  Consequently the final stream function retains only the endpoints of the candidate EWCS.  This explains in advance the cancellations that will appear explicitly below, while the subsequent minimization over the split remains a separate purification optimization.
		   
		   In the following, we provide several examples in support of this claim. We restrict ourselves to Poincar\'e AdS$_3$ and the planar BTZ black brane; extension to another geometry requires a separate check of geodesic uniqueness, homology chamber and the norm bound. The proposal is supported by three independent considerations in the examples treated here:
		   \begin{itemize}
		   	\item First, the surface/state correspondence identifies the RT surface as a legitimate purifying surface for the entanglement wedge state.  
		   	\item Second, the EWCS is the RT surface for $A\cup \CE_A$ in this purified geometry.  
		   	\item Third, the PEE-thread superposition over $A\cup \CE_A$ gives a divergenceless
		   	max-flow that saturates on the EWCS in the examples considered below.
		   \end{itemize}

		   It is useful to contrast this with the standard constrained-flow formulation for the EWCS \cite{Du:2019emy}. In that formulation, one usually describes a max flow corresponding to the boundary subsystem alone, inside the entanglement wedge, whose bottleneck is the EWCS. In the present PEE-thread formulation, the same bottleneck is obtained by treating the RT surface as an auxiliary purifying boundary and by applying the ordinary PEE-thread superposition rule to the extended subsystem $A\cup \CE_A$. Thus the two viewpoints are compatible, but they organize the degrees of freedom differently. The standard formulation emphasizes the flow between the
		   original boundary regions $A$ and $B$, while the minimal purification formulation emphasizes the entropy of the purified subsystem $A^\sharp=A\cup \CE_A$.
		   
		   \subsection{PEE threads and bit threads sourced from the RT surface}
		   
%

\subsubsection{Poincar\'e AdS$_3$}

		   According to \eqref{EW-proposal}, in the minimal-purification configuration the bit-thread current must include not only the PEE-thread bundles sourced from the boundary, but also those emitted from points on the RT surface itself. The vector fields describing the PEE flow sourced from the AdS boundary were worked out in \cite{Lin:2023rxc}. Here we compute the vector field describing the PEE flow sourced from an arbitrary RT surface. Later, we will apply this result to the different RT chords in $\mathcal{E}_A$ in order to evaluate the superposition in \eqref{EW-proposal}, thereby obtaining the bit-thread flow in the minimal-purification configuration.
		   
\begin{figure}[th]
		   	\centering
		   	\includegraphics[width=0.5\linewidth]{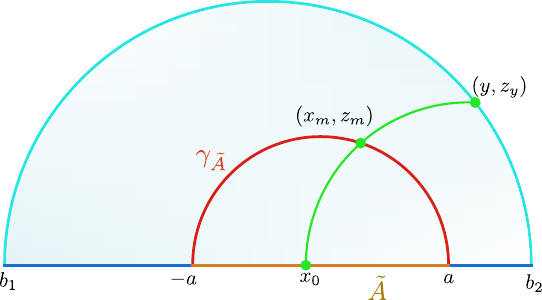}
		   	\caption{PEE threads in RT surface purification}
		   	\label{fig:rt-purification-poincare}
		   \end{figure}

		   Let us consider a single boundary interval $A$, with the purifying system defined on its RT surface $\gamma_A$, so that a pure state is defined on
		   \begin{equation}
		   	\Sigma_A=A\cup\gamma_A.
		   	\label{eq:sigma-A-single-interval}
		   \end{equation}
		   We then construct the source-labelled PEE-thread field $V_y^\mu$ emitted from a point $y\in\gamma_A$ by generalizing the flux-matching condition \eqref{eq:pee-flux-matching-review} to this configuration. For example, let us take $A=[b_1,b_2]$, then $\gamma_A$  is given by the following semicircular profile
		   \begin{align}
		   	\gamma_A:~\left(x-\frac{b_1+b_2}{2}\right)^2+z^2=\left(\frac{b_2-b_1}{2}\right)^2\label{RT-Poincare-asymm}
		   \end{align}
		   We parametrize a point on $\gamma_A$ by its horizontal coordinate $y$,
		   \begin{equation}
		   	(y,z_y)~~,~~
		   	z_y=\sqrt{(y-b_1)(b_2-y)},
		   	\qquad
		   	b_1<y<b_2 .
		   	\label{eq:Y-y-zy}
		   \end{equation}
		   We define the two-point PEE between the point $(y,z_y)$ on the RT surface and another $(x_0,\epsilon)$ on the asymptotic boundary using \eqref{eq:PEEPa}. To this end, the length of the geodesic joining these two points may be computed using \eqref{Poincare-length} as follows
		   \begin{align}
		   	\CL(x_0,y)=\log\left[\frac{(x_0-y)^2+(y-b_1)(b_2-y)}{\epsilon  \sqrt{(y-b_1) (b_2-y)}}\right]
		   \end{align}
		   Hence, the mixed boundary--RT-surface two-point PEE is obtained as
		   \begin{align}
		   	\CI_{\del\gamma_A}(x_0,y)=\frac{1}{8G}\left|\frac{\del^2\CL(x_0,y)}{\del x_0\del y}\right|=\frac{1}{4G}\frac{|(b_1-x_0)(b_2-x_0)|}{\left(x_0^2-2 x_0 y+(b_1+b_2)y-b_1 b_2\right)^2}\label{PEE-RT-purification}
		   \end{align}
		   For $x_0$ on the complementary boundary component $	A^c=(-\infty,b_1)\cup(b_2,\infty)$, the product $(b_1-x_0)(b_2-x_0)$ has a definite sign on each oriented component, and the absolute value in \eqref{PEE-RT-purification} may be implemented by choosing the appropriate orientation of the endpoint measure. The positive PEE density is the absolute value; the oriented representative is the mixed derivative itself. The corresponding PEE-thread geometry in the RT-surface purification is illustrated in figure~\ref{fig:rt-purification-poincare}.
		   
		   In order to analyze the PEE-thread flow in the RT surface purification scheme, we consider the RT surface corresponding to an auxiliary subsystem $[-a,a]$ as our reference surface $\Gamma$ and identify the point $P_m:(x_m,z_m)$ where the PEE thread joining $x_0$ and $y$ crosses the reference RT surface $\Gamma$. Note that the PEE thread, being a geodesic in Poincar\'e AdS$_3$, corresponds to a semi-circle centered at the boundary with the center and radius given by
		   \begin{align}
		   	x_c=\frac{x_m^2+z_m^2-x_0^2}{2(x_m-x_0)}~~,~~r=\frac{(x_m-x_0)^2+z_m^2}{2(x_m-x_0)}\,.\label{xc-r-RT-purification}
		   \end{align}  
		   Now, we may find the location of the anchor point $(y,z_y)$ of the PEE thread on the RT surface in terms of the crossing point $P_m$ as follows
		   \begin{align}
		   	y&=\frac{(x_0-x_m) (b_1 b_2-x_0 x_m)+x_0 z_m^2}{(x_0-x_m) (b_1+b_2-x_0-x_m)+z_m^2}
		   \end{align}
		   Solving this equation for the boundary endpoint gives
		   {\footnotesize
		   \begin{align}
		   	x_0
		   	&=
		   	\frac{
		   		x_m^2+z_m^2+b_1b_2-y(b_1+b_2)
		   		-
		   		\sqrt{
		   			\left(x_m^2+z_m^2+y(b_1+b_2-2x_m)-b_1b_2\right)^2
		   			+4z_m^2(b_1-y)(b_2-y)
		   		}
		   	}
		   	{2(x_m-y)} .
		   	\label{eq:x0-solution-poincare}
		   \end{align}} The opposite sign of the square root corresponds to the oppositely oriented complete geodesic. The branch displayed in \eqref{eq:x0-solution-poincare} is selected by demanding that the geodesic segment join the chosen component of $A^c$ to $(y,z_y)$ while crossing the reference surface $\Gamma$ once.
		   
		   From the above information, we may now find the tangent vector to the PEE thread \eqref{circle-eqn} with center and radius given by \eqref{xc-r-RT-purification} at $P_m$ as follows
		   \begin{align}
		   	\tau^\mu(x_m,z_m)&=\frac{z_m^2}{\sqrt{(x_m-x_c)^2+z_m^2}}\left(1,\frac{x_c-x_m}{z_m}\right)\notag\\
		   	&=\frac{2z_m^2\left(x_m-y,\frac{-y (b_1+b_2-2 x_m)+b_1 b_2-x_m^2+z_m^2}{2 z_m}\right)}{\sqrt{\left(x_m^2+z_m^2+y (b_1+b_2-2 x_m)-b_1 b_2\right)^2+4 z_m^2 (b_1-y) (b_2-y)}}\label{tangent-PEE-RT}
		   \end{align}
		   Direct contraction with the Poincar\'e metric gives $g_{\mu\nu}\tau^\mu\tau^\nu=1$; thus no further normalization factor is implicit in \eqref{tangent-PEE-RT}. The PEE-thread flow may now be written as $V_y^\mu(x_m,z_m)=\widehat f_y(x_m,z_m)\tau^\mu(x_m,z_m)$, where the sign of $\widehat f_y$ fixes the orientation and $|V_y|=|\widehat f_y|$. The flux-matching condition \eqref{eq:pee-flux-matching-review} states that:
		   \begin{align}
		   	\widehat f_y(x_m,z_m)(n^{}_\Gamma\cdot \tau)\d\Sigma_\Gamma=\widehat{\CI}_{\del\gamma_A}(x_0,y)\d x_0\label{proposal-RTpur}
		   \end{align}
		   Here $n_\Gamma$ is the unit normal to the auxiliary separator and $\dd\Sigma_\Gamma$ is its induced line element. Using the Jacobian $\dd x_0/\dd x_m$ obtained from
		   \eqref{eq:x0-solution-poincare}, one obtains
		   \begin{align}
		   	(n^{}_\Gamma\cdot\tau)\,\d\Sigma_\Gamma=\frac{\left(-y (b_1+b_2)+b_1 b_2+x_m^2+z_m^2\right)}{z_m \sqrt{\left(y (b_1+b_2-2 x_m)-b_1 b_2+x_m^2+z_m^2\right)^2+4 z_m^2 (b_1-y) (b_2-y)}}\d x_m 
		   \end{align}
		   This expression contains no dependence on the auxiliary interval size $a$, as it should.  Combining it with the two-point PEE \eqref{PEE-RT-purification}, we may obtain the norm of the PEE-thread flow emanating from the RT surface to be
		   \begin{align}
			   \widehat f_y(x_m,z_m)
			   &=\frac{1}{4G}\frac{z_m \left((b_1-x_m) (b_2-x_m)+z_m^2\right)}
			   {\left(y (b_1+b_2-2 x_m)-b_1 b_2+x_m^2+z_m^2\right)^2
			   +4 z_m^2 (b_1-y) (b_2-y)}\,,\notag\\
			   |V_y|&=|\widehat f_y|\,.
		   \end{align}
		   The sign of the numerator is an orientation convention. Multiplying by the coordinate direction \eqref{tangent-PEE-RT}, the complete PEE-thread current emitted from the RT-surface point $y$ is
		   given by
		   \begin{align}
		   	V^\mu_y(x_m,z_m)=\frac{1}{4G}&\frac{2 z_m^3 \left((b_1-x_m) (b_2-x_m)+z_m^2\right)}{\left[\left(y (b_1+b_2-2 x_m)-b_1 b_2+x_m^2+z_m^2\right)^2+4 z_m^2 (b_1-y) (b_2-y)\right]^{3/2}}\notag\\&\qquad\qquad\qquad\qquad\times\left(x_m-y,\frac{-y (b_1+b_2-2 x_m)+b_1 b_2-x_m^2+z_m^2}{2 z_m}\right)\label{PEE-flow-RT-purification}
		   \end{align}
		   This is the desired local RT-surface source field in Poincar\'e AdS$_3$. The auxiliary parameter $a$ has disappeared, confirming that $\Gamma$ was only
		   a book-keeping device for converting endpoint counting into flux counting. Secondly, the boundary endpoint $x_0$ has disappeared after the use of
		   \eqref{eq:x0-solution-poincare}; the field is labelled only by the source coordinate $y$ on $\gamma_A$. Finally, direct differentiation of \eqref{PEE-flow-RT-purification} gives $\nabla_\mu V_y^\mu=0$ away from the source. Flux matching fixes the normalization, while this local calculation supplies the independent divergencelessness check.

		   A decisive check is obtained by integrating the source field over the entire purifying RT surface.  Because $\Sigma_A=A\cup\gamma_A$ represents a pure state in the surface/state correspondence, the complete flow can be generated either from the boundary interval $A$ or from the purifying surface $\gamma_A$, with the appropriate orientation.  One finds
		   \begin{align}
		   	v_A^\mu(x,z)
		   	&=
		   	\int_{b_1}^{b_2}\dd y\,V_y^\mu(x,z)
		   	\nonumber\\
		   	&=
		   	\frac{1}{4G}
		   	\frac{2z^3(b_2-b_1)}{\big[(b_1-x)^2+z^2\big]
		   		\big[(b_2-x)^2+z^2\big]}
		   	\left(
		   	x-\frac{b_1+b_2}{2},
		   	\frac{z^2+(x-b_1)(b_2-x)}{2z}
		   	\right) .
		   	\label{eq:bit-thread-nonsymmetric-interval}
		   \end{align}
		   For the centered interval \(b_1=-b\), \(b_2=b\), this reduces to the standard symmetry-adapted bit-thread flow in the entanglement wedge of a single interval generated by PEE-thread superposition
		   \cite{Lin:2023rxc,Caggioli:2024uza,Agon:2018lwq}.  Equation~\eqref{eq:bit-thread-nonsymmetric-interval}
		   therefore provides a non-trivial check of the PEE-thread flow \eqref{PEE-flow-RT-purification}.
		   \subsubsection{BTZ black brane}
		   We now repeat the same construction intrinsically on a constant-time slice of
		   the BTZ black-brane geometry \eqref{BTZ-metric}. This example is important for two reasons.  First, it demonstrates that the RT-surface-sourced PEE-thread construction is not tied to the semicircular geometry of the Poincar\'e vacuum.  Second, it supplies the thermal building block used later in the BTZ realization of the minimal-purification flow.
		   
		   The RT surface for a subsystem $A=[-b,b]$ may be obtained from \eqref{geod-class-I}, with the replacements $x_1\to -b, x_2\to b$:
		   \begin{align}
		   	\gamma_A:\sqrt{1-\frac{z^2}{z_h^2}}=\cosh\left(\frac{x}{z_h}\right)\sech\left(\frac{b}{z_h}\right)\label{RT-symm-BTZ}
		   \end{align}
		   Geodesic length between an arbitrary point $y$ on the RT surface and another at $x_0$ on the asymptotic boundary may be computed using \eqref{length-BTZ} as follows
		   \begin{align}
		   	\CL(x_0,y)=\log\left[\frac{2z_h}{\epsilon}\frac{\cosh \left(\frac{x_0-y}{z_h}\right)-\sech\left(\frac{b}{z_h}\right) \cosh \left(\frac{y}{z_h}\right)}{\sqrt{1-\sech^2\left(\frac{b}{z_h}\right) \cosh ^2\left(\frac{y}{z_h}\right)}}\right]
		   \end{align}
		   Therefore, according to the extended definition of PEE in \eqref{PEE}, the two-point PEE is obtained as
		   \begin{align}
		   	\CI^{}_{\del\gamma_A}(x_0,y)=\frac{1}{8G}\frac{\sech\left(\frac{b}{z_h}\right) \cosh \left(\frac{x_0}{z_h}\right)-1}{z_h^2 \left[\cosh \left(\frac{x_0-y}{z_h}\right)-\sech\left(\frac{b}{z_h}\right) \cosh \left(\frac{y}{z_h}\right)\right]^2}\label{2-pt-PEE-RTpur-BTZ}
		   \end{align}
		   For the oriented configuration used in the flux construction, the endpoint
		   $x_0$ lies in the complementary boundary region $x_0\in(-\infty,-b)\cup(b,\infty)$.
		   On this domain,
		   \begin{equation}
		   	\sech\left(\frac{b}{z_h}\right)
		   	\cosh\left(\frac{x_0}{z_h}\right)-1\geq0,
		   	\label{eq:positivity-btz-kernel}
		   \end{equation}
		   so the density in \eqref{2-pt-PEE-RTpur-BTZ} is non-negative.  For other endpoint orderings, the absolute value of the mixed derivative gives the positive PEE density, while its sign keeps track of orientation.
		   
		   We now introduce an auxiliary symmetric interval $\widetilde A=[-a,a]$ and
		   let $\Gamma$ be its RT surface. The unit normal vector to $\Gamma$ at some point $P_m:(x_m,z_m)$ is given by
		    \begin{align}
		    n_{\Gamma,\mu}=\frac{\sqrt{1-\frac{z_m^2}{z_h^2}}}{\sqrt{\left(z_h^2-z_m^2\right) \tanh ^2\left(\frac{x_m}{z_h}\right)+z_m^2}}\left(\frac{z_h}{z_m}\tanh \left(\frac{x_m}{z_h}\right),\frac{z_h^2}{z_h^2-z_m^2}\right)\label{normal-ref-BTZ}
		   \end{align}
		   The complete BTZ geodesic through $x_0$ and $(y,z_y)$ can be parametrized by
		   the constants $x_c$ and $\hat c$ used in \eqref{general-solution-geodesics}. From the endpoint data, we find
		   \begin{align}
		   	x_c&=z_h \log \left[\frac{\cosh \left(\frac{b}{z_h}\right) \sinh \left(\frac{y-x_0}{z_h}\right)}
		   	{z_h \left(e^{-\frac{x_0}{z_h}} \cosh \left(\frac{y}{z_h}\right)-e^{-\frac{y}{z_h}} \cosh \left(\frac{b}{z_h}\right)\right)}\right]\,,\notag\\
		   	\frac{\hat c}{z_h}
		   	&=\frac{ \cosh \left(\frac{x_0-y}{z_h}\right)-\sech\left(\frac{b}{z_h}\right) \cosh \left(\frac{y}{z_h}\right)}
		   	{\sinh \left(\frac{x_0-y}{z_h}\right)}
		   \end{align}
		   Hence at the crossing point $P_m$, the geodesic has the following profile:
		   \begin{align}
		   	\sqrt{1-\frac{z_m^2}{z_h^2}}
		   	=\frac{\sech\left(\frac{b}{z_h}\right) \cosh \left(\frac{y}{z_h}\right) \sinh \left(\frac{x_0-x_m}{z_h}\right)+\sinh \left(\frac{x_m-y}{z_h}\right)}
		   	{\sinh \left(\frac{x_0-y}{z_h}\right)}
		   	\label{geodesic-at-P}
		   \end{align}
		   Solving \eqref{RT-symm-BTZ} and \eqref{geodesic-at-P}, we may find the location on the asymptotic boundary as
		   {\footnotesize
		   \begin{align}
		   	x_0=x_m+2z_h\,\textrm{arctanh}\left[
		   	\frac{\Xi+\sqrt{\Xi^2+\frac{z_m^2}{z_h^2}\sinh^2\left(\frac{x_m-y}{z_h}\right)}}
		   	{\left(1+\sqrt{1-\frac{z_m^2}{z_h^2}}\right)\sinh\left(\frac{x_m-y}{z_h}\right)}
		   	\right]\label{x0-BTZ-RTpur}
		   \end{align}
		   }
		   with
		   \begin{align}
		   	\Xi=\cosh \left(\frac{y}{z_h}\right)\sech\left(\frac{b}{z_h}\right)-\sqrt{1-\frac{z_m^2}{z_h^2}} \cosh \left(\frac{x_m-y}{z_h}\right)
		   \end{align}
		   Note that the auxiliary subsystem identifier $a$ is translated to the crossing point $P_m$, which is completely arbitrary. With all the ingredients in place, we may compute the unit tangent vector to the geodesic thread at $P_m$ as follows
		   {\footnotesize
		   \begin{align}
		   	\tau^\mu_\textrm{Thread}
		   	=\frac{z_m \left(\frac{z_m}{z_h}\sinh \left(\frac{x_m-y}{z_h}\right),
		   	\sqrt{1-\frac{z_m^2}{z_h^2}} \left(\sech\left(\frac{b}{z_h}\right) \cosh \left(\frac{y}{z_h}\right)-\sqrt{1-\frac{z_m^2}{z_h^2}} \cosh \left(\frac{x_m-y}{z_h}\right)\right)\right)}
		   	{\sqrt{\left(\sqrt{1-\frac{z_m^2}{z_h^2}} \cosh \left(\frac{x_m-y}{z_h}\right)-\sech\left(\frac{b}{z_h}\right) \cosh \left(\frac{y}{z_h}\right)\right)^2+\frac{z_m^2}{z_h^2}\sinh ^2\left(\frac{x_m-y}{z_h}\right)}}
		   \end{align}
		   }
		   Furthermore, from \eqref{2-pt-PEE-RTpur-BTZ} and \eqref{x0-BTZ-RTpur}, we may obtain
		   {\footnotesize
		   \begin{align}
		   	\CI^{}_{\del\gamma_A}(x_0,y)\d x_0
		   	=\frac{1}{4G}
		   	\frac{-\sech\left(\frac{x_m}{z_h}\right) \cosh \left(\frac{y}{z_h}\right)
		   	\left(\sqrt{1-\frac{z_m^2}{z_h^2}}-\sech\left(\frac{b}{z_h}\right) \cosh \left(\frac{x_m}{z_h}\right)\right)^2}
		   	{2 z_h^2 \left[\left(\sqrt{1-\frac{z_m^2}{z_h^2}} \cosh \left(\frac{x_m-y}{z_h}\right)-\sech\left(\frac{b}{z_h}\right) \cosh \left(\frac{y}{z_h}\right)\right)^2+\frac{z_m^2}{z_h^2}\sinh ^2\left(\frac{x_m-y}{z_h}\right)\right]^{3/2}}\d x_m
		   \end{align}
		   }
		   Now, utilizing the flux-matching condition \eqref{proposal-RTpur}, the signed amplitude of the PEE-thread flow is
		   {\small
		   \begin{align}
			   	\widehat f_y(x_m,z_m)
			   	&=\frac{1}{4G}\frac{z_m^2\left(\sqrt{1-\frac{z_m^2}{z_h^2}}
			   	-\sech\left(\frac{b}{z_h}\right)\cosh\left(\frac{x_m}{z_h}\right)\right)}
			   	{2z_h^2\left[\left(\sqrt{1-\frac{z_m^2}{z_h^2}}\cosh\left(\frac{x_m-y}{z_h}\right)
			   	-\cosh\left(\frac{y}{z_h}\right)\sech\left(\frac{b}{z_h}\right)\right)^2
			   	+\frac{z_m^2}{z_h^2}\sinh^2\left(\frac{x_m-y}{z_h}\right)\right]}\,,\notag\\
			   	|V_y|&=|\widehat f_y|\,.
		   	\label{PEE-thread-RT-purification-BTZ}
		   \end{align}
		   }
		   With the unit tangent in the preceding display, the full local field is
		   \begin{align}
		    V_y^\mu(x,z)=\widehat f_y(x,z)\,\tau_y^\mu(x,z)\,.
		   \end{align}
		   Here $\widehat f_y(x,z)$ and $\tau_y^\mu(x,z)$ are obtained from the preceding expressions by replacing $(x_m,z_m)$ with the generic bulk point $(x,z)$. The construction is independent of the auxiliary interval $\widetilde A$. The field is divergence free away from its source, checking both its normalization and the branch chosen in \eqref{x0-BTZ-RTpur}.
		   \paragraph{Bit threads:} We may obtain the bit-thread flow corresponding to the interval $A$ by integrating the PEE-thread flow over the RT surface:
		   \begin{align}
		   	v^\mu_A&=\int_{-b}^{b}\d y\,V^\mu_y(x,z)\notag\\
		   	&=\frac{1}{4G}\frac{z^2 \sinh \left(\frac{b}{z_h}\right)}{z_h^2 \left(\cosh \left(\frac{b-x}{z_h}\right)-\sqrt{1-\frac{z^2}{z_h^2}}\right) \left(\cosh \left(\frac{b+x}{z_h}\right)-\sqrt{1-\frac{z^2}{z_h^2}}\right)}\Bigg(z \sinh \left(\frac{x}{z_h}\right),\notag\\&\qquad\qquad\qquad\qquad\qquad\qquad z_h \sqrt{1-\frac{z^2}{z_h^2}} \left[\cosh \left(\frac{b}{z_h}\right)-\sqrt{1-\frac{z^2}{z_h^2}} \cosh \left(\frac{x}{z_h}\right)\right]\Bigg)
		   \end{align}
		   The above result agrees with $\textbf{V}_+$ reported in \cite{Caggioli:2024uza} and with the bit-thread configuration derived in \cite{Agon:2018lwq}, providing another consistency check of our construction of PEE threads adapted to the RT-surface purification.  The same thermal geodesic flow also appears as a benchmark in \cite{BasuWenChandra}; there it is used to separate features intrinsic to asymptotic planar BTZ from those generated by a finite radial cutoff.
		   
		   The significance of these checks is that the RT surface has been shown to support a consistent PEE source sector in both the vacuum and thermal geometries.  The fields derived in this subsection are therefore not special-purpose expressions for the entropy of a single interval.  They are universal local ingredients which may be integrated over an arbitrary purifier segment $\CE_A\subset\gamma_{AB}$.  In the following subsections, this RT-surface contribution is combined with the ordinary boundary-sourced contribution to construct the minimal-purification current whose bottleneck is the EWCS.
\subsection{Minimal purification in Poincar\'e AdS$_3$}
		   \subsubsection{Adjacent subsystems}
		   \label{subsec:adj-Poincare}
		   We now apply the minimal purification prescription \eqref{EW-proposal} to the simplest mixed state with a connected entanglement wedge, namely two adjacent intervals $A=[-b,a]$ and $B=[a,b]$, with $0<a<b$.The union $A\cup B$ is purified by the RT surface $\gamma_{AB}$ given in \eqref{RT-profile}. In the minimal-purification scheme adapted to the surface/state correspondence, $\gamma_{AB}$ is promoted from a passive homology surface to an auxiliary purifying boundary. The choice of splitting $\gamma_{AB}=\CE_A\cup\CE_B$ defines the purified bipartition \eqref{minimal-purification}, and the entanglement-wedge cross section corresponds to the RT surface for $A^\sharp$ inside the purified entanglement wedge. The bit-thread construction below should therefore be understood as a flow for the purified subsystem $A^\sharp$, not for $A$ alone.
		   
		   The EWCS $\Sigma_{AB}$ is a geodesic emanating from the common endpoint $(a,0)$ and ending orthogonally at $\gamma_{AB}$, at the point
		   \begin{align*}
		   	(\hat x_m,\hat z_m):=\left(\frac{2a b^2}{a^2+b^2},b-\frac{2 a^2 b}{a^2+b^2}\right)\,.
		   \end{align*} 
		   The center and radius of this circular arc are given by
		   \begin{align}
		   	x_c=\frac{a^2+b^2}{2 a}~~,~~r=\frac{b^2-a^2}{2 a}\,.
		   \end{align}
		   The segment $\CE_A$ is the portion of $\gamma_{AB}$ from $(-b,0)$ to $(\hat x_m,\hat z_m)$,
		   \begin{align}
		   	\CE_A=\{(x,z)|x^2+z^2=b^2,-b\leq x\leq \hat x_m\}
		   \end{align}
		   This is the precise analogue, in the minimal-purification geometry, of the brane component ${\rm I}_A$ in the AdS/BCFT construction; the latter has an island interpretation only in a doubly holographic completion. A similar construction was provided in \cite{Chandra:2024bkn} for the path integral optimized purification \cite{Caputa:2017urj,Caputa:2017yrh}, in a slightly different context.
		   
\begin{figure}
		   	\centering
		   	\includegraphics[width=0.5\linewidth]{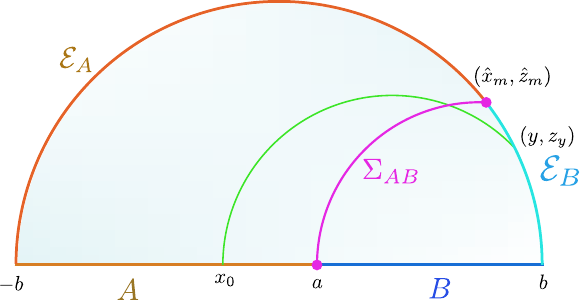}
		   	\caption{Schematic PEE threads for the EWCS with two adjacent intervals $A=[-b,a]$ and $B=[a,b]$ in Poincar\'e AdS$_3$.}
		   	\label{fig:adj-poincare}
		   \end{figure}

		   The PEE-thread flow emanating from the RT surface has the form \eqref{PEE-flow-RT-purification} with the replacements $b_1\to -b$ and $b_2\to b$. Integrating the PEE threads over the portion $\CE_A$ of the RT surface, we obtain the following expression
		   \begin{align}
		   	&\int_{\CE_A}\d y\,V^\mu_y(x,z)=\int_{-b}^{\hat x_m}\d y\,V^\mu_y(x,z)\notag\\
		   	=&\frac{1}{4G}\frac{z^2}{(b+x)^2+z^2}\left(\frac{(b+x)^2-z^2}{2 z},b+x\right)+\frac{1}{4G}\frac{z^2\left(-\frac{(x-\hat x_m)^2+\hat z_m^2-z^2}{2 z},\hat x_m-x\right)}{\sqrt{\left((x-\hat x_m)^2+\hat z_m^2-z^2\right)^2+4z^2(x-\hat x_m)^2}}
		   \end{align}
		   On the other hand, the collection of the PEE threads emanating from $A$ is given by
			   \begin{align}
			   	\int_{a}^{-b}\d x_0V^\mu_{x_0}(x,z)
			   	=\frac{z^2}{4G}\Bigg(&z \left(\frac{1}{(b+x)^2+z^2}-\frac{1}{(a-x)^2+z^2}\right),\notag\\
			   	&\frac{x-a}{(a-x)^2+z^2}-\frac{b+x}{(b+x)^2+z^2}\Bigg)
			   \end{align}
		   A schematic of the PEE threads contributing to this adjacent-interval minimal purification is shown in figure~\ref{fig:adj-poincare}.
		   Hence, according to the proposal \eqref{EW-proposal}, we obtain the bit-thread flow whose bottleneck is the EWCS for the mixed state $\rho_{AB}$:
		   \begin{align}
		   	v^\mu_{AB}=\frac{z^2}{4G}\left(\frac{1}{2z}\left[\frac{(x-a)^2-z^2}{D_a}-\frac{(x-\hat x_m)^2+\hat z_m^2-z^2}{\Delta}\right],\frac{x-a}{D_a}-\frac{x-\hat x_m}{\Delta}\right)\label{bit-thread-EW-adj}
		   \end{align}
		   where we have introduced the shorthand notations
		   \begin{align}
		   	D_a=(x-a)^2+z^2~~,~~\Delta=\sqrt{\left((x-\hat x_m)^2+\hat z_m^2-z^2\right)^2+4z^2(x-\hat x_m)^2}
		   \end{align}
		   Note that the dependence on the auxiliary endpoint $(-b,0)$ cancels between the boundary-sourced and the RT-surface-sourced pieces.  This is precisely the relative-chain cancellation described in appendix~\ref{app:distance-difference-potential}: after the two source sectors are combined, the current remembers only the common boundary endpoint $(a,0)$ and the finite point $(\hat x_m,\hat z_m)$ on the purifying RT surface. 
		   On the EWCS, we find
		   \begin{align}
		   	v^\mu_{AB}\Big|_{\Sigma_{AB}}=\frac{1}{4G(b^2-a^2)}\left(\left(a^2-2 a x+b^2\right) \sqrt{\left(\frac{x}{a}-1\right) \left(b^2-a x\right)},2 (a-x) \left(a x-b^2\right)\right)
		   \end{align}
		   It follows that this vector field is normal to the EWCS and its norm satisfies $\left|v_{AB}\right|_{\Sigma_{AB}}=\frac{1}{4G}$. Away from the EWCS, eliminating the square roots in the coordinate components produces different algebraic branches in different subregions, so a single unsigned radical is not a reliable global expression for the norm. The distance-difference representation below instead gives an exact invariant norm and proves the bound without a branch ambiguity. A candidate bit-thread flow must also be divergenceless. Here this follows locally from the stream-function representation displayed below, not merely from flux matching. We additionally investigate whether the integral lines constitute a regular non-crossing foliation of the entanglement wedge.
		   
		   \paragraph{Integral curves of the flow:} We now investigate the integral lines of the bit-thread flow \eqref{bit-thread-EW-adj} in the entanglement wedge $\CW_{AB}$.
		   The integral curves are given by the solutions to the differential equations
		   \begin{align}
		   	\frac{\dd x(s)}{\dd s}=v^x~~,~~\frac{\dd z(s)}{\dd s}=v^z\label{integral-curve-ODE}
		   \end{align}
		   Instead of solving these first order equations directly, it is more convenient to find a stream function first. A first integral of this system of differential equations is given by
		   \begin{align}
		   	\Psi(x,z)=-\frac{1}{2}\log\frac{(x-a)^2+z^2}{z\ell_\star}+\frac{1}{2}{\rm arccosh}\left[\frac{(x-\hat x_m)^2+\hat z_m^2+z^2}{2\hat z_mz}\right]
		   \end{align}
		   where $\ell_\star$ is an arbitrary reference scale that only shifts $\Psi$. As in the AdS/BCFT cases, the two terms in the stream function have very simple geometric meaning. The first term is minus one half of the Busemann function associated with the boundary point $(a,0)$ (see e.g. \eqref{Busemann}), while the second one is one half of the geodesic distance from the bulk point $(x,z)$ to the point $(\hat x_m,\hat z_m)$ on $\gamma_{AB}$:
		   \begin{align}
		   	\CL_m(x,z)=\operatorname{arccosh}\left[1+\frac{(x-\hat x_m)^2+(z-\hat z_m)^2}{2\hat z_m z}\right]
		   \end{align}
		   Therefore, the stream function is built out of the two endpoints of the EWCS. The same geometric representation gives the exact invariant norm.  In the terminology of appendix~\ref{app:fermi-extended-flows}, this is a finite--ideal endpoint pair.  Fermi coordinates adapted to the candidate EWCS therefore also furnish the normal-geodesic comparator $v_{\rm geo}=(4G)^{-1}\operatorname{sech}\lambda\,\partial_\lambda$.  It calibrates the same fixed cross section, but need not induce the same endpoint map or the same streamlines as the PEE-selected current away from the bottleneck. Both the Busemann function $\Phi_a=\log[D_a/(z\ell_\star)]$ and the geodesic distance $\CL_m$ have unit hyperbolic gradient norm, and hence
		   \begin{align}
		   	|v_{AB}|^2
		   	&=\frac{1}{32G^2}\left(1-\nabla\Phi_a\cdot\nabla\CL_m\right)
		   	\leq\frac{1}{16G^2}\,.
		   	\label{norm-adj-EoP}
		   \end{align}
		   The inequality is saturated when the two unit gradients become anti-parallel, which occurs precisely on the geodesic connecting $(a,0)$ to $(\hat x_m,\hat z_m)$, namely on the EWCS $\Sigma_{AB}$.	 Hence, we have
		   \begin{align}
		   	\left.v^\mu_{AB}\right|_{\Sigma_{AB}}=\frac{1}{4G}n^\mu_{\Sigma_{AB}}\,.
		   \end{align}
		   Therefore, the flow saturates on the correct bottleneck.
		    
\begin{figure}
		   	\centering
		   	\includegraphics[width=0.75\linewidth]{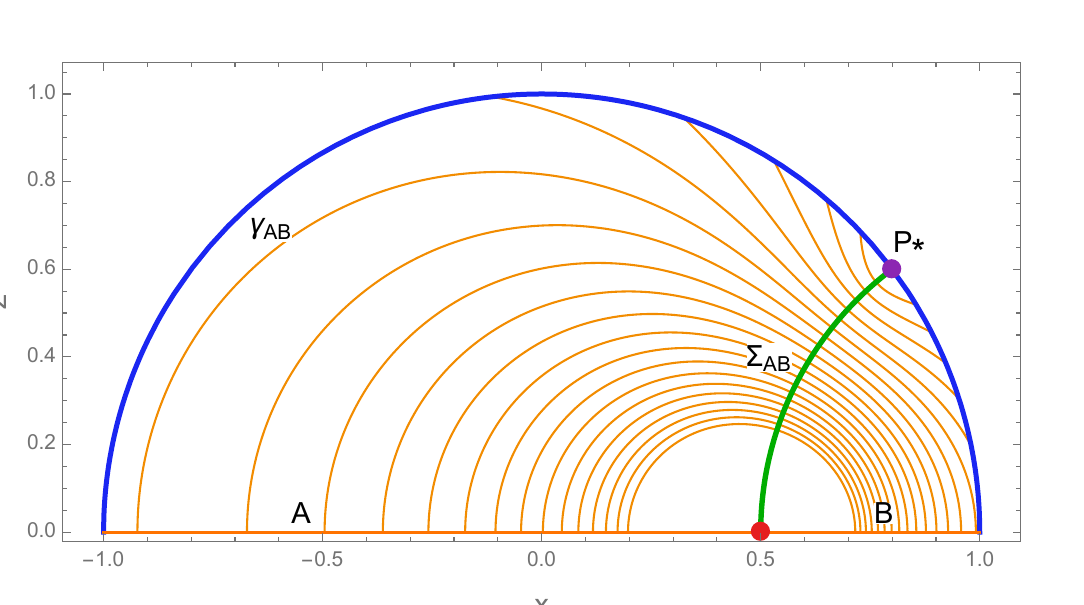}
		   	\caption{Integral curves of the non-geodesic bit-thread field \eqref{bit-thread-EW-adj} inside the entanglement wedge. The RT surface of $A\cup B$ is shown in blue, while the EWCS is depicted by the green curve. We have set $a=0.5\,,\,b=1$ and $4G=1$.}
		   	\label{fig:adj-ew-bit-plot}
		   \end{figure}

		   Furthermore, it is easy to verify that\footnote{The divergenceless nature of the bit-thread field follows directly from these relations.}
		   \begin{align}
		   	v^x=\frac{z^2}{4G}\del_z\Psi(x,z)~~,~~v^z=-\frac{z^2}{4G}\del_x\Psi(x,z)
		   \end{align}
		   so that $\frac{\dd\Psi}{\dd s}=\del_x\Psi\, \dot x(s)+\del_z\Psi \,\dot z(s)=0$. Therefore, the bit-thread integral curves are the level sets
		   \begin{align}
		   	\Psi(x,z)=\Psi_0~~,~~(x,z)\in \CW_{AB}^0
		   \end{align} 
		   where $\CW_{AB}^0=\{(x,z)|z>0,x^2+z^2<b^2\}$ is the ``open'' entanglement wedge.
		   These curves may be written in a more algebraic form as follows. Defining $\lambda=\hat z_m e^{2\Psi_0}/\ell_\star$ and using a hyperbolic trigonometric identity, we have
		   \begin{align}
		   	\lambda^2D_a^2-\lambda D_a\left[(x-\hat x_m)^2+\hat z_m^2+z^2\right]+\hat z_m^2z^2=0\,.
		   \end{align}
		   This is probably the most compact implicit equation for the integral curves of the bit-thread field \eqref{bit-thread-EW-adj}. In general these are not simple semicircular geodesics; rather, they are the streamlines of the superposed PEE-thread flow. This is surprising at first glance, since in light of the findings in \cite{Lin:2023rxc} one might expect the superposition of microscopic PEE threads to yield a geodesic bit-thread flow. However, the vector field obtained after integrating over the endpoints on $A\cup \CE_A$ is a coarse-grained current. Its integral curves are determined by the density-weighted average of all geodesic directions passing through a given bulk point. In the ordinary interval case this average closes on a geodesic congruence because of the residual conformal symmetry of the interval. In the minimal purification scheme, $A^\sharp=A\cup \CE_A$ is a piecewise extended boundary region with one component lying on the RT surface. The corresponding superposition is not constrained by the same symmetry, and the integral curves of the resulting bit-thread flow are generically not geodesics, although the flow itself remains a valid max-flow configuration. Representative integral curves of this non-geodesic flow are shown in figure~\ref{fig:adj-ew-bit-plot}.
		   
		   \paragraph{Regular non-crossing foliation:} The non-crossing property follows from the stream function. In the open wedge, $D_a>0$ and $\Delta>0$. The only point at which $\Delta=0$ is $(\hat x_m,\hat z_m)$, which lies on $\CE_{AB}$ and hence outside the open entanglement wedge. Therefore, $\Psi(x,z)$ is smooth and single valued in $\CW^0_{AB}$, so two curves with different labels $\Psi_1$ and $\Psi_2$ cannot intersect inside $\CW^0_{AB}$. The only remaining possibility for a crossing of level-set branches would be a critical point of $\Psi(x,z)$, namely $\nabla \Psi(x,z)=0$, or equivalently $(v^x,v^z)=0$. Equating the vector field components to zero and eliminating $\Delta$, one finds the constraint
		   \begin{align}
		   	x-\frac{a}{a^2+b^2}(x^2+z^2+b^2)=0\,,
		   \end{align}
		   which is precisely the semicircular EWCS. However, the portion of this semicircle inside the entanglement wedge has $a<x<\hat x_m$, which prevents setting both $v^x$ and $v^z$ equal to zero simultaneously. Therefore no critical point exists inside $\CW_{AB}^0$. The streamlines form a smooth ordered non-crossing foliation of the open entanglement wedge. Therefore the vector field $v^\mu_{AB}$ satisfies all the properties of a bit-thread flow and constitutes a legitimate configuration whose bottleneck is the EWCS.
		   
		   \subsubsection*{Flux through the EWCS, boundary, and RT-surface purifier}
		   We now compute the relevant fluxes of the bit-thread flow \eqref{bit-thread-EW-adj} through various surfaces. Since the flow saturates on the EWCS, its flux through $\Sigma_{AB}$ is simply the regulated length of the EWCS in Planck units \cite{Takayanagi:2017knl},
		   \begin{align}
		   	\Phi\left({\Sigma_{AB}}\right)&=\frac{1}{4G}\int_{a_\epsilon}^{\hat x_m}\d x\,\frac{b^2-a^2}{2 (x-a) \left(b^2-a x\right)}=\frac{1}{4G}\log\left(\frac{b^2-a^2}{\epsilon\,b}\right)\,.
		   \end{align}
		   In the intermediate steps, we have employed the entanglement wedge regularization to implement the UV cutoff near the boundary endpoint:
		   \begin{align}
		   	a_\epsilon=\frac{a^2+b^2}{2a}-\sqrt{\frac{(b^2-a^2)^2}{4a^2}-\epsilon^2}=a+\frac{a}{b^2-a^2}\epsilon^2+\CO(\epsilon^4)
		   \end{align}
		   
		   Next, consider the flux through the extended subsystem $A^\sharp$. The asymptotic-boundary contribution is obtained from the near-boundary limit of the flow:
		   \begin{align}
		   	\Phi_{AB}(A)&=\frac{1}{4G}\int_{a-\epsilon}^{-b}\d x\left(\frac{1}{x-a}+\frac{\hat x_m-x}{b^2-2\hat x_m x+x^2}\right)=\frac{1}{8G}\log\frac{(b^2-a^2)^2}{2b^2\epsilon^2}\,,
		   \end{align}
		   On the other hand, the RT surface contribution is obtained by evaluating the normal flux through the segment $\CE_A\subset \gamma_{AB}$. Using $x_m$ as a coordinate along $\gamma_{AB}$, we find
		   \begin{align}
		   	\Phi_{AB}(\CE_A)&=\frac{1}{4G}\int_{-b}^{\hat x_m}\d x_m\frac{(a-b)^2}{2 (b-x_m) \left(a^2-2 a x_m+b^2\right)}=\frac{1}{8G}\log 2\,.
		   \end{align}
		   From the above expressions, it is easy to verify the relation $\Phi_{AB}(\Sigma_{AB})=\Phi_{AB}(A)+\Phi_{AB}(\CE_A)$. This equality is the central consistency check of the construction. The flow through $A$ alone does not lead to the EWCS; it is reproduced only after the contribution from $\CE_A$ is included. This is precisely what one expects from the minimal purification: the entropy being computed is $S_{A^\sharp}$, not $S_A$.
		   
		   These results provide a microscopic PEE-thread realization of holographic entanglement of purification. The EWCS appears as the bottleneck for a max-flow inside the entanglement wedge, while the source region of the flow is the purified subsystem $A^\sharp=A\cup \CE_A$. The RT-surface segment $\CE_A$ therefore has a concrete operational meaning: it supplies the part of the geometric purifier assigned to $A$. The flux through $\CE_A$ measures the contribution of these purifying degrees of freedom to the minimal purification. This interpretation also clarifies why the PEE-superposed streamlines are non-geodesic. The physical max-flow is not a single family of elementary PEE geodesics; it is the coarse-grained current obtained after summing over all sources on a piecewise extended boundary. The defining requirements are divergencelessness, the norm bound, the appropriate boundary conditions, and saturation on the bottleneck. The regular non-crossing foliation is an additional property of this representative, not a separate max-flow axiom.

		   \subsubsection{Disjoint subsystems}
		   Next we consider two disjoint subsystems $A=[b_1,b_2]$ and $B=[b_3,b_4]$ in the vacuum state of the CFT$_2$. For the connected entanglement wedge, the RT surfaces are given by the following semi-circles centered on the asymptotic boundary
		   \begin{align}
		   	\gamma_{14}:~~&\left(x-\frac{b_1+b_4}{2}\right)^2+z^2=\left(\frac{b_4-b_1}{2}\right)^2\,,\notag\\
		   	\gamma_{23}:~~&\left(x-\frac{b_2+b_3}{2}\right)^2+z^2=\left(\frac{b_3-b_2}{2}\right)^2\,.
		   \end{align}
		   The connected wedge condition may be written in terms of the conformal cross-ratio $\eta$ as
		   \begin{align}
		   	\eta=\frac{(b_2-b_1)(b_4-b_3)}{(b_3-b_1)(b_4-b_2)}>\frac{1}{2}\,.
		   \end{align}
		   When this condition is violated, the classical entanglement wedge is disconnected and the leading holographic EWCS vanishes. The construction below should therefore be understood as applying only in the connected phase.
		   
\begin{figure}[ht]
		   	\centering
		   	\includegraphics[width=0.55\linewidth]{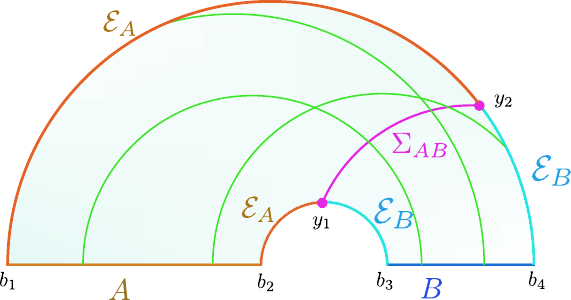}
		   	\caption{Minimal purification for disjoint intervals. The candidate EWCS is the circular segment joining the split points $y_1$ and $y_2$; representative PEE threads are shown in green.}
		   	\label{fig:disj-rt}
		   \end{figure}

		   In the minimal purification, the RT surface $\gamma_{AB}=\gamma_{14}\cup\gamma_{23}$ is promoted to auxiliary purifying boundaries. A choice of purification is specified by choosing a point $P_1\in \gamma_{23}$ on the inner RT surface and another point $P_2\in\gamma_{14}$ on the outer RT surface. These two points determine a candidate for the EWCS, $\Sigma(P_1,P_2)$, defined as the bulk geodesic connecting $P_1$ and $P_2$. Unlike the adjacent interval case, where geometry fixes the two endpoints of the EWCS, the disjoint-interval problem contains a genuine minimization over two internal purification endpoints $P_1$ and $P_2$. See figure~\ref{fig:disj-rt} for a schematic representation of the setup. We parametrize the two candidate endpoints as
		   \begin{align*}
		   	P_1=(y_1,z_1)~~,~~z_1=\sqrt{(b_3-y_1)(y_1-b_2)}~~,~~b_2<y_1<b_3\,,
		   \end{align*}
		   and
		   \begin{align}
		   	P_2=(y_2,z_2)~~,~~z_2=\sqrt{(b_4-y_2)(y_2-b_1)}~~,~~b_1<y_2<b_4\,.
		   \end{align}
		   In the splitting \eqref{minimal-purification}, $\CE_A$ therefore corresponds to two segments $\CE_A^{(1)}\subset\gamma_{23}$ and $\CE_A^{(2)}\subset\gamma_{14}$:
		   \begin{align*}
		   	\CE_A^{(1)}&=\{(x,z)|z=\sqrt{(b_3-x)(x-b_2)},b_2<x<y_1\}\\
		   	\CE_A^{(2)}&=\{(x,z)|z=\sqrt{(b_4-x)(x-b_1)},b_1<x<y_2\}
		   \end{align*}
		   The physical EWCS is obtained only after minimizing over the choice of $y_1$ and $y_2$. The candidate EWCS is the semicircular geodesic
		   \begin{align}
		   	\Sigma(P_1,P_2):~(x-x_c)^2+z^2=r^2
		   \end{align}   
		   The center and radius of this semicircular geodesic are given by
			   {\small
				   \begin{align}
				   	x_c&=\frac{b_2 b_3-b_1 b_4+y_2 (b_1+b_4)-y_1 (b_2+b_3)}{2 (y_2-y_1)}\,,\notag\\
				   	r^2&=x_c^2-\frac{(b_1-b_2-b_3+b_4)y_1 y_2-b_1 b_4 y_1+b_2 b_3 y_2}{y_2-y_1}\label{xc-r-EW-dj}
			   \end{align}
			   }
		   
		   For a fixed choice of $P_1,P_2$, the minimal-purification proposal \eqref{EW-proposal} says that the bit-thread flow should be obtained by superposing PEE-thread flows sourced from the extended subsystem $A^\sharp=A\cup \CE_A$. The PEE-thread flows emanating from the RT surfaces are given by \eqref{PEE-flow-RT-purification}, with appropriate relabeling of the endpoints. More explicitly,
		   \begin{align}
		   	v_{AB}^{\mu}(P_1,P_2)
		   	=
		   	\int_{A}\dd x_0\,V_{x_0}^\mu
		   	+
		   	\int_{\CE_A^{(1)}}d\lambda\,V_{\lambda,[b_2,b_3]}^\mu
		   	+
		   	\int_{\CE_A^{(2)}}d\lambda\,V_{\lambda,[b_1,b_4]}^\mu \,,
		   \end{align}
		   where we have chosen $\lambda$ to parametrize the RT segments, and the subscript on $V_{\lambda,[u,v]}^\mu$ reminds us that the source point $\lambda$ lies on the RT surface of the interval $[u,v]$. Explicitly performing the integrals, we find
		   \begin{align}
		   	v_{AB}^\mu(P_1,P_2)&=\int_{b_1}^{b_2}\d x_0\,V^\mu_{x_0}(x,z)+\int^{b_1}_{y_2}\dd y\,V^\mu_{y,[b_1,b_4]}(x,z)+\int^{y_1}_{b_2}\dd y\,V^\mu_{y,[b_2,b_3]}(x,z)\notag\\
		   	&=\frac{z^2}{4G}\Bigg(\frac{1}{2z}\Big(\frac{z^2-(x-y_2)^2-z_2^2}{R_2}-\frac{z^2-(x-y_1)^2-z_1^2}{R_1}\Big),-\frac{x-y_2}{R_2}+\frac{x-y_1}{R_1}\Bigg)\label{bit-thread-EW-dj}
		   \end{align}
		   where we have defined $R_i(x,z)=\sqrt{\left((x-y_i)^2+z^2+z_i^2\right)^2-4z_i^2z^2}$. This expression shows that the candidate current is controlled only by the two endpoints of the candidate EWCS. The role of the PEE-thread integrals over $A\cup\CE_A$ is to select these endpoints and to fix the normalization.  The disappearance of all other source endpoints is the finite--finite version of the endpoint-reduction identity in appendix~\ref{app:distance-difference-potential}.  In particular, it is a structural consequence of the oriented purified subsystem rather than an accidental simplification of the explicit Poincar\'e integrals. On the candidate EWCS, we find
			   {\small
			   \begin{align}
			   	v^\mu_{AB}\big|_{\Sigma(P_1,P_2)}=\frac{1}{4G}\left(\left(b_1 b_4-b_2 b_3-y_2 (b_1+b_4-2 x)+y_1(b_2+b_3-2 x)\right)\sqrt{\frac{\mathcal P}{y_2-y_1}},2\mathcal P\right)
			   \end{align}
			   }
			   with 
			   {\small
			   \begin{align}
			   	\mathcal P=(x-y_1-y_2) (b_2 b_3-b_1 b_4)-y_1 y_2 (b_1-b_2-b_3+b_4)+y_2 (x-b_1) (b_4-x)-y_1 (x-b_2) (b_3-x)
			   \end{align}
			   }
		   The unit tangent and normal vectors to the candidate EWCS are given by
		   \begin{align}
		   	\tau_{AB}^\mu=\frac{z}{r}\left(z,x_c-x\right)~~,~~n^\mu_{AB}=\frac{z}{r}\left(x-x_c,z\right)
		   \end{align}
		   Utilizing \eqref{xc-r-EW-dj}, it is straightforward to verify that
		   \begin{align}
		   	v^\mu_{AB}\big|_{\Sigma(P_1,P_2)}=\frac{1}{4G}n_{\Sigma(P_1,P_2)}^\mu\label{v-on-EW}
		   \end{align} 
		   Therefore, we have established that the vector field \eqref{bit-thread-EW-dj} is normal to the candidate EWCS on which its norm is given by $\left|v_{AB}\right|=\frac{1}{4G}$.
		   As in the adjacent case, eliminating the endpoint distances from the coordinate components produces branch-dependent radicals and does not give a reliable single global norm formula. The exact invariant norm follows below directly from the distance-difference representation. We next consider the integral curves of the bit-thread flow \eqref{bit-thread-EW-dj}.
		   
		   \paragraph{Integral curves of the flow:} From \eqref{bit-thread-EW-dj}, a stream function may be obtained from the first integral of motion corresponding to the set of differential equations \eqref{integral-curve-ODE}. From the AdS/BCFT analyses and the adjacent intervals case in subsection \ref{subsec:adj-Poincare}, it is natural to expect that the stream function is related to hyperbolic geodesic distances. Indeed, from a direct integration, we find
		   \begin{align}
		   	\Psi(x,z)=\frac{1}{2}\left(\CL_2(x,z)-\CL_1(x,z)\right)\,,\label{Psi-dj-Poincare}
		   \end{align}
		   where $\CL_i(x,z)$ correspond to the geodesic distances from an arbitrary bulk point $(x,z)$ to the candidate EWCS endpoints $(y_i,z_i)$:
		   \begin{align}
		   	\CL_i(x,z)=\operatorname{arccosh}\left[1+\frac{(x-y_i)^2+(z-z_i)^2}{2z\,z_i}\right]\,.\label{aux-lengths-dj-EoP}
		   \end{align}
		   The corresponding bit-thread field is
		   \begin{align}
		   	v_{AB}&=\frac{1}{4G}\star\dd\Psi\,,\notag\\
		   	v_{AB}^x&=\frac{z^2}{4G}\,\partial_z\Psi\,,\qquad
		   	v_{AB}^z=-\frac{z^2}{4G}\,\partial_x\Psi.
		   \end{align}
		   Once again, this representation makes the properties of the flow transparent. It immediately implies
		   \begin{align}
		   	 \nabla_\mu v_{AB}^\mu=0
		   \end{align}
		   inside the entanglement wedge. Moreover, since the distance functions $\CL_i$ have unit hyperbolic gradient norm, one finds the exact relation
		   \begin{align}
		   	\left|v_{AB}\right|^2
		   	=\frac{1}{32G^2}\left(1-\nabla\CL_1\cdot\nabla\CL_2\right)
		   	\leq\frac{1}{16G^2}\,.
		   \end{align}
		   The inequality is saturated when the two unit gradients are antiparallel,
		   which happens precisely on the geodesic segment connecting \(P_1\) and \(P_2\).
		   Therefore
		   \begin{align}
		   	v_{AB}^{\mu}\big|_{\Sigma(P_1,P_2)}=\frac{1}{4G}\,n_{\Sigma(P_1,P_2)}^{\mu}~~,~~|v_{AB}|_{\Sigma(P_1,P_2)}=\frac{1}{4G}.
		   \end{align}
		   Thus, for every fixed splitting of the RT surface, the PEE-superposed current is a valid bit-thread representative for the corresponding purified subsystem $A^\sharp=A\cup \CE_A$, with bottleneck $\Sigma(P_1,P_2)$.  Appendix~\ref{app:fermi-extended-flows} shows that the same fixed geodesic $P_1P_2$ admits a normal-geodesic comparator in Fermi coordinates.  The existence of these two calibrations should be kept separate from the optimization over $P_1$ and $P_2$: only the latter selects the physical EWCS as the common perpendicular of the two RT components.
		   
		   Since $v_{AB}$ is a rotated gradient of $\Psi(x,z)$, its integral curves are
		   precisely the level sets $\Psi(x,z)=\Psi_0$. Equivalently,
		   \begin{align}
		   	\CL_2(x,z)-\CL_1(x,z)=2\Psi_0\,.
		   \end{align}
		   These are hyperbolic Apollonius curves: the loci of constant difference of
		   distances to the two foci $P_1$ and $P_2$. From \eqref{aux-lengths-dj-EoP}, these integral curves may be given the following algebraic form
		   \begin{align}
		   	e^{2\Psi_0}
		   	=\frac{z_1\left[\CA_2+\sqrt{\CA_2^2-4z_2^2z^2}\right]}
		   	{z_2\left[\CA_1+\sqrt{\CA_1^2-4z_1^2z^2}\right]}\,,
		   \end{align}
		   One may eliminate the square roots to obtain the following algebraic equation
		   \begin{align}
		   	z_1^2\CA_2^2+z_2^2\CA_1^2
		   	-2\cosh(2\Psi_0)\,z_1z_2\CA_1\CA_2
		   	+4\sinh^2(2\Psi_0)\,z_1^2z_2^2z^2=0 
		   \end{align}
		   where we have defined $\CA_i=(x-y_i)^2+z^2+z_i^2$. The correct branch is fixed by imposing 
		   \begin{align}
		   	\sinh(2\Psi_0)\left[\frac{\CA_2}{z_2}-\cosh(2\Psi_0)\,\frac{\CA_1}{z_1}\right]\geq 0\,. 
		   \end{align}
		   with the obvious limiting interpretation when $\Psi_0=0$. The special case $\Psi_0=0$ is useful for comparison. Then $\CL_1=\CL_2$ and the level set reduces to
		   \begin{align}
		   	\frac{\CA_2}{z_2}=\frac{\CA_1}{z_1}.
		   \end{align}
		   This is the hyperbolic perpendicular bisector of the two foci.  For $\Psi_0\neq 0$, however, the curves are generically quartic rather than geodesic semicircles. Thus, as in the adjacent minimal-purification case, the elementary PEE threads are geodesics, while the integral curves of the coarse-grained bit-thread current are not generally geodesics.
		   
		   \paragraph{Regular non-crossing foliation:} We now prove the non-crossing property. The stream function $\Psi$ is smooth in the open entanglement wedge $\mathcal W_{AB}^{\circ}$, because its only
		   singularities are at the foci $P_1$ and $P_2$, which lie on the boundary of the wedge.  Therefore different level sets of $\Psi$ cannot intersect unless $\nabla\Psi=0$ somewhere in the interior.  In the following, we show that this never happens.
		   
		   From \eqref{Psi-dj-Poincare}, a critical point would satisfy $\nabla \CL_2=\nabla\CL_1$.
		   For any bulk point $P$, the vector $\nabla \CL_2(P)$ is the unit tangent at $P$ to the geodesic from $P_2$ to $P$, oriented away from $P_2$. Similarly, $\nabla \CL_1(P)$ is the unit tangent at $P$ to the geodesic from $P_1$ to $P$, oriented away from $P_1$.  Equality of these two unit vectors implies that the geodesics $P_2P$ and $P_1P$ have the same tangent at $P$.  By uniqueness of geodesics in the hyperbolic plane, $P_1,P_2$, and $P$ must lie on one complete geodesic.
		   
		   Moreover, $P$ cannot lie on the segment between $P_1$ and $P_2$, because on that segment the two unit vectors point in opposite directions, $\nabla \CL_2=-\nabla\CL_1$.
		   Hence any critical point of $\Psi$ would have to lie on one of the two rays obtained by extending the EWCS geodesic beyond $P_1$ or beyond $P_2$. However, these two rays lie outside the open entanglement wedge.  To see this, note that $\mathcal W_{AB}^{\circ}$ is the intersection of two geodesic half-planes: the half-plane bounded by $\gamma_{14}$ containing the boundary intervals and the half-plane bounded by $\gamma_{23}$ excluding the central gap.  Geodesic half-planes in $\mathbb H^2$ are convex, and therefore their intersection is convex.  The complete geodesic supporting the candidate EWCS intersects the boundary of this convex region at
		   exactly $P_1$ and $P_2$, and the segment between them lies inside
		   $\mathcal W_{AB}^{\circ}$. Since the intersection of a complete geodesic with a convex set is connected, the intersection is precisely the closed segment between $P_1$ and $P_2$. The two exterior rays are therefore outside the wedge. Hence there are no
		   points in \(\mathcal W_{AB}^{\circ}\) where $\nabla\Psi=0$. By the implicit-function theorem, each level set $\Psi(x,z)=\Psi_0$ is a smooth one-dimensional curve in the open wedge.  Since $v_{AB}$ is tangent to these level sets, the integral curves of the bit-thread flow are smooth. Furthermore, two distinct integral curves cannot intersect: an intersection point would have to carry two different values of the single-valued function $\Psi$, which is impossible. Equivalently, uniqueness of solutions to the first-order system \eqref{integral-curve-ODE} forbids two distinct flow lines from crossing.
		   
\begin{figure}[ht]
		   	\centering
		   	\includegraphics[width=0.75\linewidth]{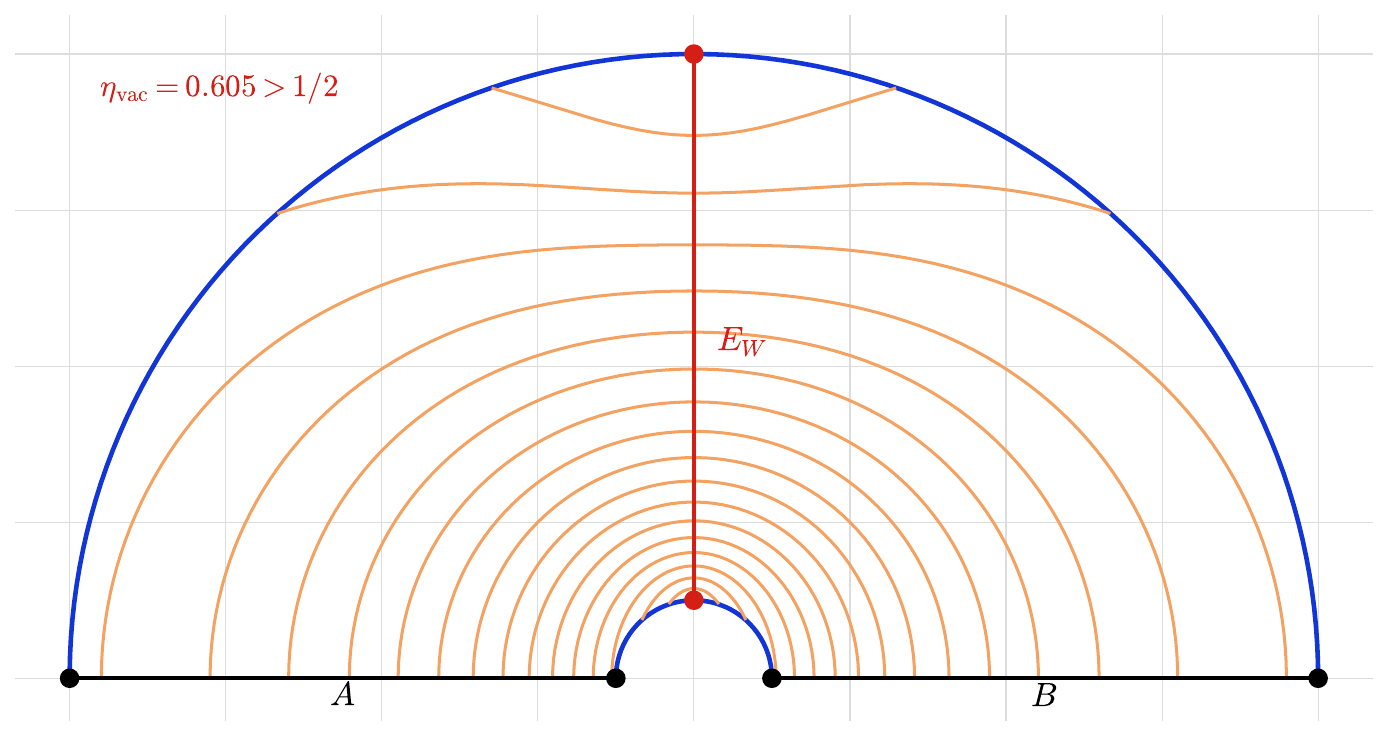}
		   	\caption{Integral curves of the disjoint-interval minimal-purification flow \eqref{bit-thread-EW-dj} in the connected entanglement wedge. The plot uses the symmetric configuration $A=[-4,-0.5]$, $B=[0.5,4]$, for which $\eta_{\rm vac}=0.605>1/2$. The two foci lie on the inner and outer RT components and the red segment is the extremized EWCS.}
		   	\label{fig:disj-eop-bit}
		   \end{figure}

		   Thus the integral curves form a regular non-crossing foliation of the
		   connected entanglement wedge, ordered by the parameter $\Psi_0$. This is a statement about one representative flow and should not be confused with simultaneous locking for a nested family of boundary regions. The resulting foliation for a representative connected disjoint-interval configuration is displayed in figure~\ref{fig:disj-eop-bit}.
		   
		   \paragraph{EWCS bit threads:} For arbitrary
		   $P_1,P_2$, the flow above is not the physical EWCS flow; it is the bit-thread representative of a non-minimal purification. The EWCS flow is obtained only after minimizing over the splitting of $\gamma_{AB}$. This is the disjoint-interval analogue of minimizing over the auxiliary purifying Hilbert-space factor in the definition of
		   entanglement of purification. From \eqref{v-on-EW}, the flux of the bit-thread vector through $\Sigma_{AB}$ is its length in Planck units,
		   \begin{align}
		   	\Phi(\Sigma_{AB})&=\frac{1}{4G}\int_{\Sigma_{AB}}\d\Sigma=\frac{1}{4G}\int_{y_1}^{y_2}\d x\,\frac{r}{r^2-(x-x_c)^2}\notag\\
		   	&=\frac{1}{4G}\textrm{arccosh}\left[\frac{b_1 (y_2-b_4)+b_2 (y_1-b_3)+b_3 y_1+b_4 y_2-2 y_1 y_2}{2 \sqrt{(b_1-y_2) (y_2-b_4)} \sqrt{(b_2-y_1) (y_1-b_3)}}\right]\label{flux-dj}
		   \end{align}
		   Now extremizing the above flux over the endpoints $y_{1,2}$, we obtain the following solutions
		   \begin{align}
		   	y_1&=\frac{b_1 b_4 (b_2+b_3)-2 b_1 b_2 b_3+b_2 b_3 (b_2+b_3-2 b_4)}{-b_1 (b_2+b_3-2 b_4)+b_2^2-b_4 (b_2+b_3)+b_3^2}\,,\notag\\
		   	y_2&=\frac{b_1^2 b_4+b_1 \left(-2 b_4 (b_2+b_3)+b_2 b_3+b_4^2\right)+b_2 b_3 b_4}{b_1^2-b_1 (b_2+b_3)-b_4 (b_2+b_3)+2 b_2 b_3+b_4^2}\,.
		   \end{align}
		   Finally substituting these back in the expression \eqref{flux-dj}, we obtain the extremal flux to be
		   \begin{align}
		   	\Phi_\textrm{ext}(\Sigma_{AB})=\frac{1}{4G}\textrm{arccosh}\left[1+2\frac{(b_1-b_2) (b_3-b_4)}{(b_1-b_4) (b_2-b_3)}\right]=\frac{1}{4G}\log\left(\frac{1+\sqrt\eta}{1-\sqrt\eta}\right)
		   \end{align}
		   where $\eta=\frac{(b_1-b_2) (b_3-b_4)}{(b_1-b_3) (b_2-b_4)}$ is the invariant cross-ratio. This result is in conformity with earlier literature \cite{Takayanagi:2017knl}.
		   
		   Recall that the construction is meaningful only in the connected entanglement-wedge
		   phase.  When the wedge disconnects, there is no cross section connecting the
		   two RT components.  At leading order in the holographic large-$c$ limit the
		   EWCS, and hence the holographic $E_W$, vanishes in that phase. The
		   intersection-weighted or chamber-projected PEE-thread viewpoint is then the
		   appropriate way to describe the transition between the connected and
		   disconnected saddles. In summary, the disjoint-interval example gives a sharp realization of the minimal-purification proposal.  The RT surfaces $\gamma_{14}\cup\gamma_{23}$ supply the purifying degrees of freedom, a choice of endpoints $(P_1,P_2)$ specifies a purification, the PEE-thread superposition over $A\cup \CE_A$ gives a valid max-flow for that purification, and minimizing the bottleneck over $(P_1,P_2)$ reproduces the standard
		   holographic entanglement of purification.
		   
		    
		    		   \subsection{Minimal purification in the planar BTZ black brane}
		    We now repeat the logic of the two Poincar\'e examples directly on the
		    constant-time BTZ slice \eqref{BTZ-metric}.  The boundary- and horizon-resolved
		    planar-BTZ ingredients used here overlap with the construction in
		    \cite{BasuWenChandra}, where the emphasis is instead on the intrinsic BTZ
		    endpoint sectors and their finite-cutoff extension.  Here they are combined
		    with RT-surface source sectors in order to implement the minimal purification.
		    We first specify a split of the RT surface and the corresponding candidate EWCS.
		    We then integrate the boundary- and RT-surface-sourced PEE-thread fields
		    explicitly.  Only after obtaining the integrated vector field do we
		    identify its stream function, analyze its integral curves and prove their
		    regular non-crossing property.  Finally, we optimize the flux over the RT
		    split and substitute the minimizing split into the candidate current.
		    
		    \subsubsection{Adjacent intervals}
		    \label{subsec:adjacent-BTZ-EoP}
		    Consider the adjacent intervals $A=[-b,a]$ and $B=[a,b]$, where $-b<a<b$.  Their union is
		    $[-b,b]$, and its RT surface is
		    \begin{align}
		    	\gamma_{AB}:\qquad
		    	\sqrt{1-\frac{z^2}{z_h^2}}
		    	=\cosh\left(\frac{x}{z_h}\right)
		    	\sech\left(\frac{b}{z_h}\right)\,.
		    \end{align}
		    A point on this surface will be denoted by
		    \begin{align}
		    	P_y=(y,z_y)\,,\qquad
		    	z_y=z_h\sqrt{1-\sech^2\left(\frac{b}{z_h}\right)
		    		\cosh^2\left(\frac{y}{z_h}\right)}\,.
		    	\label{eq:BTZ-adjacent-split-point}
		    \end{align}
		    The split $P_y$ assigns to $A$ the oriented RT segment $\CE_A(y)$
		    running from $(-b,0)$ to $P_y$, and defines the purified subsystem
		    $A^\sharp(y)=A\cup\CE_A(y)$.  The candidate cross section
		    $\Sigma_y$ is the unique BTZ geodesic joining the common boundary point
		    $(a,0)$ to $P_y$.  Directly solving the BTZ geodesic first integrals gives
		    its profile
		    \begin{align}
		    	F_y(x,z)
		    	=&\sqrt{1-\frac{z^2}{z_h^2}}
		    	-\frac{
		    		\sech\left(\frac{b}{z_h}\right)\cosh\left(\frac{y}{z_h}\right)
		    		\sinh\left(\frac{a-x}{z_h}\right)
		    		+\sinh\left(\frac{x-y}{z_h}\right)}
		    	{\sinh\left(\frac{a-y}{z_h}\right)}=0\,.
		    	\label{eq:BTZ-adjacent-EWCS-profile}
		    \end{align}
		    At this stage $P_y$ is arbitrary: $\Sigma_y$ is a candidate EWCS and
		    not yet the physical one.
		    
		    \paragraph{Direct integration of the PEE-thread fields:}
		    With the orientations used in the Poincar\'e adjacent-interval calculation,
		    the candidate current is
		    \begin{align}
		    	v_{AB}^\mu(x,z;y)
		    	=\int_a^{-b}\d x_0\,V^\mu_{x_0}(x,z)
		    	+\int_{-b}^{y}\d y'\,V^\mu_{y',[-b,b]}(x,z)\,.
		    	\label{eq:BTZ-adjacent-candidate-source}
		    \end{align}
		    The first integrand is the intrinsic boundary-source field
		    \eqref{eq:BTZ-boundary-source-flow}; the second is the intrinsic
		    RT-surface-source field constructed in subsection~\ref{sec:eop}.  To show
		    the endpoint cancellation explicitly, it is useful to record the two
		    elementary BTZ endpoint functions that appear as antiderivatives:
		    \begin{align}
		    	\Phi_{x_*}(x,z)
		    	&=\log\left[\frac{z_h}{z}\left(
		    	\cosh\left(\frac{x-x_*}{z_h}\right)
		    	-\sqrt{1-\frac{z^2}{z_h^2}}\right)\right]\,,
		    	\label{eq:BTZ-boundary-Busemann-EoP}\\
		    	\CL_y(x,z)
		    	&=\operatorname{arccosh}\left[
		    	\frac{z_h^2}{z z_y}\left(
		    	\cosh\left(\frac{x-y}{z_h}\right)
		    	-\sqrt{\left(1-\frac{z^2}{z_h^2}\right)\left(1-\frac{z_y^2}{z_h^2}\right)}
		    	\right)\right]\,.
		    	\label{eq:BTZ-bulk-distance-adjacent}
		    \end{align}
		    Substitution of \eqref{eq:BTZ-boundary-source-flow} and direct integration
		    with respect to $x_0$ give
		    \begin{align}
		    	\int_a^{-b}\d x_0\,V^x_{x_0}(x,z)
		    	&=\frac{z^3}{8Gz_h^2}\left[
		    	\frac{1}{\cosh\left(\frac{x+b}{z_h}\right)-\sqrt{1-\frac{z^2}{z_h^2}}}
		    	-\frac{1}{\cosh\left(\frac{x-a}{z_h}\right)-\sqrt{1-\frac{z^2}{z_h^2}}}
		    	\right]\,,\notag\\
		    	\int_a^{-b}\d x_0\,V^z_{x_0}(x,z)
		    	&=\frac{z^2\sqrt{1-\frac{z^2}{z_h^2}}}{8Gz_h}\left[
		    	\frac{\sinh\left(\frac{x-a}{z_h}\right)}
		    	{\cosh\left(\frac{x-a}{z_h}\right)-\sqrt{1-\frac{z^2}{z_h^2}}}
		    	-\frac{\sinh\left(\frac{x+b}{z_h}\right)}
		    	{\cosh\left(\frac{x+b}{z_h}\right)-\sqrt{1-\frac{z^2}{z_h^2}}}
		    	\right]\,.
		    	\label{eq:BTZ-boundary-integrated-adjacent}
		    \end{align}
		    The RT-surface integral is evaluated in the same way.  Its primitive may
		    be verified directly by differentiating with respect to the upper endpoint
		    $y$ and comparing with the explicit RT-source field.  One obtains
		    \begin{align}
		    	\int_{-b}^{y}\d y'\,V^x_{y',[-b,b]}(x,z)
		    	&=\frac{z^2\sqrt{1-\frac{z^2}{z_h^2}}}{8G}
		    	\left[\partial_z\CL_y(x,z)-\partial_z\Phi_{-b}(x,z)\right]\,,\notag\\
		    	\int_{-b}^{y}\d y'\,V^z_{y',[-b,b]}(x,z)
		    	&=\frac{z^2\sqrt{1-\frac{z^2}{z_h^2}}}{8G}
		    	\left[\partial_x\Phi_{-b}(x,z)-\partial_x\CL_y(x,z)\right]\,.
		    	\label{eq:BTZ-RT-integrated-adjacent}
		    \end{align}
		    Adding \eqref{eq:BTZ-boundary-integrated-adjacent} and
		    \eqref{eq:BTZ-RT-integrated-adjacent}, every term involving the internal
		    endpoint $(-b,0)$ cancels.  The candidate current is therefore
		    \begin{align}
		    	v_{AB}^\mu(x,z;y)
		    	&=\frac{z^2\sqrt{1-\frac{z^2}{z_h^2}}}{8G}\left(\partial_z\CL_y(x,z)-\partial_z\Phi_a(x,z),\partial_x\Phi_a(x,z)-\partial_x\CL_y(x,z)\right)\,.
		    	\label{eq:BTZ-adjacent-integrated-components}
		    \end{align}
		    Several properties may already be checked directly from the integrated
		    components. Using
		    $\sqrt{g}=1/(z^2\sqrt{1-z^2/z_h^2})$, direct differentiation gives
		    \begin{align}
		    	\nabla_\mu v_{AB}^\mu
		    	=z^2\sqrt{1-\frac{z^2}{z_h^2}}
		    	\left[\partial_x\left(
		    	\frac{v_{AB}^x}{z^2\sqrt{1-\frac{z^2}{z_h^2}}}\right)
		    	+\partial_z\left(
		    	\frac{v_{AB}^z}{z^2\sqrt{1-\frac{z^2}{z_h^2}}}\right)\right]=0\,.
		    	\label{eq:BTZ-adjacent-divergence}
		    \end{align}
		    The cancellation is simply the equality of the two mixed derivatives in
		    \eqref{eq:BTZ-adjacent-integrated-components}.  Moreover, both $\CL_y$ and $\Phi_a$ obey the
		    unit eikonal equation in the BTZ metric.  Consequently the norm computed
		    directly from the integrated field is
		    \begin{align}
		    	|v_{AB}|^2
		    	=\frac{1}{32G^2}\left[1-z^2\partial_x\CL_y\partial_x\Phi_a
		    	-z^2\left(1-\frac{z^2}{z_h^2}\right)
		    	\partial_z\CL_y\partial_z\Phi_a\right]
		    	\leq\frac{1}{16G^2}\,.
		    	\label{eq:BTZ-adjacent-direct-norm}
		    \end{align}
		    Equality holds when the two unit gradients are antiparallel.  This occurs
		    precisely on the BTZ geodesic $F_y=0$.  With
		    \begin{align}
		    	n_{y,\mu}
		    	=\frac{\partial_\mu F_y}
		    	{\sqrt{g_{\rm BTZ}^{\alpha\beta}\partial_\alpha F_y
		    			\partial_\beta F_y}}\,,
		    	\label{eq:BTZ-adjacent-EWCS-normal}
		    \end{align}
		    direct substitution gives
		    \begin{align}
		    	v_{AB}^\mu\big|_{\Sigma_y}=\pm\frac{1}{4G}n_y^\mu\,,\qquad
		    	|v_{AB}|_{\Sigma_y}=\frac{1}{4G}\,.
		    	\label{eq:BTZ-adjacent-candidate-saturation}
		    \end{align}
		    The sign is fixed by choosing the orientation of the candidate cross
		    section; it has no effect on the absolute flux.
		    Thus every fixed split $P_y$ produces a valid candidate max flow for the
		    corresponding geometric purification.
		    
		    \paragraph{Integral curves and stream function:}
		    The integral curves solve
		    \begin{align}
		    	\frac{\d x(s)}{\d s}=v_{AB}^x(x(s),z(s);y)\,,\qquad
		    	\frac{\d z(s)}{\d s}=v_{AB}^z(x(s),z(s);y)\,.
		    \end{align}
		    Reading off a first integral from
		    \eqref{eq:BTZ-adjacent-integrated-components}, we find the stream function
		    \begin{align}
		    	\Psi_y(x,z)=\frac12\left[\CL_y(x,z)-\Phi_a(x,z)\right]\,,
		    	\label{eq:BTZ-adjacent-stream}
		    \end{align}
		    and the bit threads field may be written as
		    \begin{align}
		    	v_{AB}&=\frac{1}{4G}\star\dd\Psi_y\,,
		    	\label{eq:BTZ-adjacent-EoP-flow}
		    \end{align}
		    Hence $\d\Psi_y/\d s=0$, and the streamlines are the level sets
		    \begin{align}
		    	&\operatorname{arccosh}\left[
		    	\frac{z_h^2}{z z_y}\left(
		    	\cosh\left(\frac{x-y}{z_h}\right)
		    	-\sqrt{\left(1-\frac{z^2}{z_h^2}\right)\left(1-\frac{z_y^2}{z_h^2}\right)}
		    	\right)\right]
		    	\notag\\[-2mm]
		    	&\hspace{25mm}
		    	-\log\left[\frac{z_h}{z}\left(
		    	\cosh\left(\frac{x-a}{z_h}\right)
		    	-\sqrt{1-\frac{z^2}{z_h^2}}\right)\right]=2\Psi_0\,.
		    	\label{eq:BTZ-adjacent-level-sets}
		    \end{align}
		    The cancellation of $(-b,0)$ in \eqref{eq:BTZ-boundary-integrated-adjacent}--\eqref{eq:BTZ-RT-integrated-adjacent} is the planar-BTZ realization of the endpoint-reduction identity in appendix~\ref{app:distance-difference-potential}.  The candidate cross section has one ideal and one finite focus, so appendix~\ref{app:fermi-extended-flows} gives the same finite--ideal Fermi classification as in the Poincar\'e adjacent case.

		    The non-geodesic character can be established intrinsically, rather than
		    inferred from the Poincar\'e analysis.  The BTZ time slice is locally the
		    hyperbolic plane.  If $\vartheta_y$ is the angle between
		    $\nabla\CL_y$ and $\nabla\Phi_a$, the standard Hessian identities for a
		    bulk distance and a boundary Busemann function are
		    \begin{align}
		    	\nabla_\mu\nabla_\nu\CL_y
		    	&=\coth\CL_y\left(g_{\mu\nu}
		    	-\partial_\mu\CL_y\partial_\nu\CL_y\right)\,,\notag\\
		    	\nabla_\mu\nabla_\nu\Phi_a
		    	&=g_{\mu\nu}-\partial_\mu\Phi_a\partial_\nu\Phi_a\,.
		    \end{align}
		    Applying these identities to the level sets of \eqref{eq:BTZ-adjacent-stream}
		    gives their geodesic curvature\footnote{For a smooth curve $X^\mu(s)$ on the constant-time Riemannian bulk slice, parametrized by proper length $s$, let
		    	$t^\mu=dX^\mu/ds$ be its unit tangent and $n^\mu$ a chosen unit normal. Its signed geodesic curvature is
		    	\begin{align}
		    		k_{\mathrm g}
		    		=n_\mu t^\nu\nabla_\nu t^\mu ,
		    	\end{align}
		    	while the orientation-independent curvature is
		    	\begin{align}
		    		|k_{\mathrm g}|
		    		=\sqrt{
		    			g_{\mu\nu}
		    			\bigl(t^\rho\nabla_\rho t^\mu\bigr)
		    			\bigl(t^\sigma\nabla_\sigma t^\nu\bigr)
		    		} .
		    	\end{align}
		    	The sign of $k_{\mathrm g}$ depends on the choice of $n^\mu$, whereas
		    	$|k_{\mathrm g}|$ does not. In particular, the curve is geodesic precisely when
		    	$k_{\mathrm g}=0$.}, up to the orientation of the normal,
		    \begin{align}
		    	|k_{\mathrm g}|
		    	=\frac{\sqrt{2\left(1-\cos\vartheta_y\right)}}{4}
		    	\left(\coth\CL_y-1\right)\,.
		    	\label{eq:BTZ-adjacent-geodesic-curvature}
		    \end{align}
		    At every regular interior point $\CL_y$ is finite and
		    $\cos\vartheta_y\neq1$, so \eqref{eq:BTZ-adjacent-geodesic-curvature}
		    is non-zero.  The streamlines are therefore not BTZ geodesics.  They
		    become asymptotically geodesic only where $\CL_y\rightarrow\infty$.
		    This makes precise the distinction already encountered in the Poincar\'e
		    analysis: the elementary PEE threads are geodesics, whereas the integral
		    curves of their density-weighted sum need not be.
		    
		    \paragraph{Regular non-crossing foliation:}
		    It remains to exclude critical points of $\Psi_y$ in the open
		    entanglement wedge $\CW_{AB}^{\circ}$.  A critical point would require
		    \begin{align}
		    	\nabla\CL_y=\nabla\Phi_a\,.
		    \end{align}
		    The first vector is tangent to the geodesic from $P_y$ to the bulk point,
		    directed away from $P_y$; the second is tangent to the geodesic with ideal
		    endpoint $(a,0)$, directed away from that endpoint.  Equality is possible
		    only on the continuation of $\Sigma_y$ beyond $P_y$.  The closed
		    entanglement wedge is a geodesically convex BTZ half-plane bounded by
		    $\gamma_{AB}$.  The segment from $(a,0)$ to $P_y$ lies in this half-plane,
		    while its continuation beyond $P_y$ crosses $\gamma_{AB}$ and lies outside
		    it.  Hence $\nabla\Psi_y$ never vanishes in $\CW_{AB}^{\circ}$.
		    
		    The implicit-function theorem now shows that each
		    $\Psi_y(x,z)=\Psi_0$ is a smooth one-dimensional curve.  Level sets with
		    different values of $\Psi_0$ cannot intersect, and two distinct branches
		    of the same regular level set cannot cross.  Equivalently, uniqueness of
		    solutions to the first-order flow equations forbids a crossing.  The
		    streamlines therefore form a regular ordered non-crossing foliation.  This
		    ordering may be called nesting of the streamlines of this representative,
		    but it should not be confused with the stronger bit-thread nesting theorem,
		    which concerns simultaneous locking for a family of nested boundary
		    regions.
		    
\begin{figure}[t]
		    	\centering
		    	\includegraphics[width=0.75\linewidth]{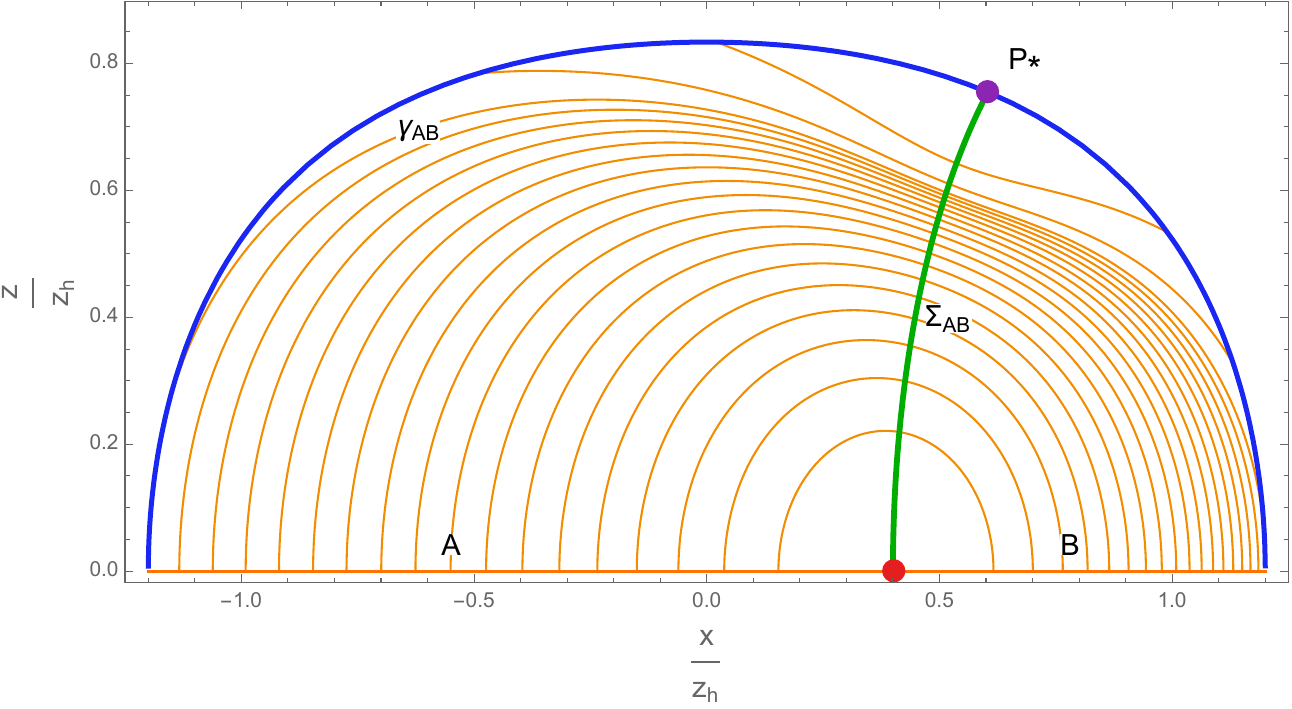}
		    	\caption{Optimized adjacent-interval minimal-purification flow in the planar BTZ black brane. The blue curve is the RT surface $\gamma_{AB}$ of the union, the orange point $P_\star$ is the minimizing split point, and the thick purple segment is the EWCS $\Sigma_{AB}$. The thin curves are representative level sets of the optimized distance-difference stream function. Norm saturation on $\Sigma_{AB}$ follows analytically from the endpoint-potential construction.}
		    	\label{fig:BTZ-adjacent-EWCS-flow}
		    \end{figure}

		    \paragraph{Flux optimization and the EWCS flow:}
		    Since \eqref{eq:BTZ-adjacent-candidate-saturation} holds, the flux through
		    the candidate EWCS equals its intrinsic BTZ length:
		    \begin{align}
		    	\Phi_y(\Sigma_y)
		    	&=\frac{1}{4G}\int_{a_\epsilon}^{y}
		    	\frac{\d x}{z(x;y)}
		    	\sqrt{1+\frac{\left[\partial_xz(x;y)\right]^2}
		    		{1-\frac{z(x;y)^2}{z_h^2}}}\notag\\
		    	&=\frac{1}{4G}\log\left[
		    	\frac{2z_h}{\epsilon}
		    	\frac{
		    		\cosh\left(\frac{a-y}{z_h}\right)
		    		-\sech\left(\frac{b}{z_h}\right)
		    		\cosh\left(\frac{y}{z_h}\right)}
		    	{\sqrt{1-\sech^2\left(\frac{b}{z_h}\right)
		    			\cosh^2\left(\frac{y}{z_h}\right)}}\right]+\CO(\epsilon^2)\,.
		    	\label{eq:BTZ-adjacent-candidate-flux}
		    \end{align}
		    Here $z(x;y)$ is the branch of
		    \eqref{eq:BTZ-adjacent-EWCS-profile} from $(a,\epsilon)$ to $P_y$, and
		    its regulated endpoint is
		    \begin{align}
		    	a_\epsilon
		    	=a-\frac{\epsilon^2}{2z_h}
		    	\frac{\sinh\left(\frac{a-y}{z_h}\right)}
		    	{\cosh\left(\frac{a-y}{z_h}\right)
		    		-\sech\left(\frac{b}{z_h}\right)
		    		\cosh\left(\frac{y}{z_h}\right)}+\CO(\epsilon^4)\,.
		    	\label{eq:BTZ-adjacent-cutoff}
		    \end{align}
		    Extremizing the finite part of the candidate flux gives the following unique solution in $-b<y<b$ is
		    \begin{align}
		    	y_*
		    	&=z_h\operatorname{arctanh}\left[
		    	\frac{\sinh\left(\frac{a}{z_h}\right)
		    		\sinh\left(\frac{b}{z_h}\right)
		    		\tanh\left(\frac{b}{z_h}\right)}
		    	{\cosh\left(\frac{a}{z_h}\right)
		    		\cosh\left(\frac{b}{z_h}\right)-1}\right]\,,\notag\\
		    	z_*&=z_h\sqrt{1-\sech^2\left(\frac{b}{z_h}\right)
		    		\cosh^2\left(\frac{y_*}{z_h}\right)}\,.
		    	\label{eq:BTZ-adjacent-optimal-point}
		    \end{align}
		    The EWCS bit-thread field is therefore not the whole family
		    \eqref{eq:BTZ-adjacent-integrated-components}, but the optimized member
		    \begin{align}
		    	v_{E_W}^\mu(x,z)
		    	&=\frac{z^2\sqrt{1-\frac{z^2}{z_h^2}}}{8G}\left(\partial_z\CL_{y_*}(x,z)-\partial_z\Phi_a(x,z),\partial_x\Phi_a(x,z)-\partial_x\CL_{y_*}(x,z)\right)
		    	\,.
		    	\label{eq:BTZ-adjacent-optimized-flow}
		    \end{align}
		    Its maximal flux is
		    \begin{align}
		    	\Phi_{E_W}
		    	=\frac{1}{4G}\log\left[
		    	\frac{4z_h}{\epsilon}
		    	\frac{\sinh\left(\frac{b-a}{2z_h}\right)
		    		\sinh\left(\frac{a+b}{2z_h}\right)}
		    	{\sinh\left(\frac{b}{z_h}\right)}\right]\,.
		    	\label{eq:BTZ-adjacent-EoP}
		    \end{align}
		    Thus the flux optimization selects both the physical EWCS and the EWCS
		    representative of the candidate PEE-superposed flows. The optimized adjacent-interval flow and its EWCS bottleneck are shown in figure~\ref{fig:BTZ-adjacent-EWCS-flow}.

		    \subsubsection{Symmetric disjoint intervals}
		    \label{subsec:disjoint-BTZ-EoP}
		    Finally, we consider the simple setup of symmetrically placed disjoint intervals,
		    $A=[-b,-a]$ and $B=[a,b]$, with $0<a<b$.  The thermal
		    cross-ratio is
		    \begin{align}
		    	\eta_\beta
		    	=\frac{\sinh^2\left(\frac{b-a}{2z_h}\right)}
		    	{\sinh^2\left(\frac{a+b}{2z_h}\right)}\,.
		    \end{align}
		    We restrict first to $\eta_\beta>1/2$, for which the dominant RT surface
		    is $\gamma_{AB}=\gamma_{[-b,b]}\cup\gamma_{[-a,a]}$ and the
		    entanglement wedge is connected.  A general split is specified by
		    \begin{align}
		    	P_1&=(y_1,z_1)\in\gamma_{[-a,a]}\,,&
		    	z_1&=z_h\sqrt{1-\sech^2\left(\frac{a}{z_h}\right)
		    		\cosh^2\left(\frac{y_1}{z_h}\right)}\,,\notag\\
		    	P_2&=(y_2,z_2)\in\gamma_{[-b,b]}\,,&
		    	z_2&=z_h\sqrt{1-\sech^2\left(\frac{b}{z_h}\right)
		    		\cosh^2\left(\frac{y_2}{z_h}\right)}\,.
		    	\label{eq:BTZ-disjoint-split-points}
		    \end{align}
		    The candidate cross section $\Sigma(P_1,P_2)$ is the unique BTZ geodesic
		    joining these two points.  The split assigns to $A$ the outer RT segment
		    from $P_2$ to $(-b,0)$ and the inner RT segment from $(-a,0)$ to $P_1$.
		    
		    \paragraph{Direct integration of the PEE-thread fields:}
		    The oriented source representation is
		    \begin{align}
		    	v_{AB}^\mu(x,z;y_1,y_2)
		    	=&\int_{-b}^{-a}\d x_0\,V^\mu_{x_0}(x,z)
		    	+\int_{y_2}^{-b}\d y\,V^\mu_{y,[-b,b]}(x,z)\notag\\
		    	&+\int_{-a}^{y_1}\d y\,V^\mu_{y,[-a,a]}(x,z)\,.
		    	\label{eq:BTZ-disjoint-candidate-source}
		    \end{align}
		    For $i=1,2$, the intrinsic BTZ distance from $(x,z)$ to $P_i$ is
		    \begin{align}
		    	\CL_i(x,z)=\operatorname{arccosh}\left[
		    	\frac{z_h^2}{zz_i}\left(
		    	\cosh\left(\frac{x-y_i}{z_h}\right)
		    	-\sqrt{\left(1-\frac{z^2}{z_h^2}\right)\left(1-\frac{z_i^2}{z_h^2}\right)}
		    	\right)\right]\,.
		    	\label{eq:BTZ-disjoint-bulk-distances}
		    \end{align}
		    Direct integration gives the candidate current as
		    \begin{align}
		    	v_{AB}^\mu(x,z;y_1,y_2)
		    	&=\frac{z^2\sqrt{1-\frac{z^2}{z_h^2}}}{8G}\left(\partial_z\CL_1(x,z)-\partial_z\CL_2(x,z),\partial_x\CL_2(x,z)-\partial_x\CL_1(x,z)\right)
		    	\,.
		    	\label{eq:BTZ-disjoint-integrated-components}
		    \end{align}
		    Directly differentiating \eqref{eq:BTZ-disjoint-integrated-components}
		    gives $\nabla_\mu v_{AB}^\mu=0$.  Its norm is
		    \begin{align}
		    	|v_{AB}|^2
		    	=\frac{1}{32G^2}\left\{1-z^2\partial_x\CL_1\partial_x\CL_2
		    	-z^2\left(1-\frac{z^2}{z_h^2}\right)
		    	\partial_z\CL_1\partial_z\CL_2\right\}
		    	\leq\frac{1}{16G^2}\,.
		    	\label{eq:BTZ-disjoint-direct-norm}
		    \end{align}
		    On the geodesic segment $\Sigma(P_1,P_2)$, the two unit distance
		    gradients are antiparallel.  Therefore the inequality is saturated and
		    \begin{align}
		    	v_{AB}^\mu\big|_{\Sigma(P_1,P_2)}
		    	=\frac{1}{4G}n_{\Sigma(P_1,P_2)}^\mu\,.
		    \end{align}
		    
		    \paragraph{Integral curves and stream function:}
		    A first integral of \eqref{eq:BTZ-disjoint-integrated-components} is
		    \begin{align}
		    	\Psi_{12}(x,z)=\frac12\left[\CL_1(x,z)-\CL_2(x,z)\right]\,,
		    	\label{eq:BTZ-disjoint-stream}
		    \end{align}
		    Thus the integral curves are the intrinsic BTZ Apollonius curves
		    \begin{align}
		    	\CL_1(x,z)-\CL_2(x,z)=2\Psi_0\,.
		    	\label{eq:BTZ-disjoint-level-sets}
		    \end{align}
		    This is the finite--finite endpoint potential of appendix~\ref{app:distance-difference-potential}; all internal endpoints of the boundary and RT-surface source integrals have cancelled before \eqref{eq:BTZ-disjoint-stream} is reached.  Locally, the same candidate bottleneck is described by the universal Fermi chart of appendix~\ref{app:fermi-extended-flows}; globally, the BTZ geodesic chamber still has to be fixed before the endpoint distances are chosen.

		    Let $\vartheta_{12}$ be the angle between $\nabla\CL_1$ and
		    $\nabla\CL_2$.  Applying the hyperbolic distance Hessian identity to
		    \eqref{eq:BTZ-disjoint-stream} gives the geodesic curvature of a regular
		    streamline,
		    \begin{align}
		    	|k_g|=\frac{\sqrt{2\left(1-\cos\vartheta_{12}\right)}}{4}
		    	\left|\coth\CL_1-\coth\CL_2\right|\,.
		    	\label{eq:BTZ-disjoint-geodesic-curvature}
		    \end{align}
		    The special level $\Psi_0=0$ has $\CL_1=\CL_2$ and is the geodesic
		    perpendicular bisector of $P_1P_2$.  For $\Psi_0\neq0$ the right-hand
		    side of \eqref{eq:BTZ-disjoint-geodesic-curvature} is non-zero, so the
		    remaining integral curves are not BTZ geodesics.  Again, geodesicity of
		    the elementary PEE threads does not imply geodesicity of their integrated
		    current.
		    
		    \paragraph{Regular non-crossing foliation:}
		    The only singularities of $\Psi_{12}$ are $P_1$ and $P_2$, which lie on
		    the boundary of the open entanglement wedge.  A critical point would obey
		    $\nabla\CL_1=\nabla\CL_2$.  By uniqueness of BTZ geodesics, this equality
		    requires the bulk point to lie on the complete geodesic through $P_1$ and
		    $P_2$.  It cannot lie between the foci, where the two gradients are
		    antiparallel, and can therefore lie only on one of the two exterior rays.
		    The connected entanglement wedge is the intersection of the BTZ geodesic
		    half-planes bounded by $\gamma_{[-b,b]}$ and $\gamma_{[-a,a]}$, and is
		    geodesically convex.  Its intersection with the complete geodesic
		    supporting $\Sigma(P_1,P_2)$ is exactly the segment $P_1P_2$; both exterior
		    rays lie outside the wedge.  Hence $\nabla\Psi_{12}\neq0$ in the open wedge.
		    
		    It follows that the level sets of $\Psi_{12}$ are smooth and mutually
		    disjoint.  They form an ordered non-crossing foliation, and uniqueness of
		    the first-order flow equations gives the equivalent local proof that two
		    integral curves cannot cross.  As in the adjacent case, this is streamline
		    nesting for a single flow representative, not a claim of simultaneous
		    locking for every nested boundary subregion.
		    
\begin{figure}[t]
		    	\centering
		    	\includegraphics[width=0.72\linewidth]{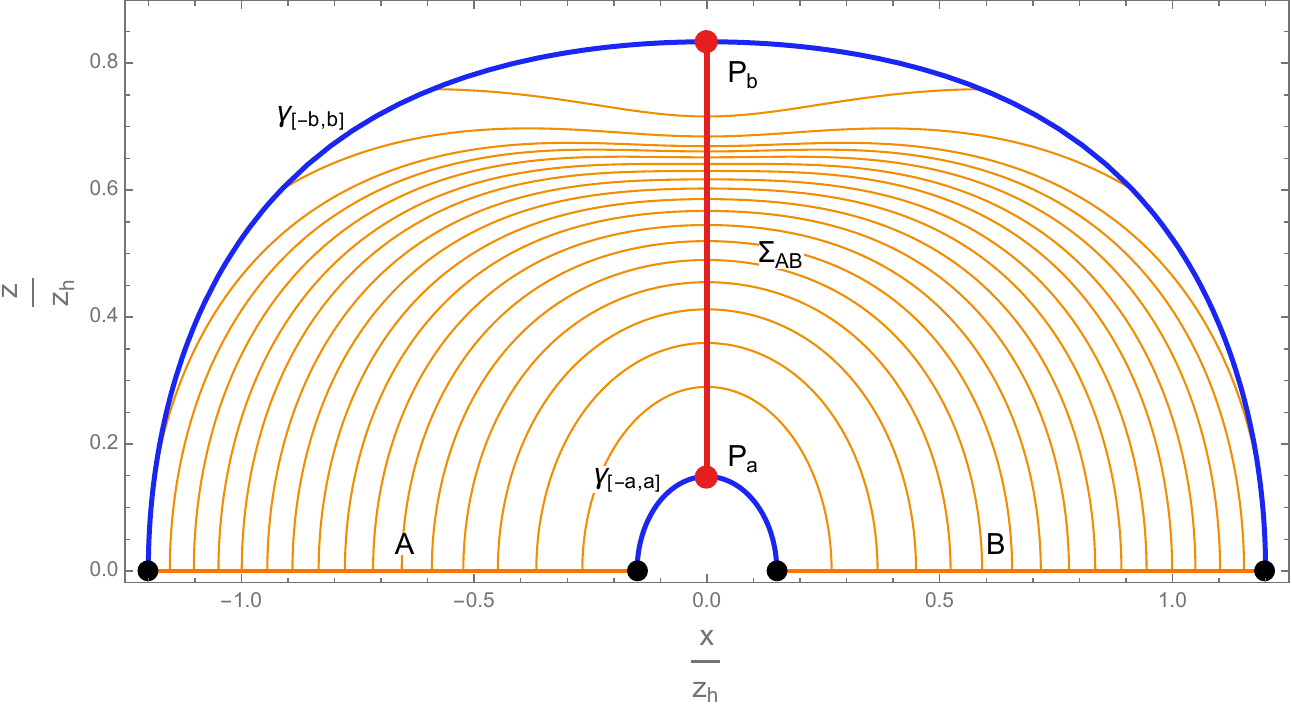}
		    	\caption{Optimized symmetric disjoint-interval minimal-purification flow in the connected planar-BTZ entanglement wedge. The outer and inner RT components are shown in blue and orange, the optimized finite foci $P_b$ and $P_a$ lie at $x=0$, and the thick green segment is their common perpendicular $\Sigma_{AB}$. The thin curves are representative level sets of the finite--finite distance-difference potential.}
		    	\label{fig:BTZ-disjoint-EWCS-flow}
		    \end{figure}

		    \paragraph{Flux optimization and the EWCS flow:}
		    For arbitrary $P_1,P_2$, the candidate flux is
		    \begin{align}
		    	\Phi(P_1,P_2)
		    	=\frac{1}{4G}\operatorname{arccosh}\left[
		    	\frac{z_h^2}{z_1z_2}\left(
		    	\cosh\left(\frac{y_1-y_2}{z_h}\right)
		    	-\sqrt{\left(1-\frac{z_1^2}{z_h^2}\right)\left(1-\frac{z_2^2}{z_h^2}\right)}
		    	\right)\right]\,.
		    	\label{eq:BTZ-disjoint-candidate-flux}
		    \end{align}
		    Here $z_1,z_2$ retain their full $y_1,y_2$ dependence in
		    \eqref{eq:BTZ-disjoint-split-points}.  Because $\operatorname{arccosh}$ is
		    monotone, differentiating its argument gives
		    \begin{align}
		    	\partial_{y_1}\Phi(P_1,P_2)=0\,,\qquad
		    	\partial_{y_2}\Phi(P_1,P_2)=0
		    	\qquad\Longrightarrow\qquad y_1=y_2=0\,.
		    	\label{eq:BTZ-disjoint-stationarity}
		    \end{align}
		    Geometrically these equations impose orthogonality to both RT components.
		    The two disjoint RT geodesics have a unique common perpendicular in the
		    negatively curved BTZ slice, and reflection symmetry $x\mapsto-x$ forces it
		    to lie on $x=0$; hence the stationary point is the unique global minimum.
		    The optimized endpoints are
		    \begin{align}
		    	P_a=(0,z_a)\,,\qquad z_a=z_h\tanh\left(\frac{a}{z_h}\right)\,,
		    	\qquad
		    	P_b=(0,z_b)\,,\qquad z_b=z_h\tanh\left(\frac{b}{z_h}\right)\,.
		    \end{align}
		    Thus the physical EWCS is the vertical BTZ geodesic $x=0$ between the
		    depths $z_a$ and $z_b$.  Substituting these minimizing endpoints into the
		    integrated candidate current gives the EWCS bit-thread field
		    \begin{align}
		    	v_{E_W}^\mu(x,z)
		    	&=\frac{z^2\sqrt{1-\frac{z^2}{z_h^2}}}{8G}\left(\partial_z\CL_a(x,z)-\partial_z\CL_b(x,z),\partial_x\CL_b(x,z)-\partial_x\CL_a(x,z)\right)
		    	\,,
		    	\label{eq:BTZ-disjoint-optimized-flow}
		    \end{align}
		    where $\CL_{a,b}$ are obtained from
		    \eqref{eq:BTZ-disjoint-bulk-distances} by setting
		    $(y_1,z_1)=(0,z_a)$ and $(y_2,z_2)=(0,z_b)$.  The optimized flux is
		    \begin{align}
		    	\Phi_{E_W}
		    	&=\frac{1}{4G}\int_{z_a}^{z_b}
		    	\frac{\d z}{z\sqrt{1-\frac{z^2}{z_h^2}}}=\frac{1}{4G}\log\left[
		    	\frac{\tanh\left(\frac{b}{2z_h}\right)}
		    	{\tanh\left(\frac{a}{2z_h}\right)}\right]\,.
		    	\label{eq:BTZ-disjoint-EoP}
		    \end{align}
		    This is the symmetric planar-BTZ EWCS \cite{Takayanagi:2017knl}. The corresponding optimized symmetric disjoint-interval flow is shown in figure~\ref{fig:BTZ-disjoint-EWCS-flow}.

		    When
		    $\eta_\beta\leq1/2$, the dominant entanglement wedge is disconnected;
		    there is no cross section joining the two components and the leading
		    classical EWCS vanishes.  The candidate connected flow above
		    may still be evaluated as a subdominant saddle, but it is not the physical
		    EWCS flow in that phase.
		    
		   \section{Summary and discussion}
		   \label{sec:summary}
   
		   \subsection*{Summary}
		   \paragraph{PEE structure on an extended state-supporting surface.}
		   In \cref{sec:two-point-pee} we organized the two-point data into
		   boundary--boundary, boundary--brane, brane--brane, boundary--horizon and
		   brane--horizon sectors.  The construction is fixed geometrically by
		   mixed endpoint derivatives of regulated geodesic lengths.  A necessary
		   conceptual distinction is that the positive density $\CI$ appearing in
		   an entropy decomposition and the oriented kernel
		   $\widehat{\CI}$ appearing in a source current are not the same
		   mathematical object.  Reversing the orientation of a boundary, brane or
		   horizon component changes the sign of the latter but cannot turn a
		   positive PEE into a negative entanglement measure.  With this
		   distinction, the boundary, brane and horizon contributions obey the
		   expected normalization identities in every connected chamber examined
		   in this paper.  The vanishing of the smooth horizon--horizon mixed
		   kernel at separated points is also consistent with the fact that the
		   equal-time horizon distance is locally linear; possible contact data at
		   coincident points are not resolved by the smooth kernel used here.

		   \paragraph{From PEE threads to extended bit-thread flows.}
		   In \cref{sec:pee-threads} we constructed source-labelled PEE-thread
		   fields and integrated them over the part of the extended surface
		   assigned to a chosen subsystem.  For the Poincar\'e AdS$_3$/BCFT$_2$
		   configurations, the boundary- and brane-sourced contributions combine
		   into an explicitly divergenceless current obeying
		   $|v|\leq 1/(4G)$ and saturating this bound on the brane-ending RT
		   geodesic.  For the planar BTZ black brane, the brane- and horizon-sourced fields are derived intrinsically from BTZ geodesics and their first integrals.  The boundary-sourced field is reproduced here by pushing forward the Poincar\'e source field under an explicit spatial isometry, while an independent intrinsic derivation is given in \cite{BasuWenChandra}.  These constructions lead to the same local current and provide complementary checks of its normalization.  The resulting currents have the correct flux through each component of the oriented exterior boundary and reproduce the RT entropy in the chambers under consideration.

		   Our explicit max-flow formulae in AdS/BCFT apply to the adjacent brane-ending and connected chambers.  For disconnected BCFT phases, the intersection-weighted PEE normalization is under control, but an explicit PEE-generated max-flow representative is left open.

		   We found that the bit threads constructed from the PEE threads are generically non-geodesic.  This explicitly separates the microscopic carriers used to represent the two-point PEE from the integral curves of the macroscopic vector field. The latter nevertheless form a regular non-crossing streamline foliation wherever the current is smooth and non-vanishing.  Thus a
		   geodesic microscopic network can generate a non-geodesic max flow
		   without violating any of the bit-thread constraints
		   \cite{Freedman:2016zud,Headrick:2020gyq,Headrick:2022nbe}.

		   \paragraph{Endpoint reduction and calibration.}
		   The explicit source integrals repeatedly collapse to a current of the
		   form
		   \begin{align}
		   	v=\frac{1}{4G}\star\dd\Psi\,,
		   	\qquad
		   	\Psi=\frac12\left(\CL_{p_-}-\CL_{p_+}\right),
		   	\label{eq:summary-distance-difference}
		   \end{align}
		   where $p_\pm$ are the endpoints of the relevant RT surface or candidate
		   EWCS and $\CL_{p_\pm}$ are suitably regulated distance functions.  As
		   explained in \cref{app:distance-difference-potential}, this form makes
		   divergencelessness manifest and the triangle inequality gives the local
		   norm bound.  On the geodesic joining $p_-$ and $p_+$, the gradients of
		   the two endpoint functions are oppositely directed, so the bound is
		   saturated and the current calibrates the bottleneck.  In the examples
		   computed here, this endpoint reduction is not merely an alternative
		   ansatz: it agrees with the direct integral of the corresponding
		   source-labelled PEE-thread fields.

		   Appendix~\ref{app:fermi-extended-flows} supplies a complementary local description of the same bottlenecks.  In Fermi normal coordinates $(\eta,\lambda)$ around any connected geodesic bottleneck, the normal congruence carries the universal current $v_{\rm geo}=(4G)^{-1}\operatorname{sech}\lambda\,\partial_\lambda$.  For an ordinary vacuum interval, whose two foci are ideal, this normal current coincides with the PEE-selected distance-difference flow.  For the brane-ending AdS/BCFT surface and the adjacent minimal purification one focus is finite, while for the disjoint minimal purification both are finite; the PEE-selected current then generically develops a tangential Fermi component away from the bottleneck.  This provides a geometric explanation of the non-geodesic streamlines and makes their representative dependence explicit: the entropy fixes the bottleneck flux, whereas the PEE data additionally select an endpoint-calibrated continuation of that flux.

		   It is useful to separate the role of the present analysis from that of \cite{BasuWenChandra}.  In \cite{BasuWenChandra}, we study the same endpoint-resolved logic for the planar BTZ black brane and then extend it to finite-cutoff AdS$_3$, where the cutoff makes the nonuniqueness of max-flow representatives particularly explicit: the PEE-selected and normal-geodesic flows saturate the same RT bottleneck while differing in endpoint pairing and off-bottleneck calibration.  In the present work the planar-BTZ construction instead serves as one component of a broader extended-boundary framework involving EOW branes, horizons and RT-surface purifiers.  The two analyses therefore provide complementary tests of the same distance-difference and Fermi-coordinate structures.

		   \paragraph{Minimal geometric purification and the EWCS.}
		   In \cref{sec:eop}, the RT surface bounding an entanglement wedge was
		   treated as an auxiliary state-supporting component.  A choice of split
		   of this surface defines a member of the surface/state class of geometric
		   purifications, and the PEE-thread flow is sourced from the extended
		   purified subsystem $A^\sharp=A\cup\CE_A$.  For adjacent and disjoint
		   intervals in Poincar\'e AdS$_3$, for adjacent intervals in the planar
		   BTZ black brane and for the symmetric connected disjoint configuration
		   in planar BTZ, the source integral reduces to
		   \eqref{eq:summary-distance-difference}.  The flux equals the length of
		   the candidate cross section, and extremizing over the purifying split
		   selects the EWCS.  These examples provide a concrete flow realization
		   of the holographic conjecture $E_P=E_W$
		   \cite{Takayanagi:2017knl,Nguyen:2017yqw,Du:2019emy} within the
		   specified class of geometric purifications.

		   The examples studied above involve asymptotic boundaries, EOW branes, horizons and RT surfaces used as geometric purifiers.  We constructed the corresponding endpoint densities, PEE-thread networks and bit-thread flows and used them to test a common geometric picture.  Kinematic-space data provide the fine-grained geodesic network that enters Crofton reconstruction; under the surface/state interpretation, the same endpoint measure acquires an entanglement interpretation on the boundary of a gravitational subregion; and its oriented source superposition produces a norm-bounded macroscopic current.  The resulting constructions provide nontrivial tests of the subregion-reconstruction framework of \cite{Basu:2026hbg} and concrete realizations of the surface/state proposal in the chambers treated explicitly here.
		   
		   \subsection*{Future directions}
		   So far our thermal analysis has concerned the planar BTZ black brane, whose spatial direction is noncompact; the same noncompact setup and its finite-cutoff extension are also analyzed in \cite{BasuWenChandra}.  A genuinely new problem arises for the compact BTZ black hole, where the boundary spatial direction is periodically identified.  On the universal cover, a fixed pair of quotient endpoints is represented by infinitely many image pairs and hence by geodesics with different winding numbers.  The relevant image can change from one homology chamber to another, and in the disconnected RT phase the flow must also route a globally fixed amount of flux through the horizon.  Consequently a naive periodic sum of the planar PEE-thread fields need not define a single-valued, globally maximal flow on the quotient.  A promising strategy is to construct the distance-difference or Busemann flow chamber by chamber on the universal cover and then impose equivariance under the BTZ identification; if a globally single-valued scalar potential is too restrictive, the natural object may instead be an equivariant closed one-form.  The remaining horizon-routing constraint may also be more naturally formulated as a locking multiflow problem.  Developing this quotient-compatible construction, in particular for the horizon-winding disconnected phase, is an interesting future direction.
		   
		   A controlled higher-dimensional extension is most immediate when the relevant RT surface is totally geodesic. For a generic non-totally-geodesic RT surface, however, the subregion kinematic space contains PEE chords connecting points on the RT surface that need not touch the external boundary $\partial\mathbf a$; see \cite{Basu:2026hbg}. Understanding the physical meaning and source-current description of these intrinsically internal chords requires additional input.

		   Promoting our analysis to covariant, or even time-dependent, configurations is highly non-trivial; see \cite{Headrick:2022nbe} for a discussion of covariant bit threads. One can also incorporate quantum corrections in the bulk by allowing PEE threads to terminate at bulk sites, thereby describing a bulk PEE structure; a similar discussion for bit threads has appeared in \cite{Rolph:2021hgz,Agon:2021tia,Headrick:2025awv}.

	\section*{Acknowledgements}
	The authors are supported by NSFC Grant No. 12447108 and the Shing-Tung Yau Center of Southeast University. Q.~Wen thanks the Institute of Theoretical Physics, Chinese Academy of Sciences, for hospitality during the preparation of this work.

		   \appendix
		    \section{Distance-difference potentials for extended holographic flows}
		   \label{app:distance-difference-potential}
		  		   
		   In the main text, the bit-thread currents relevant to AdS/BCFT and to the
		   minimal-purification construction were obtained by explicitly superposing the
		   appropriate boundary-, brane-, horizon- and RT-surface-sourced PEE-thread
		   currents.  In each example, the resulting vector field admits a stream
		   function built from the two endpoints of the corresponding RT surface or
		   entanglement wedge cross section.  The purpose of this appendix is to explain
		   the general geometric origin of this simplification.  The same endpoint-reduction
		   mechanism is used in \cite{BasuWenChandra} for planar BTZ and for finite-cutoff
		   AdS$_3$, where it is also compared directly with an independent normal-geodesic
		   max-flow representative.
		   We formulate a coordinate-independent distance-difference
		   construction which simultaneously explains
		   
		   \begin{enumerate}
		   	\item why the final current depends only on the endpoints of the relevant
		   	bottleneck, even though it is obtained by integrating PEE threads over
		   	several components of an extended state-supporting surface;
		   	
		   	\item why the current is automatically divergenceless and obeys the bit
		   	thread norm bound;
		   	
		   	\item why the norm bound is saturated precisely on the RT surface or
		   	entanglement wedge cross section;
		   	
		   	\item why the integral curves of the coarse-grained current are generally
		   	non-geodesic, despite the geodesic nature of the elementary PEE threads;
		   	
		   	\item and why the construction of a valid flow must be distinguished from
		   	the separate extremization over a choice of minimal purification.
		   \end{enumerate}
		   
		   Throughout this appendix, $\mathcal M$ denotes an oriented two-dimensional
		   Riemannian bulk slice with metric $g_{\mu\nu}$ and volume form
		   $\varepsilon_{\mu\nu}$.  We write $\star$ for the Hodge dual on
		   $\mathcal M$.  The discussion applies directly to the constant-time slices
		   considered in the main text, after restricting to a geodesic chamber in which
		   the relevant geodesics are unique.
		   
		   \subsection*{Oriented endpoint functions}
		   \label{app:oriented-endpoint-functions}
		   
		   Let $\Sigma_\star$ be a geodesic segment which is to serve as the bottleneck
		   of a candidate bit-thread flow.  Its endpoints will be denoted by
		   $p_-$ and $p_+$, and may lie on any component of the extended holographic
		   state-supporting surface.  In particular, an endpoint may lie
		   
		   \begin{itemize}
		   	\item at the asymptotic conformal boundary;
		   	\item on an end-of-the-world brane;
		   	\item on a horizon;
		   	\item or on an RT surface promoted to a purifying boundary.
		   \end{itemize}
		   
		   If $p$ is a finite point in the metric completion of $\mathcal M$, define
		   \begin{equation}
		   	r_p(X)=\CL(X,p),
		   	\label{eq:finite-endpoint-distance}
		   \end{equation}
		   where $\CL$ is the geodesic distance on $\mathcal M$.  Away from the focus
		   $p$ and its cut locus, the distance function satisfies the eikonal equation
		   \begin{equation}
		   	g^{\mu\nu}\partial_\mu r_p\,\partial_\nu r_p=1.
		   	\label{eq:finite-distance-eikonal}
		   \end{equation}
		   For an ideal endpoint $\xi$ on the asymptotic boundary, the ordinary
		   distance diverges.  The appropriate replacement is a Busemann function.
		   Choose a unit-speed geodesic ray $\gamma_\xi(s)$ approaching $\xi$ and set
		   \begin{equation}
		   	\beta_\xi(X)
		   	=
		   	\lim_{s\rightarrow\infty}
		   	\left[\CL\bigl(X,\gamma_\xi(s)\bigr)-s\right].
		   	\label{eq:Busemann-definition}
		   \end{equation}
		   The result is defined up to an additive constant, corresponding to the
		   choice of regulating surface near the asymptotic boundary.  This ambiguity
		   has no effect on the current, which depends only on
		   $\mathrm{d}\beta_\xi$.  The Busemann function also obeys an eikonal equation,
		   \begin{equation}
		   	|\nabla\beta_\xi|^2=1.
		   	\label{eq:Busemann-eikonal}
		   \end{equation}
		   
		   It is useful to treat both cases uniformly.  We associate to every endpoint
		   $p$ an eikonal function $\rho_p(X)$, with
		   \begin{equation}
		   	\rho_p(X)
		   	=
		   	\begin{cases}
		   		\CL(X,p), & p\ \text{finite},\\[2mm]
		   		\beta_p(X), & p\ \text{an ideal boundary point}.
		   	\end{cases}
		   	\qquad
		   	|\nabla\rho_p|=1.
		   	\label{eq:generalized-endpoint-distance}
		   \end{equation}
		   The additive constant in an ideal-endpoint function will always be
		   irrelevant below.
		   
		   Orient $\Sigma_\star$ from $p_-$ to $p_+$ and denote its unit tangent by
		   $t^\mu$.  On the interior of the segment, the unsigned endpoint functions
		   obey
		   \begin{equation}
		   	\left.\nabla_\mu\rho_{p_-}\right|_{\Sigma_\star}
		   	=t_\mu,
		   	\qquad
		   	\left.\nabla_\mu\rho_{p_+}\right|_{\Sigma_\star}
		   	=-t_\mu.
		   	\label{eq:endpoint-gradients-on-segment}
		   \end{equation}
		   The same statement holds when one or both foci are ideal, with the
		   appropriate limiting interpretation.
		   
		   We define the normalized distance-difference potential by
		   \begin{equation}
		   	\Psi_{p_-p_+}(X)
		   	=
		   	\frac{1}{2}
		   	\left[
		   	\rho_{p_-}(X)-\rho_{p_+}(X)
		   	\right].
		   	\label{eq:normalized-distance-difference}
		   \end{equation}
		   An overall sign in \eqref{eq:normalized-distance-difference} merely reverses
		   the orientation of the flow.  The relative sign, however, is fixed by the
		   requirement
		   \begin{equation}
		   	\left.\mathrm{d}\Psi_{p_-p_+}\right|_{\Sigma_\star}
		   	=t.
		   	\label{eq:potential-sign-diagnostic}
		   \end{equation}
		   Equation \eqref{eq:potential-sign-diagnostic} provides a convenient
		   coordinate-independent diagnostic for the relative signs of distance and
		   Busemann terms.  In particular, when different Busemann conventions are
		   used, the signs should always be adjusted so that the potential increases
		   with unit rate along the oriented bottleneck.
		   The main text occasionally uses stream functions which differ by a factor
		   of two.  In this appendix, $\Psi$ will always denote the normalized
		   dimensionless potential for which the current is
		   \begin{equation}
		   	v_{p_-p_+}
		   	=
		   	\frac{1}{4G}\star\mathrm{d}\Psi_{p_-p_+}.
		   	\label{eq:distance-difference-flow}
		   \end{equation}
		   
		   \subsection*{Divergencelessness, norm bound and saturation}
		   \label{app:distance-potential-max-flow}
		   
		   The divergenceless property follows without any coordinate calculation.
		   Using $\star^2=-1$ on one-forms in two Euclidean dimensions, one finds
		   \begin{equation}
		   	\nabla_\mu v^\mu_{p_-p_+}
		   	=
		   	-\star\mathrm{d}\star v_{p_-p_+}
		   	=
		   	\frac{1}{4G}\star\mathrm{d}^{\,2}\Psi_{p_-p_+}
		   	=0.
		   	\label{eq:distance-potential-divergence}
		   \end{equation}
		   Equivalently, the flux one-form
		   $\iota_v\varepsilon$ is proportional to the exact form
		   $\mathrm{d}\Psi$ and is therefore closed.
		   The norm bound follows solely from the two eikonal equations.  From
		   \eqref{eq:normalized-distance-difference},
		   \begin{align}
		   	(4G)^2|v_{p_-p_+}|^2
		   	&=
		   	|\mathrm{d}\Psi_{p_-p_+}|^2
		   	=
		   	\frac{1}{4}
		   	\left|
		   	\mathrm{d}\rho_{p_-}-\mathrm{d}\rho_{p_+}
		   	\right|^2
		   	=
		   	\frac{1}{2}
		   	\left(
		   	1-
		   	\nabla\rho_{p_-}\cdot\nabla\rho_{p_+}
		   	\right)
		   	\leq 1.
		   	\label{eq:universal-distance-norm-bound}
		   \end{align}
		   On $\Sigma_\star$, the two unsigned distance gradients are oppositely
		   directed, as in \eqref{eq:endpoint-gradients-on-segment}.  Therefore,
		   \begin{equation}
		   	\left.
		   	\mathrm{d}\Psi_{p_-p_+}
		   	\right|_{\Sigma_\star}
		   	=t,
		   	\qquad
		   	\left.
		   	v_{p_-p_+}
		   	\right|_{\Sigma_\star}
		   	=
		   	\frac{1}{4G}\star t.
		   	\label{eq:distance-potential-saturation}
		   \end{equation}
		   If $n^\mu$ is the unit normal obtained by rotating $t^\mu$ according to
		   the chosen orientation, this gives
		   \begin{equation}
		   	\left.v^\mu_{p_-p_+}\right|_{\Sigma_\star}
		   	=
		   	\frac{1}{4G}n^\mu,
		   	\qquad
		   	\left|v_{p_-p_+}\right|_{\Sigma_\star}
		   	=
		   	\frac{1}{4G}.
		   	\label{eq:distance-potential-normality}
		   \end{equation}
		   Hence the current is normal to the candidate bottleneck and saturates the
		   norm bound there.
		   Its flux through $\Sigma_\star$ is consequently
		   \begin{equation}
		   	\int_{\Sigma_\star}v_{p_-p_+}\cdot n\,
		   	\mathrm{d}\Sigma
		   	=
		   	\frac{\operatorname{Length}(\Sigma_\star)}{4G}.
		   	\label{eq:distance-potential-flux}
		   \end{equation}
		   When $\Sigma_\star$ is the RT surface, this reproduces the holographic
		   entropy.  When it is a candidate entanglement wedge cross section for a fixed geometric purification, it reproduces the purified entropy $S(A^\sharp)$ associated with that split.  Only after minimizing over the allowed split does this become the geometric quantity $E_{\rm geom}=E_W$; identifying the latter with the full quantum-information-theoretic $E_P$ is the separate holographic conjecture.
		   
		   The same argument may be phrased in the language of calibrations.  The
		   one-form
		   \begin{equation}
		   	\omega=\mathrm{d}\Psi_{p_-p_+}
		   	\label{eq:distance-calibration}
		   \end{equation}
		   is closed, has pointwise norm $|\omega|\leq1$, and restricts to the
		   unit arclength form on $\Sigma_\star$. Therefore, for any curve
		   $\widetilde\Sigma$ with the same regulated endpoints as
		   $\Sigma_\star$,
		   \begin{equation}
		   	\operatorname{Length}(\widetilde\Sigma)
		   	\geq
		   	\int_{\widetilde\Sigma}\omega
		   	=
		   	\int_{\Sigma_\star}\omega
		   	=
		   	\operatorname{Length}(\Sigma_\star).
		   	\label{eq:distance-calibration-minimality}
		   \end{equation}
		   Thus the distance-difference potential simultaneously supplies a max flow
		   and a calibration of the fixed-endpoint geodesic. If an endpoint is
		   asymptotic, the argument is first applied on the regulated manifold
		   $\mathcal M_\epsilon$, and the universal endpoint counterterms are
		   subtracted before taking $\epsilon\to0$.

		   A free-boundary RT or EWCS problem requires one additional variational
		   step. Equation~\eqref{eq:distance-calibration-minimality} alone does not
		   compare curves whose endpoint is allowed to slide on an EOW brane,
		   horizon, or purifying RT surface. One must first determine
		   \begin{align}
		   	q_\star=\underset{q\in\mathcal B}{\operatorname{arg\,min}}\,
		   	\CL(p,q),
		   	\label{eq:free-boundary-endpoint-minimization}
		   \end{align}
		   where $\mathcal B$ is the allowed endpoint component. The first variation
		   gives
		   \begin{align}
		   	t_\mu(q_\star)\,u^\mu_{\mathcal B}(q_\star)=0,
		   	\label{eq:free-boundary-orthogonality}
		   \end{align}
		   with $u^\mu_{\mathcal B}$ tangent to $\mathcal B$. Only after this
		   orthogonality condition has fixed $q_\star$ do we insert it as a focus in
		   the distance-difference potential and apply the fixed-endpoint
		   calibration. Equivalently, a genuine relative calibration with sliding
		   endpoints must include the appropriate boundary pullback condition or
		   the endpoint term in the first variation. The main text uses the explicit
		   two-step procedure: the brane endpoint of an AdS/BCFT RT surface is fixed
		   by extremality, and the candidate EWCS endpoints are optimized before the
		   physical EWCS flow is selected.
		   
		   \subsection*{Endpoint reduction of PEE-thread superpositions}
		   \label{app:endpoint-reduction}
		   
		   We next explain why the explicit PEE-thread integrals in the main text collapse to expressions involving only two geometric foci.  The distance-difference calibration in the preceding subsections only uses eikonal endpoint functions and is correspondingly more general.  By contrast, the identification below of an \emph{elementary PEE source current} with an endpoint derivative of a distance function is asserted only for the locally hyperbolic, locally AdS$_3$ slices used in this paper, where it follows from the explicit Poincar\'e/BTZ kernels and isometry covariance.  Extending this source-level identity to a generic two-dimensional Riemannian manifold would require a separate Jacobi-field analysis and is not assumed here.

		   Let $\mathcal C$ be an oriented component of an extended state-supporting
		   surface, parametrized by $s$, and let $p(s)\in\mathcal C$.  The
		   oriented endpoint kernel and the flux one-form of the elementary current are related by
		   \begin{align}
		   	\widehat{\CI}(s,q)\,\mathrm ds\wedge\mathrm dq
		   	&=\frac{1}{8G}\partial_s\partial_q\CL\bigl(p(s),q\bigr)\,
		   	\mathrm ds\wedge\mathrm dq\,,\notag\\
		   	\iota_{V_s}\varepsilon\,\mathrm ds
		   	&=-\frac{1}{8G}\mathrm d_X
		   	\left[\partial_s\rho_{p(s)}(X)\right]\mathrm ds\,.
		   \end{align}
		   The second line is the differential flux-matching relation. If a
		   transverse separator is parametrized by $q(\lambda)$, its pullback is
		   \begin{align}
		   	\left.\iota_{V_s}\varepsilon\right|_{q(\lambda)}
		   	=-\frac{1}{8G}
		   	\partial_s\partial_\lambda
		   	\CL\bigl(p(s),q(\lambda)\bigr)\,\mathrm d\lambda.
		   	\label{eq:endpoint-kernel-pullback}
		   \end{align}
		   This is precisely the oriented mixed endpoint kernel, including the
		   separator Jacobian and the orientation sign. Since
		   $\iota_{V_s}\varepsilon$ is exact, the endpoint-resolved PEE-thread
		   current is represented by the elementary stream function
		   \begin{equation}
		   	\psi_s(X)
		   	=
		   	-\frac{1}{8G}\del_s
		   	\rho_{p(s)}(X),
		   	\label{eq:elementary-endpoint-stream-function}
		   \end{equation}
		   where the sign follows the chosen orientation of $\mathcal C$.  The
		   corresponding elementary current is
		   \begin{equation}
		   	V_s=\star\mathrm{d}\psi_s.
		   	\label{eq:elementary-endpoint-current}
		   \end{equation}
		   This is the stream-function form of the endpoint-resolved PEE currents
		   constructed explicitly in the main text.
		   Integrating over an oriented segment
		   $\mathcal C_{12}=\{p(s)\,|\,s_1\leq s\leq s_2\}$ gives
		   \begin{align}
		   	\int_{s_1}^{s_2}\psi_s(X)\,\mathrm{d}s
		   	&=
		   	\frac{1}{8G}
		   	\left[
		   	\rho_{p(s_1)}(X)-\rho_{p(s_2)}(X)
		   	\right].
		   	\label{eq:endpoint-telescoping-single}
		   \end{align}
		   Thus the integration over a continuous family of PEE-thread sources is a
		   boundary term in the source coordinate.
		   More generally, let the purified subsystem be represented by an oriented
		   relative one-chain
		   \begin{equation}
		   	\mathcal R=\sum_{\alpha}\mathcal C_\alpha
		   \end{equation}
		   whose components may lie on the asymptotic boundary, an EOW brane, a
		   horizon or an RT surface.  Summing
		   \eqref{eq:endpoint-telescoping-single} over all components yields
		   \begin{equation}
		   	\psi_{\mathcal R}(X)
		   	=
		   	\frac{1}{8G}
		   	\sum_{p\in\partial\mathcal R}
		   	\sigma_p\,\rho_p(X),
		   	\qquad
		   	\sigma_p=\pm1.
		   	\label{eq:relative-chain-endpoint-formula}
		   \end{equation}
		   Every auxiliary endpoint shared by two oriented components occurs twice
		   with opposite signs and therefore cancels.  Only the relative boundary of
		   the complete extended subsystem survives.
		   When
		   \begin{equation}
		   	\partial\mathcal R=p_+-p_-,
		   \end{equation}
		   equation \eqref{eq:relative-chain-endpoint-formula} reduces to
		   \begin{equation}
		   	\psi_{\mathcal R}
		   	=
		   	\frac{1}{8G}
		   	\left(
		   	\rho_{p_-}-\rho_{p_+}
		   	\right)
		   	=
		   	\frac{1}{4G}\Psi_{p_-p_+}.
		   	\label{eq:relative-chain-distance-difference}
		   \end{equation}
		   Consequently,
		   \begin{equation}
		   	v_{\mathcal R}
		   	=
		   	\star\mathrm{d}\psi_{\mathcal R}
		   	=
		   	\frac{1}{4G}
		   	\star\mathrm{d}\Psi_{p_-p_+}.
		   	\label{eq:relative-chain-final-flow}
		   \end{equation}
		   This endpoint-reduction identity is the structural reason for several
		   cancellations encountered in the main text.  In particular, it explains
		   why the auxiliary endpoints introduced when decomposing a purified
		   subsystem into boundary and purifying-surface pieces disappear from the
		   final answer.  Their cancellation is not an accidental algebraic
		   simplification; it follows from the fact that the complete purified
		   subsystem is an oriented relative chain and that the PEE source integral
		   is an exact derivative in endpoint space.

		   The different constructions in the main text are special cases of the
		   same endpoint formula.
		   
		   \paragraph{AdS/BCFT.}
		   
		   For an interval adjacent to the BCFT boundary in the brane-ending RT
		   phase, the extended subsystem is
		   \begin{equation}
		   	\mathcal R_A=A\cup I_A,
		   \end{equation}
		   where $I_A$ is the corresponding brane component assigned to $A$. In a doubly holographic interpretation this component may be called an island segment. Its
		   relative boundary consists of the asymptotic endpoint of the RT surface
		   and its endpoint $q$ on the EOW brane.  The normalized stream function is
		   therefore an oriented difference between a Busemann function associated with the asymptotic endpoint and the geodesic distance to $q$.
		   The explicit Poincaré and BTZ stream functions found in the main text are
		   coordinate representations of this statement.  Their apparently
		   different functional forms arise only from the different expressions for
		   Busemann and distance functions in the two geometries.
		   The location of $q$ is fixed by the AdS/BCFT extremality condition, or
		   equivalently by orthogonality of the RT surface to the EOW brane.  Once
		   $q$ has been fixed, the distance-difference potential constructs the
		   corresponding max flow without any further optimization.
		   
		   \paragraph{Adjacent intervals in the minimal purification.}
		   
		   For adjacent intervals, the EWCS has one endpoint at the common
		   asymptotic boundary point and one endpoint $P_m$ on the purifying RT
		   surface $\gamma_{AB}$.  The extended purified subsystem
		   $A^\sharp=A\cup \CE_A$ is a relative chain whose auxiliary endpoint on
		   $\gamma_{AB}$ cancels against the corresponding endpoint of the boundary
		   contribution.  Its stream function is consequently the oriented
		   difference between the boundary Busemann function and the finite distance
		   to $P_m$.
		   This explains directly why the explicit current obtained in the main
		   text depends only on the two endpoints of the EWCS and not on the
		   auxiliary endpoint used to parametrize the split of $\gamma_{AB}$.
		   
		   \paragraph{Disjoint intervals in the minimal purification.}
		   
		   For disjoint intervals with a connected entanglement wedge, the candidate
		   EWCS connects two finite points $P_1$ and $P_2$ lying on the two
		   components of $\gamma_{AB}$.  All asymptotic and auxiliary RT-surface
		   endpoints cancel in the complete purified subsystem.  One is therefore
		   left with
		   \begin{equation}
		   	\Psi_{P_1P_2}(X)
		   	=
		   	\frac{1}{2}
		   	\left[
		   	\CL(X,P_1)-\CL(X,P_2)
		   	\right],
		   	\label{eq:disjoint-general-distance-potential}
		   \end{equation}
		   up to an overall sign.  The explicit result derived in the main text is
		   the Poincaré-coordinate realization of
		   \eqref{eq:disjoint-general-distance-potential}.
		   The same construction applies in a BTZ entanglement wedge.  The only
		   change is that the distances are evaluated using the BTZ spatial metric
		   and within the geodesic chamber corresponding to the chosen homotopy
		   class.
		   
		   An important distinction is obscured if one focuses only on the final
		   coordinate expressions.  The distance-difference potential proves that,
		   for any admissible pair of endpoints $(P_1,P_2)$, the geodesic segment
		   joining them possesses a norm-bounded divergenceless flow which saturates
		   on that segment.  It does not by itself determine which pair
		   $(P_1,P_2)$ realizes the physical minimal purification.
		   
		   For adjacent intervals, one endpoint of the candidate EWCS is fixed at
		   the common boundary point, while the other is allowed to vary on
		   $\gamma_{AB}$.  The purification optimization is
		   \begin{equation}
		   	E_W(A:B)
		   	=
		   	\frac{1}{4G}
		   	\min_{P\in\gamma_{AB}}
		   	\CL(p_{\mathrm{adj}},P).
		   	\label{eq:adjacent-purification-optimization}
		   \end{equation}
		   Its stationary condition requires the minimizing geodesic to meet
		   $\gamma_{AB}$ orthogonally.
		   
		   For disjoint intervals, both endpoints vary on the two RT components,
		   and
		   \begin{equation}
		   	E_W(A:B)
		   	=
		   	\frac{1}{4G}
		   	\min_{\substack{
		   			P_1\in\gamma_{AB}^{(1)}\\
		   			P_2\in\gamma_{AB}^{(2)}
		   	}}
		   	\CL(P_1,P_2).
		   	\label{eq:disjoint-purification-optimization}
		   \end{equation}
		   At the minimum, the EWCS is the common perpendicular to the two RT
		   components.
		   
		   Thus the calculation naturally separates into two logically distinct
		   steps: choose a splitting of the purifying surface and hence $(P_1,P_2)$, construct the associated distance-difference max flow, then minimize the calibrated bottleneck length over the allowed splitting. The first step is a bit-thread problem, whereas the last is the
		   purification optimization.
		   
		   \subsection*{Integral curves and regularity}
		   \label{app:distance-potential-nesting}
		   
		   Because the current is a rotated gradient,
		   \begin{equation}
		   	v^\mu\nabla_\mu\Psi_{p_-p_+}=0,
		   	\label{eq:flow-tangent-level-sets}
		   \end{equation}
		   its integral curves are the level sets
		   \begin{equation}
		   	\Psi_{p_-p_+}(X)=\Psi_0.
		   	\label{eq:distance-difference-level-sets}
		   \end{equation}
		   For two finite foci, these are hyperbolic loci of constant distance
		   difference.  With one ideal focus, they are the corresponding
		   Busemann--distance loci.  They are the hyperbolic analogues of
		   Apollonius curves and are not generically geodesics. This observation resolves a possible conceptual tension. The elementary PEE threads entering the superposition are geodesics. The integral curves of the superposed current, however, are level sets of a sum or difference of endpoint functions. There is no general reason for such level sets to be geodesics. Geodesic streamlines arise only in special symmetric limits.
		   
		   A sufficient condition for a global regular non-crossing foliation can also be stated
		   geometrically.  Suppose the relevant entanglement wedge or the truncated AdS
		   manifold in AdS/BCFT $\mathcal W$ is geodesically convex and the complete
		   geodesic supporting $\Sigma_\star$ intersects $\mathcal W$ precisely in the segment between $p_-$ and $p_+$.  A critical point of $\Psi$ would obey
		   \begin{equation}
		   	\nabla\rho_{p_-}=\nabla\rho_{p_+}.
		   	\label{eq:distance-potential-critical-point}
		   \end{equation}
		   By uniqueness of geodesics, such a point must lie on one of the two
		   extensions of the supporting geodesic beyond the foci.  These extensions
		   are outside $\mathcal W$ by the convexity assumption.  Therefore, we have $\mathrm{d}\Psi\neq0$ throughout $\mathcal W$.
		   The implicit-function theorem then guarantees that every level set
		   \eqref{eq:distance-difference-level-sets} is a smooth curve in the open
		   wedge.  Distinct level sets cannot intersect because $\Psi$ is
		   single-valued. Hence the integral curves form a regular non-crossing
		   foliation. The argument extends to an asymptotic boundary endpoint by taking the appropriate Busemann limit. It also extends to a BTZ exterior after choosing a
		   geodesic chamber in which the relevant geodesics are unique. This convexity criterion is sufficient, not necessary. In particular, a nonzero-tension EOW brane is not generally a totally geodesic boundary, so the criterion does not automatically establish regularity for every AdS/BCFT chamber; those cases require the direct critical-point checks performed in the main text.
		   
		   \subsection*{Domain of validity}
		   \label{app:distance-potential-validity}
		   
		   The distance-difference construction is local to a chosen geodesic
		   chamber.  Several qualifications are therefore important.
		   
		   \begin{itemize}
		   	\item First, in a quotient geometry there may be multiple geodesics connecting
		   	the same pair of endpoints.  A separate distance function must be chosen
		   	for each homotopy class.  The physical branch is selected by comparing
		   	the corresponding extremal lengths.
		   	\item Second, at a phase transition where two geodesic branches become
		   	degenerate, the globally minimal distance is generally only piecewise
		   	smooth.  The potential should then be constructed separately in each
		   	phase.  The flow is smooth in the interior of each chamber but need not
		   	admit a single differentiable continuation across the transition locus.
		   	\item Third, the potential calibrates the geodesic segment determined by its
		   	two foci.  In the minimal-purification problem, the endpoints themselves
		   	remain variational data until the EWCS extremization has been performed.
		   	\item Finally, the distance-difference current is a canonical representative
		   	selected by the PEE-thread superposition, but it need not be the unique
		   	max flow.  Bit-thread max flows are generally non-unique.  The special
		   	role of \eqref{eq:distance-difference-flow} is that its dependence on the
		   	extended subsystem is completely encoded by the relative boundary
		   	$p_+-p_-$ and that all of its defining properties follow directly from
		   	the eikonal equations.
		   \end{itemize}
		   In this sense, the detailed AdS/BCFT and minimal-purification
		   constructions of the main text are manifestations of a single geometric
		   principle: after all internal source endpoints have cancelled, the
		   coarse-grained PEE current is the Hodge dual of an oriented difference
		   of endpoint distances.

		   \paragraph{Relation to the Fermi normal representative.}
		   The endpoint-difference current is not the only way to calibrate the geodesic joining the surviving foci.  Appendix~\ref{app:fermi-extended-flows} introduces Fermi normal coordinates around the same geodesic and constructs the independent normal congruence $v_{\rm geo}=(4G)^{-1}\operatorname{sech}\lambda\,\partial_\lambda$.  In the ideal--ideal case the two constructions coincide.  In finite--ideal and finite--finite cases they agree on the bottleneck but generically differ away from it.  The distance-difference current should therefore be viewed as the representative selected by the endpoint-resolved PEE superposition, while the Fermi current is selected by normal geodesic propagation.

\section[Geodesics in locally AdS3 geometries]{Geodesics in locally AdS$_3$ geometries}
		   \label{app:geodesics}
		   In this appendix, we collect several geodesic formulae used as
		   independent checks of the intrinsic coordinate derivations in the main
		   text.  A locally AdS$_3$ geometry may be represented as a quotient of the
		   unit hyperboloid in $\mathbb{R}^{2,2}$ with metric
		   \begin{align}
		   	\d s^2=\eta_{AB}\,\d X^A\d X^B~~,~~
		   	\eta=\textrm{diag}(-1,-1,1,1)\,,
		   \end{align}
		   subject to the quadratic constraint $X^2=-1$.  If $X_a^A$ and
		   $X_b^A$ are null representatives of two boundary points, the
		   boundary-anchored geodesic has the parametrization
		   \begin{align}
		   	X_{ab}^A(\lambda)=\frac{X_a^Ae^{-\lambda}+X_b^Ae^\lambda}{\sqrt{-2X_a\cdot X_b}}
		   \end{align}
		   where $\lambda$ is proper length along the geodesic.  The null
		   representatives do not themselves obey the unit-hyperboloid constraint,
		   and their separation does not define a finite bulk distance.  By
		   contrast, for two finite bulk points $U,V$ satisfying
		   $U^2=V^2=-1$, their spacelike geodesic distance obeys
		   \begin{align}
		   	\CL(U,V)=\textrm{arccosh}\left(-U\cdot V\right)\,.
		   	\label{length-AdS}
		   \end{align}
		   Boundary-anchored lengths quoted below are obtained by applying this
		   formula to cutoff bulk representatives and then expanding in the
		   cutoff.
\subsection*{Poincar\'e AdS$_3$}
		   For example, the embedding coordinates for Poincar\'e AdS$_3$ with metric
		   \begin{align}
		   	\d s^2=\frac{-\d t^2+\d x^2+\d z^2}{z^2}
		   \end{align}
		   are given by
		   \begin{align}
		   	X_0=\frac{1+x^2-t^2+z^2}{2 z}~~&,~~X_1=\frac{t}{z}\,,\notag\\
		   	X_2=\frac{1-x^2+t^2-z^2}{2 z}~~&,~~X_4=\frac{x}{z}\,.
		   \end{align}
		   The geodesic length between two arbitrary bulk points $(x_a,t_a,z_a)$ and $(x_b,t_b,z_b)$ is obtained from \eqref{length-AdS} as follows
		   \begin{align}
		   	\CL_{ab}=\textrm{arccosh}\left[\frac{(x_a-x_b)^2-(t_a-t_b)^2+z_a^2+z_b^2}{2z_az_b}\right]\label{Poincare-length}
		   \end{align}
		   \subsection*{BTZ black brane}
		   As an independent check only, the embedding coordinates for the BTZ
		   black brane with metric \eqref{BTZ-metric} are
		   \begin{align}
		   	X_0=\frac{\sqrt{z_h^2-z^2}}{z}\sinh\left(\frac{t}{z_h}\right)~~&,~~X_1=\frac{z_h}{z}\cosh\left(\frac{x}{z_h}\right)\,,\notag\\
		   	X_2=\frac{\sqrt{z_h^2-z^2}}{z}\cosh\left(\frac{t}{z_h}\right)~~&,~~X_4=\frac{z_h}{z}\sinh\left(\frac{x}{z_h}\right)\,.
		   \end{align}
		   The geodesic length between two arbitrary bulk points $(x_i,t_i,z_i)$ and $(x_j,t_j,z_j)$ may be computed through the expression
		   \begin{align}
		   	\CL_{ij}
		   	&=\textrm{arccosh}\left[\frac{z_h^2}{z_i z_j}\cosh\left(\frac{x_i-x_j}{z_h}\right)-\frac{\sqrt{\left(z_h^2-z_i^2\right)\left(z_h^2-z_j^2\right)}}{z_iz_j}\cosh\left(\frac{t_i-t_j}{z_h}\right)\right]\label{length-BTZ}
		   \end{align}
		   For example, the geodesic length between two boundary points $z_i=z_j=\epsilon$ on a time slice ($t_i=t_j$) of the BTZ geometry, the geodesic length is given by
		   \begin{align}
		   	\CL(x_i,x_j)=\textrm{arccosh}\left[\frac{z_h^2}{\epsilon^2}\left(\cosh\left(\frac{x_i-x_j}{z_h}\right)-1\right)\right]=2 \log\left[\frac{2z_h}{\epsilon}\sinh\left(\frac{x_i-x_j}{2z_h}\right)\right]+\CO\left(\epsilon^2\right)\,.\label{length-global}
		   \end{align}
		   \paragraph{Geodesics on a fixed time slice:} The geodesics on a fixed time slice of the BTZ black brane can be found by solving the Euler--Lagrange equations. From \eqref{BTZ-metric}, a first integral of motion reads
		   \begin{align}
		   	z\sqrt{\displaystyle 1+\frac{(z^\prime)^2}{1-\frac{z^2}{z_h^2}}}=\textrm{const}.
		   \end{align}
		   Here and only in this paragraph, the prime denotes
		   $z^\prime=\dd z/\dd x$. We choose the exterior branch
		   $0<z<z_h$ with the positive square root
		   $\sqrt{1-z^2/z_h^2}\geq0$. The remaining sign of $\dd z/\dd x$
		   distinguishes the two sides of a turning point and is fixed by the
		   endpoint ordering.
		   Solving the above differential equation, we may obtain the following profile for geodesics in the constant time-slice,
		   \begin{align}
		   	\sqrt{1-\frac{z^2}{z_h^2}}=\frac{1}{2 z_h}\left(e^{\frac{x-x_0}{z_h}}+(z_h^2-\hat c^2) e^{-\frac{x-x_0}{z_h}}\right)\label{general-solution-geodesics}
		   \end{align}
		   For example, for a geodesic anchored at two boundary points $x_1$ and $x_2$, we may find
		   \begin{align}
		   	\hat c=z_h \tanh\left(\frac{x_2-x_1}{z_h}\right)~~,~~x_0=\frac{x_1+x_2}{2}+z_h\log\left[\frac{1}{z_h}\cosh\left(\frac{x_1-x_2}{2z_h}\right)\right]
		   \end{align}
		   leading to the following profile for the RT surface homologous to the subsystem $[x_1,x_2]$
		   \begin{align}
		   	\sqrt{1-\frac{z^2}{z_h^2}}=\frac{\cosh\left(\frac{x-\frac{x_1+x_2}{2}}{z_h}\right)}{\cosh\left(\frac{x_2-x_1}{2z_h}\right)}\label{geod-class-I}
		   \end{align}
		   On the other hand, for the geodesics emanating from the asymptotic boundary at spatial location $x_1$ and reaching the event horizon at $x_2$, we may find
		   \begin{align}
		   	\hat c=z_h\coth\left(\frac{x_1-x_2}{z_h}\right)~~,~~x_0=x_2+z_h\log\left[\frac{1}{z_h}\sinh\left(\frac{x_1-x_2}{z_h}\right)\right]\,.
		   \end{align}
		   leading to
		   \begin{align}
		   	\sqrt{1-\frac{z^2}{z_h^2}}=\frac{\sinh\left(\frac{x-x_2}{z_h}\right)}{\sinh\left(\frac{x_1-x_2}{z_h}\right)}\label{geod-class-II}
		   \end{align}
		   
		   \section{Two-point PEE and entanglement entropy}\label{appB}
		   \subsection*{Poincar\'e AdS$_3$: disconnected phase}
		   Here we list the two-point PEE contributions relevant to the
		   entanglement entropy of $A=[a,b]$ in the disconnected phase of
		   Poincar\'e AdS$_3$.  Whenever an integration limit is reversed, the
		   integrand is the oriented kernel $\widehat{\CI}$ introduced in
		   section~\ref{sec:two-point-pee}. The corresponding integrated quantity
		   is denoted by $\widehat{\CI}(X,Y)$ and is an oriented flux
		   contribution, not a negative physical PEE.
		   	\begin{align}
		   	\CI(A,B_1\cup B_2)=\int_{a}^{b}\d x\left(\int_{0}^{a}+\int_{b}^{\infty}\right)\d y\,\CI_{\del\del}(x,y)=\frac{1}{4G}\log\left[\frac{a(b-a)^2}{\epsilon^2b}\right]\,,
		   \end{align}
		   \begin{align}
		   	\widehat{\CI}(A,I_{B_1}\cup I_{B_2})=\int_{a}^{b}\d x\left(\int^{0}_{a}+\int^{b}_{\infty}\right)\d y\,\widehat{\CI}_{\del \CQ}(x,y)=\frac{1}{4G}\log\left[\frac{2b^2(1+\sin\theta)}{a^2+b^2+2ab\sin\theta}\right]\,,
		   \end{align}
		   \begin{align}
		   	\widehat{\CI}(I_A,B_1\cup B_2)=\int_{b}^{a}\d x\left(\int_{0}^{a}+\int_{b}^{\infty}\right)\d y\,\widehat{\CI}_{\CQ\del}(x,y)=\frac{1}{4G}\log\left[\frac{2b^2(1+\sin\theta)}{a^2+b^2+2ab\sin\theta}\right]\,,
		   \end{align}
		   \begin{align}
		   	\widehat{\CI}(I_A,I_{B_1}\cup I_{B_2})=\frac{1}{4G}\left(\textrm{arccosh}\left[1+\frac{(b-a)^2}{2ab}\sec\theta^2\right]+\log\left(\frac{a}{b}\right)\right)
		   \end{align}
		   {\footnotesize
		   \begin{align}
		   	&\widehat{\CI}(B_1\cup I_{B_1},B_2\cup I_{B_2})\notag\\
		   	&=\widehat{\CI}(B_1,B_2)+\widehat{\CI}(B_1,I_{B_2})
		   	+\widehat{\CI}(B_2,I_{B_1})+\widehat{\CI}(I_{B_1},I_{B_2})\notag\\
		   	&=\int_{0}^{a}\d x\int_{b}^{\infty}\d y\,\CI_{\del\del}(x,y)
		   	+\int_{0}^{a}\d x\int_{\infty}^{b}\d y\,\widehat{\CI}_{\del \CQ}(x,y)
		   	+\int_{b}^{\infty}\d x\int_{a}^{0}\d y\,\widehat{\CI}_{\CQ\del}(x,y)
		   	+\int_{a}^{0}\d x\int_{\infty}^{b}\d y\,\widehat{\CI}_{\CQ\CQ}(x,y)\notag\\
		   	&=\frac{1}{4G}\log\left(\frac{b}{b-a}\right)
		   	+\frac{1}{4G}\log\left(\frac{a^2+b^2+2a b \sin\theta}{b^2}\right)\notag\\
		   	&\quad+\frac{1}{8G}\left(-\textrm{arccosh}\left[1+\frac{(b-a)^2}{2ab}\sec\theta^2\right]
		   	+\log\left(\frac{b\sec^2\theta}{a}\right)\right)
		   \end{align}
		   }
		   It is instructive to compare the following combinations with \eqref{EE}:
		   {\small
		   \begin{align}
		   	&\CI(A,B)+2\CI(B_1,B_2)=\frac{1}{4G}\log\left(\frac{ab}{\epsilon^2}\right),\notag\\
		   	&\widehat{\CI}(A,I_{B_1}\cup I_{B_2})+\widehat{\CI}(B_1\cup B_2,I_A)
		   	+2\widehat{\CI}(B_1,I_{B_2})+2\widehat{\CI}(B_2,I_{B_1})\notag\\
		   	&\hspace{1cm}=\frac{1}{2G}\log\left[2(1+\sin\theta)\right]\\
		   	&\widehat{\CI}(I_A,I_{B_1}\cup I_{B_2})+2\widehat{\CI}(I_{B_1},I_{B_2})=\frac{1}{2G}\log\sec\theta\notag
		   \end{align}
		   }
		   Adding these three lines gives
		   \begin{align}
		   	S_A^{\rm disc}
		   	&=\frac{1}{4G}\log\left(\frac{4ab}{\epsilon^2}\right)
		   	+\frac{1}{2G}\log\left(\sec\theta+\tan\theta\right),
		   \end{align}
		   in exact agreement with \eqref{EE}. The factor of four in the final
		   entropy is therefore generated only after the mixed and brane--brane
		   sectors are included; it is not present in the first line by itself.
		   \subsection*{BTZ black brane}
		   We now list the contributions to the entanglement entropy of
		   $A=[0,b]$ in the thermal BCFT$_2$ dual to the BTZ black brane.  The
		   formulas in this subsection use the $\sigma=-1$ brane branch. In
		   integrals over the brane, $x$ and $y$ denote the ambient BTZ spatial
		   coordinate restricted to the brane; this is why the endpoint limits are
		   $x_\CQ$ and $x_h$. The
		   brane-coordinate Jacobian is already included in the oriented kernels
		   displayed in section~\ref{sec:PEE-BTZ}. As above, every reversed integral
		   uses an oriented kernel.
		   \begin{align}
		   	\widehat{\CI}(A,B)=\int_{0}^{b}\d x\int_{b}^{\infty}\d y\,\widehat{\CI}_{\del\del}(x,y)=\frac{1}{8G}\left(2 \log \left[\frac{2 z_h}{\epsilon}\sinh \left(\frac{b}{2 z_h}\right)\right]-\frac{b}{z_h}\right)
		   \end{align}
		   \begin{align}
		   	\widehat{\CI}(A,h)=\int_{0}^{b}\d x\int_{\infty}^{x_h}\d y\,\widehat{\CI}_{\del h}(x,y)=\frac{1}{8G}\left(\log \left[\cosh \left(\frac{b}{z_h}\right)-\frac{\sinh \left(\frac{b}{z_h}\right)}{\sqrt{1+\kappa^2}}\right]+\frac{b}{z_h}\right)
		   \end{align}
		   {\small
		   \begin{align}
		   	\widehat{\CI}(A,I_B)
		   	&=\int_{0}^{b}\d x\int_{x_h}^{x_\CQ}\d y\,\widehat{\CI}_{\del \CQ}(x,y)\notag\\
		   	&=\frac{1}{8G}\log \left[\frac{\left(-1+\sqrt{1+\kappa^2}\right) \left(\cosh \left(\frac{b}{z_h}\right)+1\right)}
		   	{\sqrt{\kappa ^2+1} \cosh \left(\frac{b}{z_h}\right)-\sinh \left(\frac{b}{z_h}\right)}\right]
		   \end{align}
		   \begin{align}
		   	\widehat{\CI}(I_A,B)&=\int_{x_\CQ}^{0}\d x\int_{b}^{\infty}\d y\,\widehat{\CI}_{\CQ\del}(x,y)\notag\\
		   	&=\frac{1}{8G}\log \left[\frac{\left(-1+\sqrt{1+\kappa ^2}\right) \left(1+\sech\left(\frac{b}{z_h}\right)\right)}
		   	{\sqrt{\sech^2\left(\frac{b}{z_h}\right)+\kappa ^2}}\right]\notag\\
		   	&\quad+\frac{1}{8G}\textrm{arctanh}\left[\frac{1}{\sqrt{1+\kappa^2}}\tanh\left(\frac{b}{z_h}\right)\right]
		   \end{align}
		   \begin{align}
		   	\widehat{\CI}(I_A,h)&=\int_{x_\CQ}^{0}\d x\int_{\infty}^{x_h}\d y\,\widehat{\CI}_{\CQ h}(x,y)\notag\\
		   	&=\frac{1}{8G}\log \left[\frac{2\sqrt{1+\kappa ^2}}{\kappa ^2}
		   	\left(\sqrt{\kappa ^2+1} \coth \left(\frac{b}{z_h}\right)-1\right)\right]\notag\\
		   	&\quad-\frac{1}{8G}\textrm{arccosh}\left[
		   	\frac{\left(\kappa ^2+1\right) \coth \left(\frac{b}{z_h}\right)-1}{\kappa ^2}\right]
		   \end{align}
		   }
		   \begin{align}
		   	\widehat{\CI}(I_A,I_B)&=\int_{x_\CQ}^{0}\d x\int_{x_h}^{x_\CQ}\d y\,\widehat{\CI}_{\CQ\CQ}(x,y)\notag\\&=\frac{1}{8G}\cosh ^{-1}\left[\frac{\left(\kappa ^2+1\right) \coth \left(\frac{b}{z_h}\right)-1}{\kappa ^2}\right]+\frac{1}{8G}\log \left[\tanh \left(\frac{b}{2 z_h}\right)\right]
		   \end{align}
		   Summing the six pairwise terms gives the selected-branch entropy:
		   \begin{align}
		   	&\widehat{\CI}(A,B)+\widehat{\CI}(A,h)+\widehat{\CI}(A,I_B)+\widehat{\CI}(I_A,B)
		   	+\widehat{\CI}(I_A,h)+\widehat{\CI}(I_A,I_B)\notag\\
		   	&\hspace{1cm}=
		   	\frac{1}{4G}\log\left[\frac{2z_h}{\epsilon}
		   	\sinh\left(\frac{b}{z_h}\right)\right]
		   	-\frac{1}{4G}\operatorname{arcsinh}\left(\frac{1}{\kappa}\right)
		   	=S_A\big|_{\sigma=-1}\,.
		   \end{align}


\section{Fermi normal coordinates for extended holographic flows}
\label{app:fermi-extended-flows}

The distance--difference construction of appendix~\ref{app:distance-difference-potential}
provides a coordinate-independent description of the PEE-selected flows used
throughout the main text.  In this appendix we place the same flows in Fermi
normal coordinates adapted to their geodesic bottleneck.  This serves two
purposes.  First, it gives a universal form for an independent local
\emph{normal-geodesic} calibration/comparator.  Second, it makes precise why
this comparator agrees with the PEE-selected flow for the ordinary
ideal--ideal vacuum interval but is generically different when an endpoint of
the bottleneck lies at a finite bulk location, as happens in AdS/BCFT and in
the minimal-purification construction.

Throughout this appendix the AdS radius is set to one.  The discussion is local
to a fixed geodesic chamber of a static slice of locally AdS$_3$, and therefore
applies directly to the Poincar\'e and planar-BTZ geometries considered in the
main text.  A finite-cutoff realization of the same comparison between the
PEE-selected flow and the normal-geodesic comparator is developed in
\cite{BasuWenChandra}.

\subsection*{Universal Fermi chart around a geodesic bottleneck}
\label{app:fermi-chart}

Let $\gamma_\star$ be the complete geodesic containing the connected bottleneck under consideration, either an RT segment or a candidate EWCS.  Parametrize $\gamma_\star$ by proper length $\eta$ and denote by $P(\eta)$ its embedding-space position.  If $N$ is a unit spacelike normal to the plane containing $\gamma_\star$, the normal exponential map may be written as
\begin{equation}
 X(\eta,\lambda)
 =\cosh\lambda\,P(\eta)+\sinh\lambda\,N,
 \label{eq:fermi-embedding}
\end{equation}
where changing $N\to-N$ only reverses the sign of $\lambda$.  The induced metric
is the standard Fermi form
\begin{equation}
 \mathrm ds^2=\mathrm d\lambda^2+\cosh^2\!\lambda\,\mathrm d\eta^2\,.
 \label{eq:fermi-metric}
\end{equation}
The bottleneck itself is $\lambda=0$, while every curve $\eta=\text{constant}$
is a unit-speed geodesic orthogonal to it.  For a complete geodesic in
$\mathbb H^2$ this chart is global; in the physical AdS/BCFT or entanglement
wedge geometry we simply restrict it to the appropriate domain and homology
chamber.

A vector field tangent to the normal geodesics has the form
\begin{equation}
 v_{\rm geo}=\frac{1}{4G}\,f(\eta,\lambda)\,\partial_\lambda .
 \label{eq:fermi-normal-ansatz}
\end{equation}
Its divergence is
\begin{equation}
 \nabla_\mu v_{\rm geo}^{\mu}
 =\frac{1}{4G\cosh\lambda}
 \partial_\lambda\!\left(\cosh\lambda\,f\right).
 \label{eq:fermi-normal-divergence}
\end{equation}
Demanding divergencelessness and pointwise saturation on the bottleneck,
$f(\eta,0)=1$, fixes
\begin{equation}
 v_{\rm geo}
 =\frac{1}{4G}\operatorname{sech}\lambda\,\partial_\lambda\,.
 \label{eq:fermi-universal-normal-flow}
\end{equation}
Thus the familiar dilution of a normal bit-thread bundle is simply the
Jacobian of the normal exponential map.  Indeed, a transverse strip has
proper width $\cosh\lambda\,\mathrm d\eta$, and therefore carries the constant
flux
\begin{equation}
 \mathrm d\Phi
 =\lvert v_{\rm geo}\rvert\cosh\lambda\,\mathrm d\eta
 =\frac{\mathrm d\eta}{4G}.
 \label{eq:fermi-flux-strip}
\end{equation}
Equation~\eqref{eq:fermi-universal-normal-flow} is unique after one has fixed
the normal-geodesic congruence and imposed saturation on $\gamma_\star$; it is
not a uniqueness statement for the full bit-thread max-flow problem.

\subsection*{Distance--difference flows in the same chart}
\label{app:fermi-distance-difference}

We now rewrite the endpoint construction of
appendix~\ref{app:distance-difference-potential} in the Fermi chart.  If a finite focus
$p_i$ lies on $\gamma_\star$ at Fermi coordinate $\eta_i$, its distance from a
generic point $(\eta,\lambda)$ is
\begin{equation}
 \CL_i(\eta,\lambda)
 =\operatorname{arccosh}\!\left[
 \cosh\lambda\,\cosh(\eta-\eta_i)
 \right].
 \label{eq:fermi-finite-distance}
\end{equation}
Writing
\begin{equation}
 R_i
 =\sinh \CL_i
 =\sqrt{\cosh^2\!\lambda\,\cosh^2(\eta-\eta_i)-1},
 \label{eq:fermi-Ri}
\end{equation}
one finds
\begin{align}
 \partial_\lambda \CL_i
 &=\frac{\sinh\lambda\,\cosh(\eta-\eta_i)}{R_i},
 &
 \partial_\eta \CL_i
 &=\frac{\cosh\lambda\,\sinh(\eta-\eta_i)}{R_i}.
 \label{eq:fermi-distance-derivatives}
\end{align}
For the two ideal endpoints of the complete geodesic, Busemann functions may
be chosen, up to additive constants, as
\begin{equation}
 \beta_+=\log\cosh\lambda-\eta,
 \qquad
 \beta_-=\log\cosh\lambda+\eta,
 \qquad
 \lvert\nabla\beta_\pm\rvert=1.
 \label{eq:fermi-busemann}
\end{equation}
Let $F_\pm$ denote the appropriate endpoint function: a finite distance $\CL_i$
for a finite endpoint, or a Busemann function $\beta_\pm$ for an ideal one.
The PEE-selected stream function is
\begin{equation}
 \Psi=\frac12\left(F_+-F_-\right),
 \qquad
 v_{\rm PEE}=\frac{1}{4G}\star\mathrm d\Psi,
 \label{eq:fermi-pee-potential}
\end{equation}
with the overall sign fixed by the desired orientation.  With volume form
$\cosh\lambda\,\mathrm d\lambda\wedge\mathrm d\eta$, this is
\begin{equation}
 v_{\rm PEE}
 =\frac{1}{4G\cosh\lambda}
 \left(
 -\partial_\eta\Psi\,\partial_\lambda
 +\partial_\lambda\Psi\,\partial_\eta
 \right).
 \label{eq:fermi-pee-vector}
\end{equation}
The eikonal equations immediately give
\begin{equation}
 4G\lvert v_{\rm PEE}\rvert
 =\lvert\nabla\Psi\rvert
 \leq\frac12\left(\lvert\nabla F_+\rvert+\lvert\nabla F_-\rvert\right)=1.
 \label{eq:fermi-pee-norm}
\end{equation}
On the geodesic segment joining the two endpoint foci the corresponding unit
gradients are antiparallel.  Equivalently, in the Fermi chart
$\partial_\lambda\Psi=0$ and $\lvert\partial_\eta\Psi\rvert=1$ on
$\lambda=0$ between the foci.  Hence the current is normal to the bottleneck
and saturates it pointwise.

Equations~\eqref{eq:fermi-universal-normal-flow} and
\eqref{eq:fermi-pee-vector} therefore provide two natural local calibrations
of the same geodesic bottleneck.  Equation~\eqref{eq:fermi-universal-normal-flow}
becomes a global admissible max flow only after one verifies that the chosen normal
bundle remains in the physical homology chamber, can be completed to the required
source/sink support, and obeys the boundary conditions on every EOW-brane, horizon,
or purifying-surface component.  We do not infer these global properties from the
local Fermi formula alone.  Their difference is entirely
visible in the off-bottleneck component $\partial_\lambda\Psi$.  The three
endpoint classes relevant to the main text make this particularly clear.

\paragraph{Ideal--ideal endpoints.}
For the ordinary vacuum interval, both bottleneck endpoints are ideal.  From
\eqref{eq:fermi-busemann},
\begin{equation}
 \Psi_{\rm ii}=\frac12(\beta_+-\beta_-)=-\eta,
 \label{eq:fermi-ideal-ideal-potential}
\end{equation}
up to an orientation and an additive constant.  Therefore
\begin{equation}
 v_{\rm PEE}=\frac{1}{4G}\operatorname{sech}\lambda\,\partial_\lambda
 =v_{\rm geo}
 \label{eq:fermi-ideal-ideal-coincidence}
\end{equation}
(up to orientation).  The geodesicity of the familiar vacuum bit-thread
streamlines is thus a special consequence of having two ideal foci.

\paragraph{Finite--ideal endpoints.}
If the finite focus is at $\eta_0$ and the ideal focus is the $+$ endpoint,
one may take
\begin{equation}
 \Psi_{\rm fi}
 =\frac12\left(\beta_+-\CL_0\right).
 \label{eq:fermi-finite-ideal-potential}
\end{equation}
Then
\begin{equation}
 \partial_\lambda\Psi_{\rm fi}
 =\frac12\left[
 \tanh\lambda
 -\frac{\sinh\lambda\,\cosh(\eta-\eta_0)}{R_0}
 \right],
 \label{eq:fermi-finite-ideal-tangential}
\end{equation}
which vanishes on the bottleneck but not at a generic bulk point.  The
PEE-selected current therefore develops a nonzero $\partial_\eta$ component
away from $\lambda=0$, whereas the normal representative
\eqref{eq:fermi-universal-normal-flow} remains tangent to the normal geodesics.
This is precisely the endpoint class relevant both to a brane-ending RT
surface in AdS/BCFT and to the adjacent-interval EWCS in the minimal
purification.

\paragraph{Finite--finite endpoints.}
If the two finite foci are located at $\eta_1$ and $\eta_2$, the potential is
\begin{equation}
 \Psi_{\rm ff}
 =\frac12\left(\CL_2-\CL_1\right).
 \label{eq:fermi-finite-finite-potential}
\end{equation}
Again $\partial_\lambda\Psi_{\rm ff}$ is generically nonzero away from the
bottleneck.  This is the class relevant to a connected disjoint-interval
minimal purification, where the EWCS joins two finite points on the two RT
components.  Thus a finite endpoint is the geometric ingredient that
separates the PEE calibration from the normal-geodesic calibration.

\subsection*{Poincar\'e AdS/BCFT: the brane endpoint as a finite Fermi focus}
\label{app:fermi-bcft}

Consider the adjacent interval $A=[0,b]$ in the vacuum AdS/BCFT geometry of
section~4.1.  The RT geodesic is
\begin{equation}
 x^2+z^2=b^2,
 \label{eq:fermi-bcft-rt}
\end{equation}
and its finite endpoint on the EOW brane is
\begin{equation}
 q=(-b\sin\theta,b\cos\theta).
 \label{eq:fermi-bcft-q}
\end{equation}
A convenient Fermi chart adapted to the complete semicircle is
\begin{align}
 x(\eta,\lambda)
 &=b\,
 \frac{\cosh\lambda\,\sinh\eta}
 {\cosh\lambda\,\cosh\eta-\sinh\lambda},
 \nonumber\\
 z(\eta,\lambda)
 &=\frac{b}
 {\cosh\lambda\,\cosh\eta-\sinh\lambda}.
 \label{eq:fermi-bcft-map}
\end{align}
At $\lambda=0$ this reduces to
\begin{equation}
 x=b\tanh\eta,
 \qquad
 z=b\operatorname{sech}\eta.
 \label{eq:fermi-bcft-rt-param}
\end{equation}
The EOW brane $x=-z\tan\theta$ intersects the RT geodesic at
\begin{equation}
 \eta_Q=-\rho_0,
 \qquad
 \rho_0=\operatorname{arctanh}(\sin\theta)
 =\operatorname{arcsinh}(\tan\theta).
 \label{eq:fermi-bcft-rho0}
\end{equation}
The asymptotic endpoint $(b,0)$ is the ideal limit $\eta\to+\infty$.
Consequently the PEE-selected stream function obtained explicitly in the main
text,
\begin{equation}
 \Psi_{\rm BCFT}
 =\frac12\left(\Phi_b-L_q\right),
 \label{eq:fermi-bcft-main-potential}
\end{equation}
is exactly the finite--ideal potential
\eqref{eq:fermi-finite-ideal-potential}: in the present chart
\begin{equation}
 \Phi_b=\log(2b)+\log\cosh\lambda-\eta,
 \qquad
 \CL_q=\operatorname{arccosh}\!\left[
 \cosh\lambda\cosh(\eta+\rho_0)
 \right].
 \label{eq:fermi-bcft-endpoint-functions}
\end{equation}
The irrelevant constant $\log(2b)$ drops out of the current.  This gives a
coordinate-level bridge between the explicit Poincar\'e expressions in
the Poincar\'e AdS/BCFT analysis in section~4 and the endpoint-reduction argument of
appendix~\ref{app:distance-difference-potential}.

The Fermi coordinate also gives a simple interpretation of the BCFT boundary
entropy.  Regulating the asymptotic endpoint at $z=\epsilon$ gives
\begin{equation}
 \eta_\epsilon=\operatorname{arccosh}\!\left(\frac{b}{\epsilon}\right),
 \label{eq:fermi-bcft-cutoff-eta}
\end{equation}
so the regulated RT length is
\begin{align}
 \operatorname{Length}(\gamma_A)
 &=\eta_\epsilon-\eta_Q
 =\operatorname{arccosh}\!\left(\frac{b}{\epsilon}\right)+\rho_0
 \nonumber\\
 &=\log\!\left(\frac{2b}{\epsilon}\right)+\rho_0+O(\epsilon^2).
 \label{eq:fermi-bcft-rt-length}
\end{align}
Thus the familiar boundary contribution is literally the finite Fermi-length
shift of the brane endpoint along the complete RT geodesic,
\begin{equation}
 S_{\rm bdy}=\frac{\rho_0}{4G}.
 \label{eq:fermi-bcft-boundary-entropy}
\end{equation}
This is the same constant obtained from the two-point PEE decomposition in
section~3.1, now expressed directly in the geometry of the bottleneck.

Finally, the large-tension limit has an immediate interpretation.  As
$\theta\to\pi/2$, one has $\rho_0\to\infty$ and therefore
$\eta_Q\to-\infty$.  The finite brane focus is pushed to the second ideal
endpoint of the complete semicircle.  The finite--ideal potential then
becomes the ideal--ideal potential, so the PEE-selected current approaches
\eqref{eq:fermi-universal-normal-flow}.  This gives a coordinate-independent
explanation of the geodesic limit displayed in the main-text flow plots.

\subsection*{Minimal purification}
\label{app:fermi-minimal-purification}

The same Fermi classification organizes the minimal-purification examples of
section~5 without introducing any new local geometry.

\paragraph{Adjacent intervals.}
For adjacent intervals the candidate EWCS has one ideal endpoint at the
common asymptotic boundary point and one finite endpoint $P_m$ on the
purifying RT surface $\gamma_{AB}$.  After the boundary-sourced and
RT-surface-sourced currents are added, their shared auxiliary endpoint
cancels.  Appendix~\ref{app:distance-difference-potential} shows that the resulting
stream function depends only on the two EWCS endpoints; in the Poincar\'e
coordinates of the adjacent-interval analysis in section~5 it is the displayed Busemann--distance
combination.  By a hyperbolic isometry, it is therefore exactly of the
finite--ideal form \eqref{eq:fermi-finite-ideal-potential}.

The corresponding normal-geodesic comparator is always given locally by
\eqref{eq:fermi-universal-normal-flow}.  Both fields saturate the same
candidate cross section, but they need not induce the same endpoint map on
the extended purified boundary.  At the minimizing split, the EWCS meets the
purifying RT surface orthogonally; this is the usual stationarity condition
for the minimal purification and is independent of which max-flow
representative is used to calibrate the chosen cross section.

\paragraph{Disjoint intervals.}
For a connected entanglement wedge of two disjoint intervals, the candidate
EWCS joins two finite points $P_1$ and $P_2$ on the two components of
$\gamma_{AB}$.  The source-integrated PEE current reduces to the finite--finite
potential
\begin{equation}
 \Psi_{P_1P_2}
 =\frac12\left[\CL(X,P_2)-\CL(X,P_1)\right]
 \label{eq:fermi-disjoint-minpur-potential}
\end{equation}
(up to orientation), while the independent normal flow is again
\eqref{eq:fermi-universal-normal-flow} in Fermi coordinates adapted to the
geodesic $P_1P_2$.  The construction of either flow is logically distinct
from the subsequent optimization
\begin{equation}
 E_W(A:B)
 =\frac{1}{4G}
 \min_{P_1\in\gamma_{AB}^{(1)},\,P_2\in\gamma_{AB}^{(2)}}
 L(P_1,P_2),
 \label{eq:fermi-minpur-optimization}
\end{equation}
which selects the common perpendicular of the two RT components.  Fermi
coordinates therefore separate cleanly the max-flow problem for a fixed
purifying split from the purification optimization itself.

\subsection*{Planar BTZ and the local hyperbolic statement}
\label{app:fermi-btz}

A constant-time slice of the noncompact BTZ black brane is locally
$\mathbb H^2$.  Consequently every connected RT segment or EWCS lying in a
fixed geodesic chamber admits the same Fermi metric
\eqref{eq:fermi-metric}, the same normal flow
\eqref{eq:fermi-universal-normal-flow}, and the same endpoint classification
as above.  The explicit BTZ Busemann functions and finite-point distances in
sections~4.2.4 and~5.3 are simply the intrinsic BTZ-coordinate expressions
of the functions $F_\pm$ entering \eqref{eq:fermi-pee-potential}.

This local statement does not remove the global information carried by the
BTZ endpoint support.  One must still specify the appropriate exterior
component, horizon or brane endpoint, and geodesic homotopy chamber before
constructing the potential.  Once that data are fixed, however, the norm
bound and bottleneck saturation are the same hyperbolic facts as in the
Poincar\'e examples.

\subsection*{Interpretation: endpoint calibration versus normal calibration}
\label{app:fermi-interpretation}

The combined distance--difference and Fermi descriptions sharpen the role of
max-flow nonuniqueness in the constructions of this paper.  The bottleneck
geometry fixes the maximal flux and therefore fixes the common value of the
current on the RT surface or EWCS.  It does not determine how the flow is
continued away from that surface.  Two particularly natural continuations
are available:
\begin{align}
 &\text{PEE-selected:}
 &&v_{\rm PEE}=\frac{1}{4G}\star\mathrm d
 \left[\frac12(F_+-F_-)\right],
 \nonumber\\
 &\text{normal-geodesic:}
 &&v_{\rm geo}=\frac{1}{4G}\operatorname{sech}\lambda\,\partial_\lambda.
 \label{eq:fermi-two-calibrations}
\end{align}
The first retains the endpoint calibration inherited from the two-point PEE
measure and from the oriented source integration.  The second retains the
geometric normal congruence of the bottleneck.  They coincide for the
ideal--ideal vacuum interval, but a finite endpoint generically separates
them off the bottleneck.  This is why the non-geodesic streamlines found in
the AdS/BCFT and minimal-purification calculations are not a failure of the
max-flow construction: they are the characteristic curves of a different,
PEE-calibrated flow representative.

The Fermi construction should therefore be viewed as a comparator for the
PEE-selected current, not as a replacement for the endpoint-resolved
construction.  The latter retains the physical information about which
components of the extended state-supporting surface act as sources and sinks,
whereas the Fermi form isolates the local geometry of a geodesic bottleneck.

	\bibliographystyle{JHEP}
	{\footnotesize
	\bibliography{Threads}
	}
\end{document}